\documentclass[aps,prd,reprint,superscriptaddress,floatfix,twocolumn,nofootinbib]{revtex4-2}

\usepackage{color}
\usepackage{amssymb,bbold}
\usepackage{graphicx,color}
\usepackage[usenames,dvipsnames]{xcolor}
\usepackage[section]{placeins}
\usepackage{amsmath}
\usepackage{subfigure}
\usepackage[normalem]{ulem}
\usepackage{booktabs}
\usepackage{xcolor}
\usepackage{epsfig,graphics,color,graphicx,amsmath}
\colorlet{darkgreen}{green!50!black}
\colorlet{brightyellow}{yellow!75!red}
\colorlet{orange}{red!50!yellow}
\colorlet{darkblue}{blue!60!black}
\colorlet{darkred}{red!80!black}

\def\bwt{\begin{widetext}}
\def\ewt{\end{widetext}}
\newcommand{\bg}[1]{\mbox{\boldmath $#1$}}

\usepackage{soul}

\usepackage{mathtools}
\usepackage{mathrsfs}
\usepackage{bm}
\usepackage{relsize}
\usepackage{indentfirst}

\usepackage{xcolor}

\usepackage{amssymb,amsmath,multirow,epsfig,graphicx,color}

\usepackage{epsfig}
\usepackage{graphicx,amsmath}
\def\be{\begin{eqnarray} &&}
\def\nonu{\nonumber \\ &&}

\def\ee{\end{eqnarray}}
\def\psla{\slash \! \!\! }
\def\Psla{\slash \! \! \!\! }

\begin{document}

\title{Unpolarized transverse-momentum dependent  distribution functions of a quark in a pion with  Minkowskian dynamics}

\author{E.~Ydrefors}
\affiliation{Institute of Modern Physics, Chinese Academy of Sciences, Lanzhou 730000, China}
\author{W. de Paula}
\affiliation{Instituto Tecnol\'ogico de Aeron\'autica,  DCTA,
12228-900 S\~ao Jos\'e dos Campos,~Brazil}
\author{T.~Frederico}
\affiliation{Instituto Tecnol\'ogico de Aeron\'autica,  DCTA,
12228-900 S\~ao Jos\'e dos Campos,~Brazil}
\author{G. Salm\`e}
\affiliation{
INFN, Sezione di Roma, P.le A. Moro 2, 00185 Rome, Italy}

\date{\today}

\begin{abstract}
The  unpolarized  twist-2 (leading) and twist-3 (subleading),  T-even,
transverse-momentum dependent quark distributions  in the pion are 
 evaluated for the first time by using  the actual  solution of a dynamical equation in
 Minkowski space. The adopted theoretical framework is  based on the   homogeneous Bethe-Salpeter integral equation  with   an interaction kernel given by  a one-gluon exchange, featuring an extended
quark-gluon vertex. The masses of quark and gluon as well as the  interaction-vertex scale have been  chosen in a range suggested by
 lattice-QCD  calculations, and calibrated to reproduce
 both pion mass and decay constant.  The sum rules to be fulfilled by the transverse-momentum dependent distributions are carefully investigated, particularly the leading-twist one, that has to match  the collinear parton distribution function, and hence can be scrutinized in terms of existing data as well as theoretical predictions.  Noteworthy, the joint use of the Fock expansion of the pion state facilitates a more in-depth analysis of the content of the pion Bethe-Salpeter amplitude, allowing for  the first time to determine the gluon contribution to the quark average longitudinal fraction, that results to be  $\sim 6\%$.   The current analysis highlights the role of the gluon exchanges through quantitative analysis of  collinear  and  transverse-momentum  distributions,  showing, e.g. for both leading and subleading-twists, an early departure from the widely adopted exponential fall-off, for $|{\bf k}_\perp|^2> m^2$, with the quark mass $\sim \Lambda_{QCD}$.
 \end{abstract}
\maketitle
\section{Introduction}
\label{Sec:intro}

Quark transverse-momentum dependent   distribution functions (TMDs for short)  are the basic ingredients for parametrizing the hadronic quark-quark correlator (see the seminal Ref.~\cite{Tangerman:1994eh} and for the complete parametrization Ref.~\cite{Goeke:2005hb}, while Refs.~\cite{Mulders:2000sh,Boer:2003cm} for  correlators involving gluons),  and represent  direct generalization of the  parton distribution functions (PDFs), so that    both  longitudinal   and   transverse degrees of freedom (dof)  can be addressed (see, e.g., Refs.~\cite{Barone:2001sp,bacchetta2007semi} for an extensive introduction to the transverse dof and related distribution functions).  Clearly, {the} access {to the}  3D imaging of hadrons  allows  us to achieve a deeper and deeper understanding of   the non-perturbative regime of QCD, also exploiting the {non-trivial} coupling to the spin dof (see, e.g., Refs.~\cite{Anselmino:2020vlp,Constantinou:2020hdm} and references  therein). Hence, by means of TMDs, one can gather unique information on QCD at work in  hard semi-inclusive reactions (both unpolarized and polarized) at low transverse-momentum, like  low-$q_\perp$ Drell-Yan (DY) processes, vector/scalar  boson productions or
semi-inclusive deep inelastic scattering (SIDIS) (see, e.g., Refs.~\cite{Angeles-Martinez:2015sea,Avakian:2016rst,pitonyak2019transverse,Constantinou:2020hdm} for a status-report on the experimental measurements). Indeed,  the extraction  of TMDs  from the experimental cross-section is a highly challenging task, as shown by the intense theoretical work on the factorization of the cross sections into transverse-momentum dependent matrix elements (see, e.g., Refs.~\cite{Collins:1989gx,Ji:2004wu,Collins:2011zzd,Rogers:2015sqa,Ji:2004wu,Echevarria:2011epo}) and the TMDs evolution that becomes a two-scale problem, since the rapidity  $\zeta$ comes into play in addition to the renormalization scale $\mu$ (see, e.g., Ref.~\cite{Collins:1989gx,Aybat:2011zv,Echevarria:2012pw,Vladimirov:2017ksc} and Ref.~\cite{Scimemi:2019mlf} 
for a recent review that covers also the factorization).  Noteworthy, one has to mention the efforts for obtaining reliable global fits (see, e.g., Refs.~\cite{Bacchetta:2017gcc,Scimemi:2019cmh,Cammarota:2020qcw,Bacchetta:2022awv} and also Ref.~\cite{Constantinou:2020hdm} for a general discussion), early-stage lattice calculations (see, e.g., Refs.~\cite{Hagler:2009mb,Musch:2010ka,Musch:2011er,Engelhardt:2015xja,Yoon:2017qzo} and also Refs.~\cite{LatticeParton:2020uhz,Constantinou:2020hdm,Schlemmer:2021aij,Constantinou:2022yye})  and, finally, the broad set of phenomenological models, that we can only partially list:  the bag model (see, e.g., Ref.~\cite{Signal:2021aum} and references therein), covariant model (see, e.g., Ref.~\cite{Bastami:2020rxn} and  references therein), light-front (LF) constituent quark models (see, e.g., Refs.~\cite{Lorce:2016ugb,Pasquini:2018oyz}) and the basis LF quantization framework~\cite{Hu:2022ctr},  the approaches based on the Nambu-Jona-Lasinio interaction (see, e.g., Refs.~\cite{Noguera:2015iia,Ninomiya:2017ggn}),  the holographic models (see,e.g., Refs.~\cite{Ahmady:2019yvo,Kaur:2020vkq}), etc.  
In view of our study, one has to separately mention the approaches developed within the so-called  continuum-QCD, that are based on  solutions (actually in Euclidean space) of dynamical equations like the homogeneous 4D Bethe-Salpeter equation (BSE)~\cite{BS51,GellMann:1951rw} in combination or not with  the quark gap-equation (see, e.g. Refs.~\cite{Shi:2018zqd,Shi:2020pqe,Zhang:2020ecj,Shi:2022erw}).

 It should be recalled that the proton is the elective target of  much experimental (see, e.g., Refs.~\cite{Angeles-Martinez:2015sea,Avakian:2016rst,pitonyak2019transverse}) and theoretical  research (see, e.g., Refs.~\cite{Bertone:2019nxa,Bacchetta:2020gko,Bury:2022czx} and references therein).  While the pion, given the experimental challenges its study poses,  has surely attracted less efforts in spite of  its  intriguing  double-nature, being both a Goldstone boson (and hence fundamental for investigating the dynamical chiral-symmetry breaking) and a quark-antiquark bound system (i.e. the simplest bound system in QCD). In particular, 
a first extraction of the pion unpolarized leading-twist TMD  from  Drell-Yan data  can be found in Ref.~\cite{Vladimirov:2019bfa}, where the results of the E615 Collaboration~\cite{ConwayPRD1989} has been used, and  in Ref.~\cite{Cerutti:2022}, where both the previous data and  the  E537 Collaboration cross-sections~\cite{Anassontzis:1987hk}  have been included. As to  the phenomenological calculations,  a broad overview, embracing different approaches, can be gained from   Refs.~\cite{Shi:2018zqd,Zhang:2020ecj,Shi:2020pqe,Shi:2022erw,Matevosyan:2011vj,Pasquini:2014ppa,Noguera:2015iia,Lorce:2016ugb,Bacchetta:2017vzh,Ahmady:2019yvo,Kaur:2020vkq,Bastami:2020asv} (see also Ref.~\cite{Meissner:2008ay} for the generalized TMDs  in a spin-0 hadron).

As a conclusion to the above schematic introduction, it has to be emphasized that the vast amount  of nowadays theoretical studies  on TMDs finds its strong motivation in the very accurate measurements that will come from   forthcoming electron-ion colliders, that promise to achieve  greatly expected milestones in the experimental investigation of non-perturbative QCD, given the planned high energy and luminosity~\cite{AbdulKhalek:2021gbh,Anderle:2021wcy}.

Our aim  is  to obtain, for the first time,  T-even leading- and subleading-twist unpolarized TMDs (uTMDs) of the pion, by solving a dynamical equation directly in Minkowski space,  namely  relying on a genuinely quantum-field theory framework based on the 4D homogeneous  BSE~\cite{BS51,GellMann:1951rw}. 
 The homogeneous  BSE is an integral equation and therefore suitable for dealing with the fundamentally non-perturbative nature  of bound states. One should not get confused by the use of an interaction kernel expressed in a perturbative series, since  an integral equation has a peculiar feature of infinitely many  times iterating  the boson exchanges contained in each term of the kernel, just  what one needs for obtaining a pole in the relevant Green's function.  
  In our approach (see Ref.~\cite{dePaula:2020qna} for details and references therein), based on the 4D homogeneous  BSE in Minkowski space and the 
 Nakanishi integral representation (NIR) of the BS-amplitude~\cite{nak63,Nakanishi:1971}, the interaction kernel is given by the exchange of a massive vector boson in the Feynman gauge,  with three input  parameters, inferred 
 from lattice QCD (LQCD) calculations (see, e.g. Refs. \cite{DuPRD14,Rojas2013,Oliveira:2020yac}):  (i) the constituent-quark 
  and gluon masses, and  (ii) a scale parameter
featuring the extended quark-gluon vertex. 
 It should be pointed out that the   ladder kernel, i.e. the first term in a perturbative series,  can be   a reliable approximation to 
  evaluate the pion bound state,
   as suggested by the suppression of the non-planar contributions for $N_c=3$  within the BS approach in a scalar QCD model~\cite{Nogueira:2017pmj}, and the presence of massive quarks and gluons, featuring the confinement effects in a relatively large system ($r_{ch} \sim 0.66~$fm). 
    There is another important consequence stemming from the use of the BS-amplitude. Although in the definition of the $q\bar q$-pair BS-amplitude there is a simple dependence upon two interacting fermionic fields, one ends up dealing with an infinite 
 content of Fock states (the use of the Fock space allows one to recover a probabilistic language within the BS framework). In particular, by exploiting the Fock expansion of the pion state, one can establish a formal link  between the LF-projected BS-amplitude (see, e.g., Refs.~\cite{FSV1,Sales:2001gk,Marinho:2008pe}),  
and the amplitude of the Fock component of the pion state with the lowest number of constituents.  Therefore,
in our approach, it is natural to call the LF-projected BS-amplitude  {\em LF valence wave function} (LFWF), to be distinguished from the valence wave function, when a $SU(3)$-flavor language is adopted. In the latter case, the pion is composed by only two fermionic constituents, suitably dressed. One should keep in mind that 
 within our framework, the pion LFWF  contributes only with 
 70\%~\cite{dePaula:2020qna}
 of the normalization, and consequently  a significant  
 role of the higher Fock components has to be highlighted,  and possibly analyzed in-depth, as illustrated in what follows. Finally, we would emphasize that the first evaluation of the uTMDs strengthens the reliability of  our approach and makes sound the ground for the next step, already  in progress, i.e. taking into account the self-energy of the quarks (see Refs.~\cite{Mello:2017mor} and~ \cite{Duarte:2022yur,Mezrag:2020iuo}).
 
  Indeed, in spirit, our approach is similar to the one developed in Ref.~\cite{Shi:2020pqe} for evaluating the leading-twist uTMD, where it was also taken into account the self-energy of the quark propagator (solving the gap equation) and a  confining interaction, but in Euclidean space. In this case, one resorts to a suitable method (based on the moments and a parametrization of the Euclidean BS-amplitude) to get the Minkowski-space distribution function. Differently, in our approach the NIR of the BS-amplitude allows one to successfully deal with the analytic structure of the BS-amplitude itself,  obtaining an integral equation formally equivalent to the initial 4D homogeneous BSE, but  more suitable for the numerical treatment. Many and relevant applications of our approach to the pion, such as the electromagnetic form factor~\cite{Ydrefors:2021dwa}, the PDF~\cite{dePaula:2022pcb} and the 3D imaging~\cite{dePaula:2020qna},  have confirmed its reliability and  encouraged to  broad the scope of our investigation.  It should be pointed out that (it will become clear in what follows) the evaluation of quantities that depend not only upon the longitudinal dof but also the transverse ones leads to sharply increase the sensitivity to the dynamical content of a given phenomenological description of the pion, namely to increase its predictive power. Furthermore, the joint use of the Fock expansion, meaningful in the Minkowski space, allows one  to resolve  the gluonic content of the pion state.

The  paper outline is as follows. In Sect.~\ref{Sec:gen}, the general formalism and the notations are introduced, highlighting the ingredients of our dynamical approach, namely i) the Bethe-Salpeter amplitude, solution of the 4D homogeneous Bethe-Salpeter equation, and ii) the Nakanishi integral representation of the BS-amplitude. In Sect.~\ref{Sec:uTMDS}, the expressions of leading- and subleading-twist  uTMDs are given in terms of the Bethe-Salpeter amplitude of the pion. In Sec.~\ref{Sec:f1} and~\ref{Sec:subl}, the leading and subleading-twist uTMDs are shown and  compared with  outcomes from other approaches.  
Finally, in Sect.~\ref{Sec:concl}, the conclusions are drawn, and the perspectives of our approach are presented.
\section{Generalities}
\label{Sec:gen}
 For  a pion  with four-momentum $P\equiv \{P^-,P^+,{\bf P}_\perp\}$ (where $P^2=P^+P^--|{\bf P}_\perp|^2=M^2$ and the LF coordinates are  $a^\pm=a^0\pm a^3$), and by adopting  both i)~a frame where 
${\bf P}_\perp=0$ and ii)~the light-cone gauge $A^+_g = 0$, the quark leading-twist  uTMD, $f^q_1(\gamma, \xi)$,  is defined   as follows
(for a general introduction see, e.g.,  Ref. \cite{Tangerman:1994eh,bacchetta2007semi}) 
\begin{small}
\begin{multline}
 f^q_1(\gamma, \xi) 
 =~{N_c\over 4}\int d\phi_{\hat {\bf k}_\perp}
 \int_{-\infty}^\infty {dy^-  d{\bf y}_\perp \over 2 (2\pi)^3}
 \\ \times
 e^{i [\xi P^+\frac{y^-}{2} -
 {\bf k}_\perp\cdot
 {\bf y}_\perp] }  
  \langle P| 
\bar{\psi}_q (- \tfrac y 2) \gamma^+  \psi_q(\tfrac  y2) 
|P \rangle\big|_{ y^+=0}\, ,
\label{Eq:f1q}    
\end{multline}
\end{small}
where $N_c$ is the number of colors, $\psi_q$ is the fermionic field, and the quark four-momentum is given in terms of LF coordinate by $p_q\equiv\{p^-_q, \xi P^+,{\bf k}_\perp+{\bf P}_\perp/2\} $, with $\gamma=|{\bf k}_\perp|^2$. The  antiquark uTMD is obtained by using the proper four-momentum $p_{\bar q}\equiv\{p^-_{\bar q}, (1-\xi) P^+,-{\bf k}_\perp+{\bf P}_\perp/2\} $,  recalling that  $P=p_q+p_{\bar q}$ and $k=(p_q-p_{\bar q})/2$. 

The normalization of $f^q_1(\gamma, \xi)$ is given by 
\begin{multline}
\int_{-\infty}^\infty d\xi\int_0^\infty d\gamma~f^q_1(\gamma, \xi)=
{N_c \over 2}\int  d{\bf p}_{q\perp }\int_{-\infty}^\infty {dp^+_q\over P^+}
\\ 
\times\int_{-\infty}^\infty {dp^-_q\over 2}
 \int_{-\infty}^\infty {d^4y   \over  (2\pi)^4}e^{i \, p_q\cdot y } 
 \langle P| 
\bar{\psi}_q (- \tfrac  y 2) \gamma^+  \psi_q(\tfrac  y 2) 
|P \rangle
\\
=N_c~{\langle P| 
\bar{\psi}_q (0) \gamma^+  \psi_q(0) 
|P \rangle\over 2P^+} =F^q_\pi(0)=1\, , 
\label{Eq:f1norm}
\end{multline} 
where $F^q_\pi(t)$  is the quark contribution to the electromagnetic (em) form factor of the pion. The latter   results  to  be  equal  to $F_\pi(t)=e_q F^q_\pi(t)+e_{\bar q}F^{\bar q}_\pi(t)$, with $t=(P'-P)^2$, and is related to the matrix element of the four-current   by  $N_c~\langle P| 
\bar{\psi}_q (0) \gamma^\mu  \psi_q(0) 
|P \rangle =2P^\mu ~F_\pi(t=0)$.  Finally, it should be pointed that inserting a complete basis in Eq. \eqref{Eq:f1q} and exploiting the good and bad components of the fermionic field one can easily demonstrate  that  $f^q_1(\gamma,\xi)\ge 0$ (see Ref. \cite{Jaffe:1991ra}).

In order to describe the pion by taking into account at some extent  the QCD  dynamics  in the non-perturbative regime, it 
is useful to resort to  the Mandelstam framework~\cite{Mandelstam:1955sd},  where the interacting quark-pion  vertex is expressed in terms of  the (reduced) BS-amplitude, i.e. the solution of the 4D homogeneous BSE, and defined by
\begin{equation}
\Phi(k,P)=\int d^4x~e^{i k\cdot x}\,
\langle 0|T\bigl\{\psi(\tfrac 
 x2)\,\bar \psi(-\tfrac x2) \bigr\}|P\rangle\,,
\end{equation}
where the fermionc field fulfills the Poincar\'e translation $\psi(x)=e^{i\hat P\cdot x} \psi(0) e^{-i\hat P\cdot x}$ (recall that  only the component $\hat P^-$ is interacting  in the LF dynamics, see, e.g., Ref. \cite{Brodsky:1997de}).

 Thus, by using the  Feynman-like diagrammatic picture inherent to the Mandelstam framework (see, e.g., Ref.~\cite{Ydrefors:2021dwa} for the application to the em form factor), one can write the following expression for   $f^q_1(\gamma, \xi)$ 
 \begin{small}
\begin{multline}
f^q_1(\gamma, \xi) =
{N_c \over 4(2\pi)^3}
\int_{-\infty}^\infty {dk^+\over 2(2\pi)} \delta\Big(k^+ + \frac{P^+}{2}- \xi P^+\Big)
\\
\times\int_{-\infty}^\infty \hspace{-.3cm} dk^- \hspace{-.1cm}
\int_0^{2\pi}\hspace{-.3cm} d\phi_{\hat {\bf k}_\perp}\hspace{-.1cm}
 \text{Tr}\left[ S^{-1}(-p_{\bar  q}) \bar \Phi(k,P)\,  \gamma^+  \,\Phi(k,P) \right]\,,
\label{Eq:f1qb}
\end{multline}
\end{small}
where 
\be
p_{q(\bar q)}\,=\, \pm\, k+\frac P2\, . 
\ee
For the sake of completeness, let us write the BSE in ladder approximation, i.e. the one we are adopting for the numerical calculations, viz.
\begin{multline}\label{Eq:BSE}
  \Phi(k; P) = S\bigl(p_q\bigr)
  \int \frac{d^4k'}{(2\pi)^4}S^{\mu\nu}(q)\Gamma_{\mu}(q) \\ \times~
  \Phi(k';P)\widehat\Gamma_{\nu}(q)S\bigl(-p_{\bar q}\bigr),
\end{multline}
where  quark  and antiquark momenta are off-shell, i.e.
$p^2_{q(\bar q)}=(\pm k +\tfrac 
P2)^2\neq m^2$,  and   $q = k - k'$  is the gluon four-momentum. 
In Eq.~\eqref{Eq:BSE},   the fermion propagator,  the gluon 
propagator in the Feynman gauge  and  
the   quark-gluon  vertex,  dressed through a simple form factor, 
are 
\be
S(p) =  {i \over \psla{p} - m + i\epsilon}~, \qquad 
    S^{\mu\nu}(q) = -i \frac{g^{\mu\nu}}{q^2 - \mu^2 + i\epsilon}~,
    \nonu 
    \Gamma^\mu=   i g\gamma^\mu~ \frac{\mu^2 - \Lambda^2}{q^2 - \Lambda^2 + i\epsilon},
\label{Eq:def1}\ee
where $g$ is the coupling constant, $\mu$ the mass of the exchanged vector-boson 
and $\Lambda$ is a scale parameter, featuring the extension of 
 the  color distribution  in the
 interaction vertex of the 
dressed constituents. Moreover,  in Eq.~\eqref{Eq:BSE}, one has  
$\widehat\Gamma_{\nu}(q)=C~\Gamma^T_{\nu}(q)
~C^{-1}$,
where  $C=i\gamma^2\gamma^0$  is the  charge-conjugation operator. The normalization of the BS-amplitude reads (cf. Refs.~\cite{Lurie} and~\cite{dePaula:2020qna} for details) 
\be
N_c~\text{Tr} \Biggl[ \int{d^4k\over (2\pi)^4}~{\partial\over \partial P^{\prime \mu}}\Biggl\{
S^{-1}\Big(k-\tfrac{P'}{2}\Big)\,\bar \Phi(k,P)
\nonu \times
\,S^{-1}\Big(k+\tfrac{P'}{2}\Big)~\Phi(k,P)\Biggr\}\Biggr]_{P'=P}=-2iP_\mu\,.
\label{Eq:norm}
\ee

The antiquark  uTMD is  given by
\begin{small}
\begin{multline}
f^{\bar q}_1(\gamma, 1-\xi)=-{N_c  \over 4(2\pi)^3}
\int_{-\infty}^\infty {dk^+\over 2 (2\pi)} \delta(k^+ + \tfrac{P^+}{2}- \xi P^+)
\\ \times
\int_{-\infty}^\infty \hspace{-.3cm} dk^- 
\int_0^{2\pi}\hspace{-.3cm} d\phi_{\hat {\bf k}_\perp}
\text{Tr}\left[ S^{-1}(p_q) \Phi(k,P)  \gamma^+  \bar \Phi(k,P) \right] \,,
\label{Eq:f1aq}   
\end{multline}
\end{small}
where the minus sign results from the property  of the normal-ordered em  current  to be odd under the action of the charge conjugation operator. It is noteworthy that in Appendix~\ref{norm_app}, it is proven the identity of the normalization condition,  Eq.~\eqref{Eq:norm}, and the   half sum of Eqs.~\eqref{Eq:f1q}  and \eqref{Eq:f1aq}.

Within  a $SU(3)$-flavor  symmetry framework, one describes a pion as a bound system of  a massive $q\bar q$ pair. This leads to introduce  the so-called valence-quark PDF in the pion,  that  is  charge symmetric (once the isospin breaking is disregarded~\cite{Londergan:2009kj})   as well as fulfills the charge conjugation. From those properties one deduces that the $SU(3)$-valence PDFs in the charged pions must verify: $u^v_{\pi^+}(\xi)=d^v_{\pi^-}(\xi)=\bar d^v _{\pi^+}(\xi)$.   
In our BS framework, in addition to the fermionic dof (still massive) one introduces also  gluonic dof, by adding an explicit  dynamical description of the binding. This amounts to the  ladder exchange of infinite  number of massive gluons. Therefore, at the initial scale,  the quark and anti-quark longitudinal-momentum fraction distributions  are not expected to be symmetric with respect to $\xi=1/2$ (as it follows from the charge symmetry), given the gluon-momentum flow in the composite pion (see Sect.~\ref{Sec:f1}). The symmetric combination of quark and anti-quark contribution allows one to fulfill the charge symmetry, and hence it is relevant in the comparison with experimental data (see Ref. \cite{dePaula:2022pcb}). In what follows, in addition to the quark distributions, symmetric and anti-symmetric combinations are introduced for all the uTMDs we are going to analyze. 

The half sum (difference) of the quark and anti-quark contributions, Eqs.~\eqref{Eq:f1q} and \eqref{Eq:f1aq}, yields
 the following charge-symmetric (anti-symmetric) expression for the leading-twist uTMD inside a $\pi^+$ meson 
\be
f^{S(AS)}_1(\gamma, \xi)={f^q_1(\gamma, \xi)
\pm  f^{\bar q}_1(\gamma, 1-\xi) \over 2}
\nonu
=
{N_c \over 8(2\pi)^3}
 \int_{-\infty}^\infty {dk^+\over 2 (2\pi)} \delta\left(p^+_q- \xi P^+\right)
 \int_{-\infty}^\infty  dk^- 
  \nonu \times \int_0^{2\pi} d\phi_{\hat {\bf k}_\perp} \text{Tr}\Big[ S^{-1}(-p_{\bar q}) \bar \Phi(k,P)~  \gamma^+  ~\Phi(k,P) 
  \nonu
 \mp \,S^{-1}(p_q) \Phi(k,P)~  \gamma^+  ~ \bar \Phi(k,P) \Big]~.
\label{Eq:f1t}   
\ee  
  Analogously to Eq.~\eqref{Eq:f1q},  one can define  the  T-even subleading quark uTMDs, starting from the decomposition of the pion correlator~\cite{Mulders:1995dh,bacchetta2007semi}.  To be specific, one has   two twist-3 uTMDs (see, e.g., Ref.~\cite{Lorce:2016ugb} for the pion case)
\be
{M\over P^+ }~e^q(\gamma,\xi)={N_c\over 4}\int d\phi_{\hat {\bf k}_\perp}
 \int_{-\infty}^\infty {dy^-  d{\bf y}_\perp \over 2 (2\pi)^3}
\\ && \times e^{i [\xi P^+\frac{y^-}{2} -
 {\bf k}_\perp\cdot
 {\bf y}_\perp] }
  \langle P| 
\bar{\psi}_q (-\tfrac y 2)\,\mathbb{1}\,\psi_q(\tfrac y  2) 
|P \rangle\big|_{ y^+=0}\, , \nonumber
\label{Eq:eq}
\ee
\be
{M\over P^+ }~f^{\perp q}(\gamma,\xi)={M\over \gamma}~{N_c \over 4}\int d\phi_{\hat {\bf k}_\perp}
 \int_{-\infty}^\infty {dy^-  d{\bf y}_\perp \over 2 (2\pi)^3}
 \\ &&
\times e^{i [\xi P^+\frac{y^-}{2} -
 {\bf k}_\perp\cdot
 {\bf y}_\perp] }
 \, \langle P| 
\bar{\psi}_q (-\tfrac y 2)\,{\bf k}_\perp\cdot{\bg \gamma}_\perp\, \psi_q(\tfrac y  2) 
|P \rangle\big|_{ y^+=0}~. \nonumber
\label{Eq:fperp}
\ee

In analogy to  Eq.~\eqref{Eq:f1norm}, one gets for the twist-3 $e^q(\xi)$  (see   Refs.~\cite{Mulders:1995dh,Jaffe:1991ra} and Ref.~\cite{Lorce:2016ugb} for the pion in  phenomenological models)
\be 
\int_{-\infty}^\infty d\xi\int_0^\infty d\gamma~e^q(\gamma, \xi)=
{N_c \over 2}\int  d{\bf p}_{q\perp }\int_{-\infty}^\infty {dp^+_q\over P^+}
\nonu  \times
\int_{-\infty}^\infty {dp^-_q\over 2}
 \int_{-\infty}^\infty {d^4y   \over  (2\pi)^4}e^{i \, p_q\cdot y } 
 ~ \langle P| 
\bar{\psi}_q (- \tfrac  y 2) \,\mathbb{1}\,  \psi_q(\tfrac y 2) 
|P \rangle
\nonu=N_c~{\langle P| 
\bar{\psi}_q (0) ~\mathbb{1}~  \psi_q(0) 
|P \rangle\over 2P^+}  ~~,
\label{Eq:e_norm}
\ee
where the matrix element $\langle P| 
\bar{\psi}_q (0) ~\mathbb{1}~  \psi_q(0) 
|P \rangle$ has to be proportional to the pion sigma term, once a QCD framework is adopted. As a matter of fact, one gets
\be
\int_0^1 d\xi \int_0^\infty d\gamma~e^q(\gamma,\xi)={\sigma_\pi\over m_{cur}}
\label{Eq:sigmat}
\ee
where  $m_{cur}$ is the quark current mass and $\sigma_{\pi}$ is the pion sigma term, that becomes $\sigma_{\pi}=M/2$, in the leading order of the chiral expansion, i.e. the Gell-Mann-Oakes-Renner relation~\cite{Gell-Mann:1968hlm}. It should be pointed that recent LQCD calculations~\cite{Bali:2016lvx} confirm, with high accuracy, the Gell-Mann-Oakes-Renner 
 relation  in the range of the explored pion masses. Indeed, the  QCD equations  of motion gives a decomposition of the collinear PDF $e(\xi)= \int d\gamma ~e(\gamma,\xi)$ in three terms. Among them, there is a singular term proportional to the pion sigma term, that  reads (see, e.g., Ref. \cite{Efremov:2002qh})
 \be 
 e_{sing}(\xi)=\delta(\xi)~{\langle P| 
\bar{\psi}_q (0) ~\mathbb{1}~  \psi_q(0) 
|P \rangle/ 2P^+}\, ,
\label{Eq:eq_sing}
\ee 
while the other two terms, one is due to   quark-antiquark-gluon correlations and the  other is proportional to the quark mass,   do not contribute to Eq.~\eqref{Eq:sigmat} (see Ref.~\cite{Efremov:2002qh}, where the issue is analyzed, taking the nucleon as actual case).  In our phenomenological model the strength is distributed over the whole range of $\xi$ (as in Ref. \cite{Lorce:2016ugb}), without the singularity at $\xi=0$, as it will be shown in Sect.~\ref{Sec:subl}. Moreover, one has  for the first moment~\cite{Efremov:2002qh}
\be
\int_0^1 d\xi \int_0^\infty d\gamma~\xi~e^q(\gamma,\xi)={m_{cur}\over M}~~,
\label{Eq:sum_rul}
\ee
where the singular term and the gluonic contribution vanish, and only the term proportional to the quark mass contributes.

 From the equations of motion of a free-quark model, one deduces the following relations between the above uTMDs (see,e.g., Ref \cite{Lorce:2014hxa,Efremov:2002qh,Lorce:2016ugb})
 \be 
\xi~e^q_{EoM}(\gamma,\xi)= \xi~\tilde e^q(\gamma,\xi)+{m\over M}~f^q_{1;EoM}(\gamma,\xi)
\nonu 
\xi~f^{q\perp}_{EoM}(\gamma,\xi)= \xi~\tilde f^{q\perp}(\gamma,\xi)+f^q_{1;EoM}(\gamma,\xi)
~,
\label{Eq:LIR}\ee 
where the   uTMDs with a tilde are  the gluonic contributions. The relevant point is the dependence of all the subleading-twist  uTMDs from only the leading one, modulo the gluonic terms.  In our fully interacting framework, one can anticipate that the relations are not  recovered, and rather heavily broken. For a derivation of the first line of Eq. \eqref{Eq:LIR}, fully consistent with QCD, one could apply the formalism presented in Ref. \cite{Efremov:2002qh}.

Following Eq.~\eqref{Eq:f1t},  one readily writes down   charge-symmetric  and the anti-symmetric combinations for   the subleading TMDs. One has to take care   how   the scalar and vector operators behave under the charge conjugation that impose a different combination of signs (cf.  below Eq.~\eqref{Eq:f1aq}). Namely, one gets
\begin{multline}
\hspace{-.3cm}{M\over P^+ }~e^{S(AS)}(\gamma,\xi)=
{N_c \over 8(2\pi)^3}
 \int_{-\infty}^\infty {dk^+\over 2 (2\pi)} \delta(p^+_q- \xi P^+)
 \\ \times 
 \int_{-\infty}^\infty\hspace{-.2cm}  dk^- 
\int_0^{2\pi} \hspace{-.2cm} d\phi_{\hat {\bf k}_\perp}
 \text{Tr}\Big[ S^{-1}(-p_{\bar q}) \bar \Phi(k,P)~  \mathbb{1}  ~\Phi(k,P) \\
 \pm \, S^{-1}(p_q) \Phi(k,P)~  \mathbb{1}  ~ \bar \Phi(k,P) \Big]~.
\label{Eq:et}
\end{multline}
\begin{multline}
{M\over P^+ }~f^{\perp S(AS)}(\gamma,\xi)={N_c M\over 8(2\pi)^3\gamma}
 \int_{-\infty}^\infty {dk^+\over 2 (2\pi)} \delta(p^+_q- \xi P^+)
 \\
 \int_{-\infty}^\infty  dk^- 
\int_0^{2\pi} d\phi_{\hat {\bf k}_\perp}
 \text{Tr}\Big[ S^{-1}(-p_{\bar q}) \bar \Phi(k,P)\,  {\bg \gamma}_\perp  \,\Phi(k,P) 
\\ 
 \pm \, S^{-1}(p_q) \Phi(k,P)\, {\bg \gamma}_\perp  \, \bar \Phi(k,P) \Big]\cdot{\bf k}_\perp~.
\label{Eq:fperpt}
\end{multline}

\subsection{The BS-amplitude and its Nakanishi integral representation}

It is useful to briefly recall some features of our approach for obtaining the actual solution of the ladder BSE given in Eq.~\eqref{Eq:BSE}. The basic ingredient is the NIR of the BS-amplitude (see Ref.~\cite{Nakanishi:1971} for the general introduction, and Refs.~\cite{Carbonell:2010zw,dePaula:2016oct,dePaula:2017ikc,dePaula:2020qna,dePaula:2022pcb} for the application to a two-fermion case), but  let us first introduce the general decomposition of  
the BS-amplitude, $\Phi(k; P)$, for a $0^-$ bound state, viz. \cite{Llewel,Carbonell:2010zw}
\be\label{Eq:BS_decomp}
  \Phi(k;P) =  S_1(k;P) \phi_1(k;P)+S_2(k;P) \phi_2(k;P)
  \nonu 
  +S_3(k;P) \phi_3(k;P)+S_4(k;P) \phi_4(k;P)~~,
\ee
 where  $\phi_i$'s are unknown scalar 
 functions, that depend upon the kinematical scalars at disposal ($k^2$, $k\cdot P$ and $P^2$), and $S_i$'s  are suitable 
 Dirac structures,  given by
\be
    S_1(k;P) = \gamma_5, \,\, S_2(k;P) = \frac{\Psla{P}}{M}{\gamma_5}, \nonu
    S_3(k;P) = \frac{k\cdot P}{M^3}\Psla{P}\gamma_5 - 
    \frac{1}{M}\psla{k}\gamma_5, 
    \nonu 
    S_4(k;P) =
    \frac{i}{M^2}\sigma^{\mu\nu}P_\mu k_\nu \gamma_5\, .
\label{EQ:def2}\ee 
The functions  $\phi_i$  must be 
 even for $i=1,2,4$ and odd for $i=3$, under the change 
 $k \rightarrow - k$, as dictated by the anti-commutation rules of the fermionic fields, and  they 
can be written in terms of the NIR as follows
\begin{multline}
  \label{Eq:NIR}
  \phi_i(k; P) = \int_{-1}^1 dz'\int_0^\infty d\gamma'
  \\ \times \frac{g_i(\gamma',z';\kappa^2)}{[k^2 + z'(P\cdot k) - \gamma' - \kappa^2 + i \epsilon]^3},     
\end{multline}
where  $\kappa^2 = m^2 -{M^2/4}$. The real functions  $g_i(\gamma',z';\kappa^2)$, the unknowns of the problem under scrutiny, are the 
 Nakanishi weight 
functions (NWFs),  and
 assumed   to be unique, following    the uniqueness theorem from Ref.~\cite{Nakanishi:1971}. The properties of the scalar functions $\phi_i$
under the exchange $k\to -k$ translate to   properties of the NWFs, but under the exchange  $z'\to -z'$. 

Finally, it should be mentioned that NWFs are determined  by solving a system of  integral equation, so that one is able to non-perturbatively embed dynamical information that characterize  the BS interaction kernel. The system of integral equations is formally deduced from the initial BSE, by exploiting 
the analytic structure of the  scalar functions $\phi_i$,   made explicit by means of the NIR.  In fact, after inserting Eqs.~\eqref{Eq:BS_decomp} and \eqref{Eq:NIR} in the BSE,
Eq.~\eqref{Eq:BSE}, and  performing both the Dirac traces and a LF projection, i.e. the integration over the $k^-=k^0-k^3$ component of the relative momentum,
one gets a  
 coupled system of integral equations for the NWFs (see details in
Ref.~\cite{dePaula:2017ikc}). Once the NWFs are known, the BS-amplitude can be fully reconstructed through  an inverse path, i.e. Eqs.~\eqref{Eq:NIR} and \eqref{Eq:BS_decomp}. 

\section{The  unpolarized TMDs and the pion BS-amplitude}
\label{Sec:uTMDS}
The evaluation of the leading- and subleading-twist uTMDs, given in Eqs.~\eqref{Eq:f1t}, \eqref{Eq:et} and \eqref{Eq:fperpt}, can be performed by inserting the decomposition of the BS-amplitude in Eq.~\eqref{Eq:BS_decomp}, obtaining
\begin{multline}
{\cal T}^{S(AS)}_i(\gamma,\xi)={N_c \over 8 (2\pi)^3}
\int_{-\infty}^\infty {dk^+\over 2 } \delta(p^+_q- \xi P^+) 
\\ \times\int_{-\infty}^\infty  {d k^{-}\over 2 \pi} 
\int_0^{2\pi} d\phi_{\hat {\bf k}_\perp}
\text{Tr} \Big[ S^{-1}(-p_{\bar q}) \, {\cal A}_i(k,P)
 \\
 +\eta^{S(AS)}_i \, S^{-1}(p_q) 
\,\bar {\cal A}_i(k,P)
 \Big]
\\
={i\,N_c \over 8 (2\pi)^2}
 \sum_{\ell j} \int_{-1}^1\hspace{-.2cm} dz \, \delta( z - \left(1- 2\xi\right)) \, F^i_{\ell j}(\gamma, z;{S(AS)})~,
\label{Eq:tmd_t1}
\end{multline}
where 
\be
{\cal A}_i(k,P)=\bar \Phi(k,P)\, {\cal O}_i \,\Phi(k,P)\, \nonu 
\bar{\cal A}_i(k,P)= \Phi(k,P)\, {\cal O}_i \,\bar\Phi(k,P)\, .
\ee
A new variable $z$ is defined as $z=-2k^+/P^+$ and the three quantities:
 i)~the functions ${\cal T}_i(\gamma,\xi)$,   ii)~the operators ${\cal O}_i$ and iii)~the phase $\eta^{S(AS)}_i$ are given by
 \bwt 
\be 
\begin{array}{lll}{\cal T}^{S(AS)}_0(\gamma,\xi)\equiv~f^{S(AS)}_1(\gamma,\xi)~, & \quad {\cal O}_0=\gamma^+~,&\quad \eta^{S(AS)}_0=\mp 1~,
\\ ~&~& ~\\
{\cal T}^{S(AS)}_1(\gamma,\xi)\equiv{M\over P^+}~e^{S(AS)}(\gamma,\xi)~, & 
\quad {\cal O}_1=\mathbb{1} ~,& \quad\eta^{S(AS)}_1=\pm 1~,
\\ ~&~& ~\\
{\cal T}^{S(AS)}_2(\gamma,\xi)\equiv{M\over P^+}~f^{\perp S(AS)}(\gamma,\xi)~, 
 & \quad {\cal O}_2={M\over |{\bf k}_\perp|^2}~{\bf k}_\perp\cdot{\bg \gamma}_\perp~,&\quad \eta^{S(AS)}_2=\pm 1~.
\end{array}
\ee
\ewt
Finally,  the integrand $F^i_{\ell j}$ in Eq.~\eqref{Eq:tmd_t1} reads
\begin{multline} 
\hspace{-.4cm} F^i_{\ell j}(\gamma, z;{S(AS)})= \int_{-\infty}^\infty {d k^{-}\over 2 \pi} ~a^i_{\ell j}(k^{-},\gamma,z;{S(AS)})\\ \times ~\phi_{\ell}(k,P) \, \phi_j(k,P)~.
\label{Eq:Filj}
\end{multline}
where $a^i_{\ell j}(k^{-}, \gamma,;S(AS))$  are polynomial in $k^-$ (up to the cubic power) and can be found  in Appendix~\ref{trace_app} for each  uTMDs, we are considering.

By exploiting the NIR,  Eq.~\eqref{Eq:NIR},  one can perform the integration on $k^-$. This integration amounts to restrict  the LF-time to $x^+=0$, and it is also known as LF-projection (see, e.g., Refs.~\cite{Sales:1999ec,Sales:2001gk,Marinho:2008pe}). After carrying out the $k^-$-integration, the expression of each ${\cal T}^{S(AS)}_i(\gamma,\xi)$ can be decomposed as follows (the details of this formal step can be found in Appendix~\ref{LFPro_app})
\begin{multline}
{\cal T}^{S(AS)}_i(\gamma,\xi)=
{3N_c \over  (2\pi)^2}~\sum_{\ell j}~\Bigl[{\cal F}^{i}_{0;\ell j}(\gamma,z;S(AS)) \\
+{\cal F}^{i}_{1;\ell j}(\gamma,z;S(AS))+{\cal F}^{i}_{2;\ell j}(\gamma,z;S(AS))\\ +{\cal F}^{i}_{3;\ell j}(\gamma,z;S(AS))\Bigr]~,
\label{Eq:tmd_t2}   
\end{multline} 
where $\xi=(1-z)/2$ and the functions ${\cal F}^{i}_{n;\ell j} (\gamma, z;S(AS))$ $(n=1,2,3,4)$ are given in Eqs.~\eqref{calF0b_app}, \eqref{calF1b_app},  \eqref{calF2b_app} and \eqref{calF3b_app}, respectively.

%
\section{The leading-twist  $f_1^{S(AS)}(\gamma,\xi)$}
\label{Sec:f1}
The symmetric and anti-symmetric combinations of the T-even leading-twist uTMD, $f_1^{S(AS)}(\gamma,\xi)$,  allow us to address the evaluation of both  quark and anti-quark contributions, $f_1^{q(\bar q)}(\gamma,\xi)$, that  
{ in the BS framework plus the Fock expansion of the pion state have interesting features, distinct from the ones of $f_1^{S(AS)}(\gamma,\xi)$}. 

After integrating the leading-twist $f_1^{q(\bar q)}(\gamma,\xi)$   on  $\gamma$, one gets the quark PDF $u^q(\xi)$, while the symmetric combination    provides the {charge-symmetric} PDF $u^S(\xi)$, { i.e. the one is expected to have relevance at the valence scale (see, e.g., Ref. \cite{Londergan:2009kj}).} Indeed, in the Mandelstam approach the quark and antiquark PDFs do not have in general a symmetry with respect to $\xi=1/2$, since each  receives contributions from states containing an infinite number of gluons, { as a consequence of the ladder-interaction kernel.} But if we restrict to the contribution from the first Fock component in  the expansion of  the pion state, one gets  the LF-valence $u^{LF}_{val}(\xi)$, that is given by the BS-amplitude projected onto the null plane~\cite{Brodsky:1997de} and is fully compliant with the charge symmetry (see below the   discussion   on the differences among   $u^q(\xi)$, $u^S(\xi)$ and $u^{LF}_{val}(\xi)$). 

To illustrate general features and relations, in this Section  we give some details, referring to Appendix~\ref{LTMD_app} for a more complete discussion.  

The symmetric and anti-symmetric leading-twist uTMDs,  can be decomposed as follow 
\be
f^{S(AS)}_1(\gamma,\xi)={\cal I}_N(\gamma,\xi;S(AS))+{\cal I}_d(\gamma,\xi;S(AS))\nonu  +{\cal I}_{2d}(\gamma,\xi;S(AS))+{\cal I}_{3d}(\gamma,\xi;S(AS))\, ,
\ee
{where  the non-vanishing symmetric contributions  are given by Eqs.~\eqref{calINcS_app}, \eqref{calIdcS_app}, \eqref{calI2dcS_app} and ${\cal I}_{3d}(\gamma,\xi;S)=0$, respectively. 
 The anti-symmetric quantities are
 shown in  Eqs.~\eqref{calI0dcAS_app}, \eqref{calI1dcAS_app}, \eqref{calI2dcAS_app} and \eqref{calI3dcAS_app}, respectively.

Two comments are in order. The symmetry properties of the above quantities with respect to the transformation
 $z\to -z$ are demonstrated in  Appendix~\ref{LTMD_app}, and can be translated into the symmetry with respect to $\xi\to 1-\xi$ {(that implements the charge-symmetry)}. A relevant feature is given by the presence in the expressions of ${\cal I}_{d,2d,3d}$ of the partial derivatives $\partial^n/\partial z^n$, that should be considered  dual of the $n$-th moment in  $k^-$ of the relevant functions, generated by the formal step of the LF-projection (cf Eq.~\eqref{Eq:Filj}). This is not a surprise since the variable $z$ is proportional to $k^+$.

\begin{figure*}[t]
\begin{center}
\includegraphics[width=8.8cm]{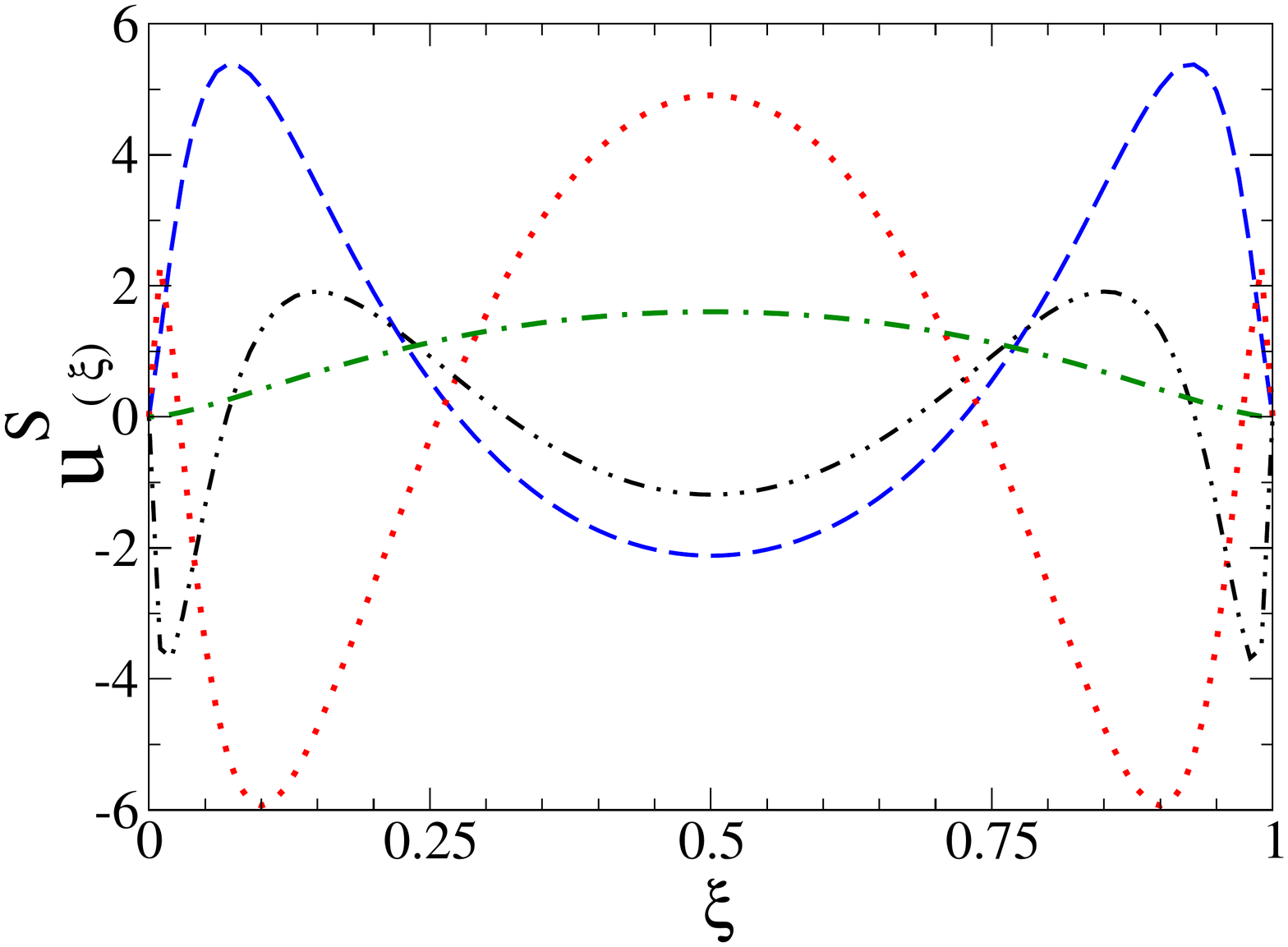}
\hspace{-0.8 cm}\includegraphics[width=8.8cm]{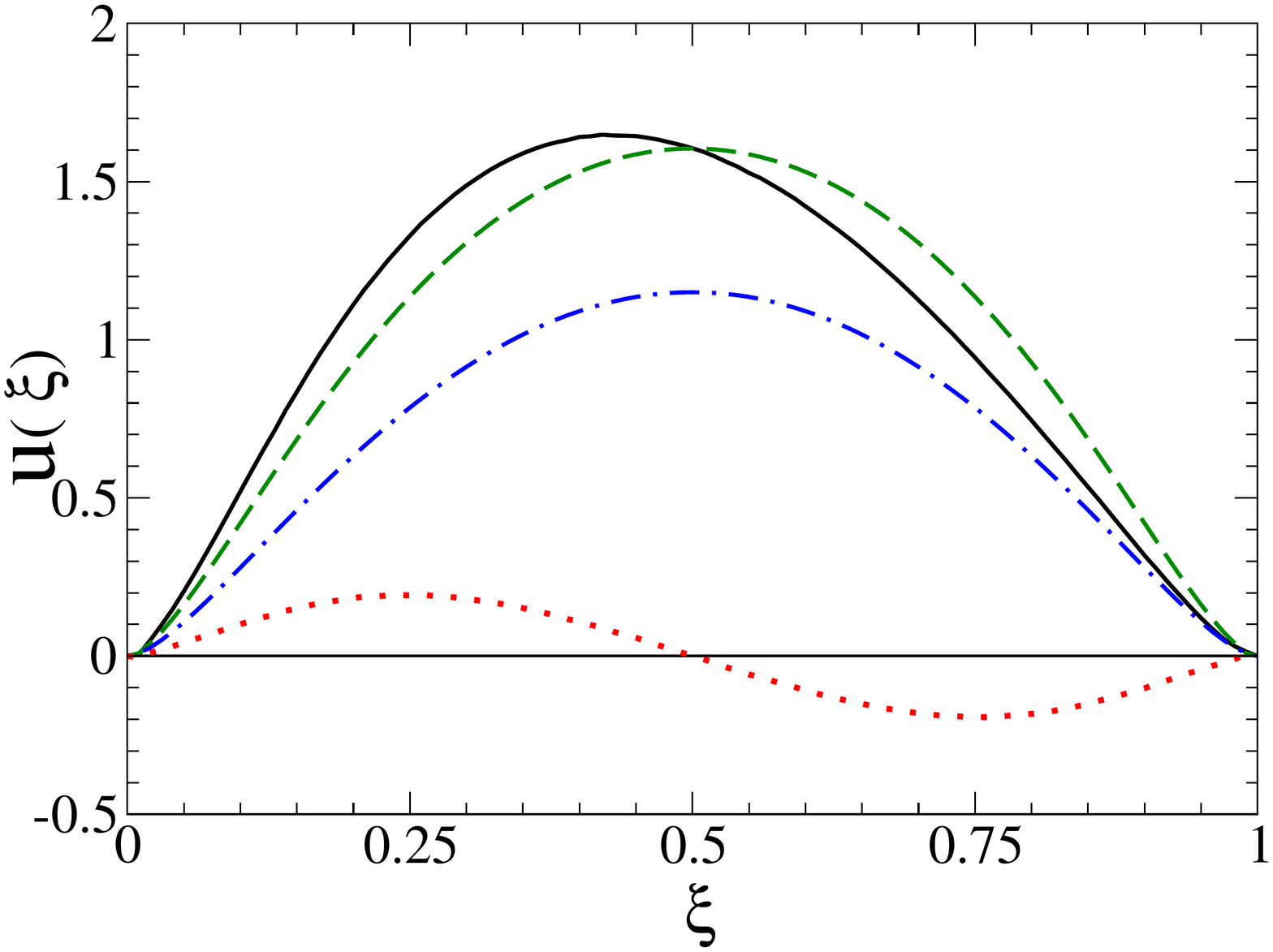}
\caption{(Color online). {\it Left panel}: the symmetric pion PDF, $u^S(\xi)$,  with its contributions $u^S_N(\xi)$, $u^S_d(\xi)$ and $u^S_{2d}(\xi)$ (cf Eq. \eqref{Eq:pdf1}). Dash-dotted  line: $u^S(\xi)$.
Dashed line: $u^S_N(\xi)$. Dotted line: $u^S_d(\xi)$. Dash-double-dotted line: $u^S_{2d}(\xi)$.
{\it Right panel}: $u^q(\xi)$, $u^S(\xi)$, $u^{AS}(\xi)$ and the LF-valence PDF of the pion, $u^{LF}_{val}(\xi)$. Solid line: quark PDF, Eq. \eqref{Eq:uxiquark}. Dashed line:
$u^S(\xi)$. Dotted line: $u^{AS}(\xi)$.
 Dash-dotted line: $u^{LF}_{val}(\xi)$ (see Ref. \cite{dePaula:2022pcb}), with normalization equal to $P_{val}=0.7$ (see text).}
\label{fig:pdf}
 \end{center}
\end{figure*}

A first  consistency check of our formalism has been carried out in Appendix~\ref{norm_app}, where it is shown that, within  the Mandelstam approach,    
$f^{S}_1(\gamma,\xi)$ and in turn  $f^q_1(\gamma,\xi)$ are normalized to $1$, as  naturally  follows  from the  canonical BS-amplitude  normalization~\cite{Lurie,Nakarev},   performed according to Eq.~\eqref{Eq:norm} (see also Ref.~\cite{dePaula:2020qna}). In particular,
 the integral on $\gamma$ and $\xi$ of ${\cal I}_N(\gamma,\xi;S)$  saturates the normalization, while the other two terms  provide vanishing contributions. Hence, one gets
\be 
\int_{0}^1 d\xi \int_0^\infty d\gamma~f^S_1(\gamma,\xi)
\nonu =\int_{0}^1 d\xi \int_0^\infty d\gamma~{\cal I}_N(\gamma,\xi;S)
\nonu=\int_{0}^1 d\xi \int_0^\infty d\gamma~f^q_1(\gamma,\xi)=1 .
\label{Eq:f1_norm}
\ee 
It should be recalled that all the calculated uTMDs vanish outside the interval $0\le \xi\le1$,  as  dictated by  the conservation of the plus components of the four-momenta of both pion and constituents (cf. Eq.~\eqref{Eq:f1t}). It is understood that the integral of $f^{AS}_1(\gamma,\xi)$ is vanishing, given the antisymmetry with respect to $\xi\to 1-\xi$.
 
\subsection{Longitudinal degree of freedom}
The symmetric and the anti-symmetric PDFs, $ u^{S(AS)}(\xi)$ (for the explicit expressions see   Appendix~\ref{pdf_app}) are defined by
\be 
u^{S(AS)}(\xi)= \int_0^\infty d\gamma~f^{S(AS)}_1(\gamma,\xi)
\\ &&=u^{S(AS)}_N(\xi)+u^{S(AS)}_d(\xi)+u^{S(AS)}_{2d}(\xi)+u^{S(AS)}_{3d}(\xi)~,
\nonumber\label{Eq:pdf1}\ee
with the normalization that follows from Eq.~\eqref{Eq:f1_norm} and the vanishing result of the  double integration  of $f^{AS}_1(\gamma,\xi)$. Finally, the quark and anti-quark PDFs are evaluated through
\be 
u^{q(\bar q)}(\xi)=u^{S}(\xi)\pm u^{AS}(\xi)~,
\label{Eq:uxiquark}\ee 
with the normalization still given by Eq.~\eqref{Eq:f1_norm}. {Within} the SU(3)-flavor symmetry, one has to implement the charge symmetry (see, e.g. Ref.~\cite{Londergan:2009kj}) at the initial scale, and therefore $u^S(\xi)$
is the PDF to be compared, after the proper evolution, with the experimental data, as it has been shown in Ref.~\cite{dePaula:2022pcb}.

In the left panel of Fig.~\ref{fig:pdf},  $u^S(\xi)$ and its three contributions (see Eqs.~\eqref{uN_app}, \eqref{ud_app} and \eqref{u2d_app}) are shown.  The calculation has been carried out by adopting the BS-amplitude obtained  by using the solution of the BSE as described in  Ref.~\cite{dePaula:2020qna},  using the following values of the three input parameters: $m= 255$ MeV, $\mu=637.5$ MeV and $\Lambda=306$ MeV, able to reproduce the pion decay constant $f^{PDG}_\pi= 130.50(1)(3)(13)\,$MeV~\cite{PDG_2018} (recall that the  pion charge radius results to be $r_{ch}=0.663\,$fm~\cite{Ydrefors:2021dwa},  in excellent  agreement with $r^{PDG}_{ch}= 0.659\pm 0.004\,$fm~\cite{Zyla:2020zbs}).
A remarkable cancellation  among the contributions  takes place,  and this  represents  a common feature for all the integrated quantities generated by the uTMDs  we are considering.  In the right panel,  one can see the comparison between  the quark PDF, 
 $u^{S(AS)}(\xi)$ and 
 the LF-valence PDF, resulting from the one-to-one relation between the LF-projected BS amplitude and the valence amplitude of the Fock expansion of the pion state. In particular,
 the LF-valence PDF (see Refs.~\cite{dePaula:2020qna,dePaula:2022pcb}), is given by
 \begin{equation}
 u^{LF}_{val}(\xi)=
\int_0^\infty \hspace{-.2cm}{d\gamma\over (4\pi)^2}~\Bigl[|\psi_{\uparrow\downarrow}(\gamma,z)|^2+|\psi_{\uparrow\uparrow}(\gamma,z)|^2\Bigr]\, ,
\label{Eq:uval}
 \end{equation}
 where $\xi=(1-z)/2$, $ \psi_{\uparrow\downarrow}(\gamma,z)$ is the anti-aligned component of the LF-valence amplitude and $ \psi_{\uparrow\uparrow}(\gamma,z)$ the aligned one (of purely relativistic nature having an orbital angular momentum equal to $1$). These amplitudes are suitable combinations of 
 the LF-projected scalar 
 functions $\phi_i(k;P)$, Eq.~\eqref{Eq:NIR}. The integral on $\xi$ of LF-valence PDF gives the probability of the valence state in the Fock expansion and amounts to
 \be
P_{val}=\int_0^1 d\xi ~u^{LF}_{val}(\xi)=0.7 \, .
 \label{Eq:norm_val}
 \ee
 { The striking feature shown in the left panel is the shift toward low $\xi$ of the quark PDF, so that for this quantity the symmetry $\xi\to 1-\xi$ is slightly violated.}

\subsection{Analysing the shift and the gluon content}

The PDF calculations based on the BS-amplitude are able to capture an explicit gluonic effect, to be taken distinct from the one responsible for the effective mass of the constituents.  In particular,  the difference between the two symmetric PDFs,  i.e. $u^S(\xi)$  and $u^{LF}_{val}(\xi)$ (recall that has $P_{val}=0.7$),  can be traced back to the non negligible probability of the higher Fock states (HFS),  where  a $q\bar q$ pair interacts by exchanging   any number of gluons. Interestingly, the difference can be effectively described only by a factor, since it turns out  that $u^{LF}_{val}(\xi)/P_{val}$ largely  overlaps $u^S(\xi)$. Finally, also the   small, but relevant, shift of the quark PDF  with respect to $u^S(\xi)$  
has  to be ascribed to the presence of HFS, as discussed in what follows.

 To get a qualitative view, we remind that the pion state can be, in principle, decomposed in Fock-components, which are schematically written in ladder approximation as
\begin{equation}\label{eq:pionstate}
|\pi\rangle=|q\bar q\rangle+ |q\bar q g\rangle+ |q\bar q \, 2g\rangle+ \cdots
\end{equation}
Due to the charge symmetry, each Fock-component is invariant by $q\leftrightarrow \bar q$, and  hence the valence state $|q\bar q\rangle$ provides a symmetric contribution to $u^q(\xi)$, identified with $u_{val}^{LF}(\xi)$. The following terms  contain gluons up to infinity.  In our model, the gluon has an effective mass about twice  the quark mass, so that the  HFS cumulative effect  results in a  small shift  of the $u^q(\xi)$ peak  at $\xi<1/2$, as shown in the right panel of Fig.~\ref{fig:pdf}.  Actually, a similar effect, related to the increasing mass of the remnant, can be also recognized  in  the nucleon, where one  has a valence parton distribution with a peak around 1/3 due to the presence of the other two constituent quarks.  In the case of the pion, the effect is small since   the valence component $|q\bar q\rangle$ has  70\% of probability (as generated by our dynamical calculation), and hence is largely dominant.

 To become   more quantitative  and illustrate this effect, we schematically write   the quark PDF  by using   the  Fock expansion  of  the pion state, Eq.~\eqref{eq:pionstate},  and inserting  LF variables~\cite{Brodsky:1997de}, one has
\be
\label{Fockpdf}
    u^q(\xi)=
    \sum_{n=2}^\infty   \,
    \Bigg\{ \prod_{i}^n \int\frac{d^2k_{i\perp}}{(2\pi)^2} \int^1_0 d\xi_i\Bigg\}\, 
    \nonu
    \times \delta\left(\xi-\xi_1\right)\delta\left(1-\sum_{i=1}^n \xi_i\right)\, \delta\left(\sum_{i=1}^n {\bf  k}_{i\perp}\right) \nonu
    \times \big|\Psi_n(\xi_1, {\bf  k}_{1\perp},\xi_2, {\bf k}_{2\perp},...)\big|^2\, ,
\ee
where $\xi_{1(2)}$ is the longitudinal-momentum  fraction of the quark (antiquark) in each Fock state, composed by a $q\bar q$ pair and $n-2$ gluons, generated by the iteration of the one-gluon exchange. Moreover, $\Psi_n(\xi_1,{\bf k}_{1\perp},\xi_2, {\bf k}_{2\perp},...)$  is  the  probability amplitude of the corresponding Fock component and  fulfills a normalization condition that follows from the one  of the pion state. In the $n$-th state one has 
\be
\xi_1=1 -\xi_2 -\sum_{g=3}^{n} \xi_{g}~~.
\label{Eq:xiconserv}\ee
Since  { $\xi_i > 0$ for massive particles},   the average value of $\xi_1$ starts to decreases while the number of gluons increases, as quantitatively shown in what follows.

Looking at the right panel of Fig.~\ref{fig:pdf}, one can realize that while the valence term, with probability $P_{val}=0.7$,  has a peak at $\xi_1=\xi_2=1/2$,  given the symmetry of $|\Psi_2(\xi_1,{\bf k}_{1\perp},\xi_2,{\bf k}_{2\perp})|^2$  all the  HFS  shift the peak to $\xi_1<1/2$, and  decrease the tail,   due to the constraint of the overall normalization.
This is reflected in the evaluation of the first moment (recall $\xi_q\equiv \xi_1$)
  \be
  \langle \xi_q\rangle=P_{val}~\langle \xi_q\rangle_{val}+
  \sum_{n>2} P_n~\langle \xi_q\rangle_n
  \nonu= P_{val}~\langle \xi_q\rangle_{val}+(1-P_{val})~\langle \xi_q\rangle_{HFS}~,
  \label{Eq:HFS}
  \ee
  where $P_n$ is the probability of the $n$-th Fock state beyond the valence one. The first term  in Eq.~\eqref{Eq:HFS} is equal to $0.35$, since   $1/2$ is weighted by $P_{val}$, and the rest  is weighted by $0.3$. Notice that 
  for each HFS, normalized to 1, one has 
  {
  \begin{eqnarray}
  \langle \xi_q\rangle_n &=& 1- \langle \xi_{\bar q}\rangle_n- \sum_{i=3}^{n}\langle \xi_{g_i}\rangle_n  \nonumber \\ &=& 1- \langle \xi_{\bar q}\rangle_n- (n-2)\langle \xi_g\rangle_n \,,
  \end{eqnarray}
  where the gluon bosonic nature leads to the factor $n-2$.}
  
 The actual value of the first moment of $u^q(\xi)$ is
\be
\langle \xi_q\rangle=\int_0^1d\xi\int_0^\infty d\gamma~\xi~f^q_1(\gamma,\xi) =0.471  
 ~,
\label{Eq:f1_firstm}\ee
that amounts to an average of $\langle\xi_q\rangle_{HFS}$   equal to $0.40$.  

\begin{figure*}[t]
\begin{center}
\includegraphics[width=8.8cm]{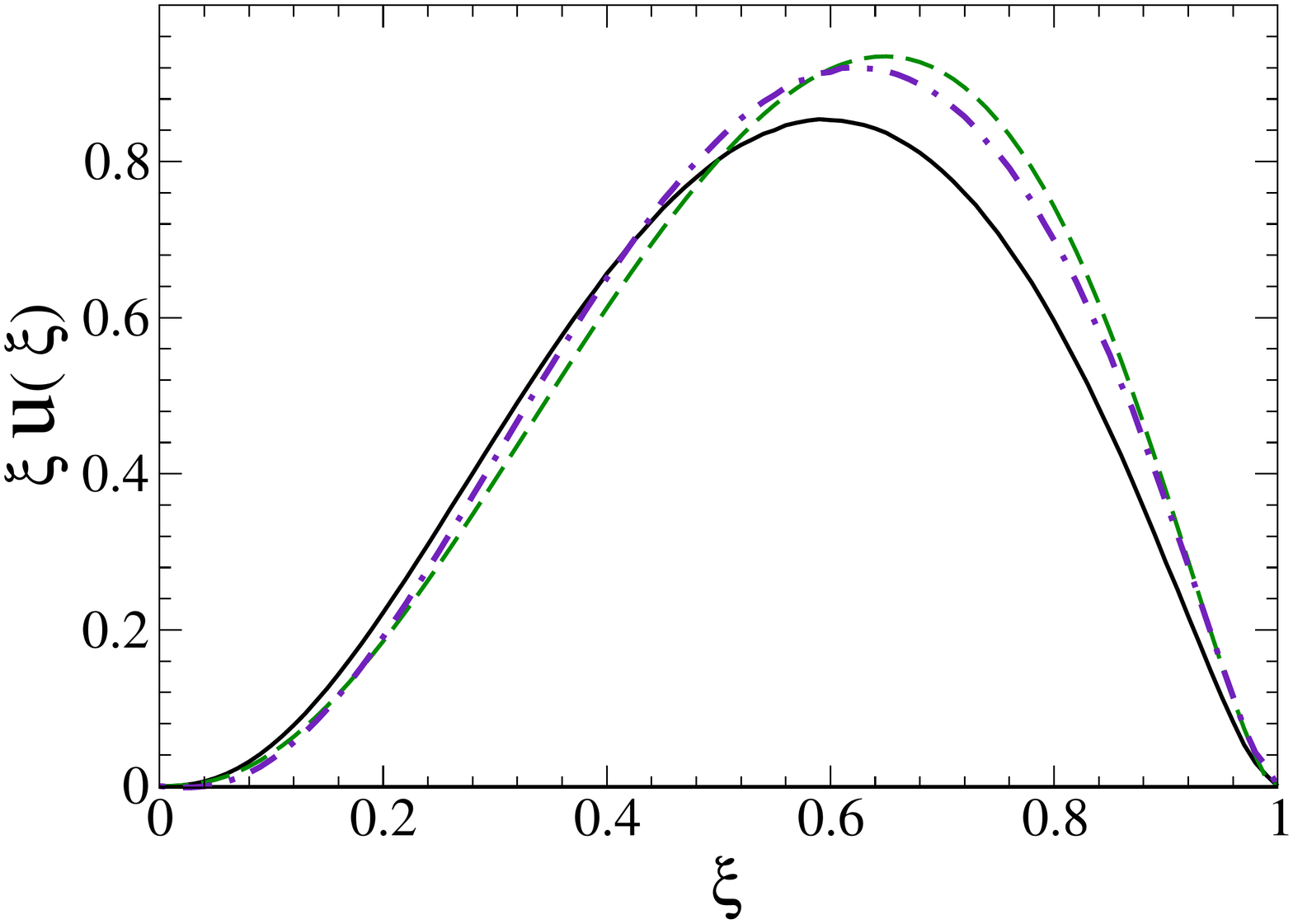}
\hspace{-0.8cm}\includegraphics[width=8.8cm]{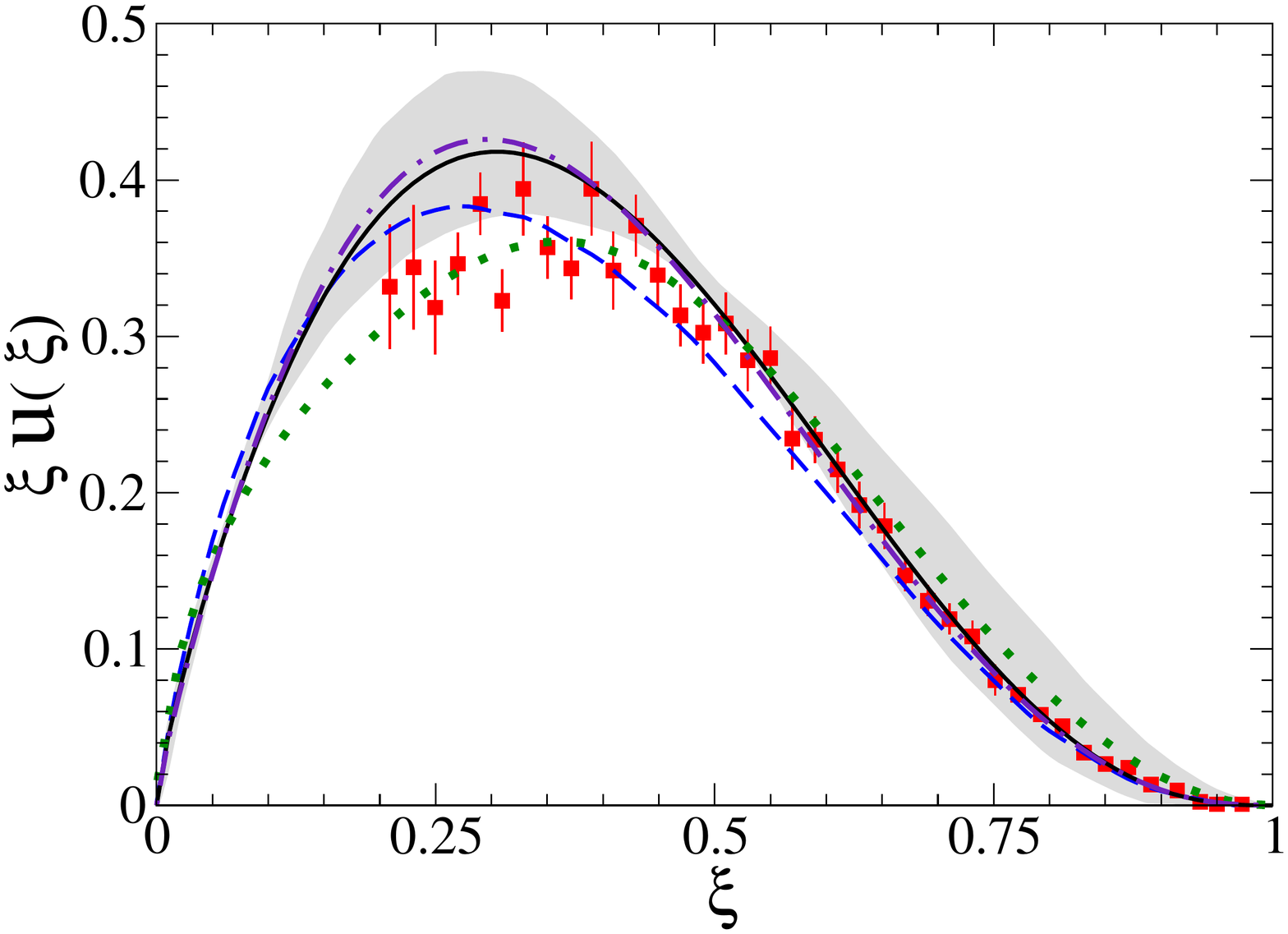}
\caption{ (Color online). {\it Left panel}. Pion longitudinal distributions, with different scales (see text for details). Dashed line: $\xi\,u^S(\xi)$, with  an assigned   initial scale equal to $360\,$MeV, and first moment equal to $0.5$. Solid  line:  $\xi\,u^q(\xi)$, with a deduced scale equal $389\,$MeV, obtained by using a backward evolution of the first moment $\langle \xi_q \rangle=0.471$. Dot-dashed line $\xi~u^q(\xi)$ backward-evolved from $389\,$MeV to $360\,$MeV.  
{\it Right panel}. Comparison with the experimental data at the scale $5.2\,$GeV. Solid line: evolved $u^S(\xi)$ starting from $360\,$MeV.  Dot-dashed line: evolved $u^q(\xi)$, starting from $389\,$MeV.
Dashed line: DSE calculation from {Fig.~5} of Ref.~\cite{Cui:2021mom}. 
Dotted line: BLFQ result at $4.0$ GeV~\cite{Lan:2019vui}. Shaded area: LQCD calculation extracted via Mellin moments from Ref.~\cite{Alexandrou:2021mmi}. Full squares: reanalyzed data by using the ratio between the fit 3 of Ref.~\cite{Aicher_PRL}, evolved to $5.2\,$GeV, and 
 the experimental data~\cite{ConwayPRD1989}, at each data point (see Ref.~\cite{dePaula:2022pcb} for details).
}
\label{fig:xpdf}
 \end{center}
\end{figure*}

We can further analyse $\langle\xi_q\rangle_{HFS}$, aiming at extracting a quantitative estimate of the exchanged-gluon contribution,  $\langle\xi_g\rangle$.  From the momentum sum rule Eq.~\eqref{Eq:HFS},   and recalling    Eq.~\eqref{Eq:xiconserv},  we get 
\begin{eqnarray}
\langle\xi_q\rangle_{HFS}&=&
{1 \over 1-P_{val}}\sum_{n>2} P_n~\langle \xi_q\rangle_n
\nonumber \\
&=& 1 -\langle\xi_{\bar q}\rangle_{HFS}-\langle\xi_g\rangle\, ,
\label{Eq:gluon_term}
\end{eqnarray}
where
\be
\langle \xi_g\rangle ={1 \over 1-P_{val}} \sum_{n\ge 3}
P_n (n-2)~\langle \xi_g\rangle_n~.
\ee
Moreover, since each Fock component fulfills the charge symmetry, i.e. $q\leftrightarrow \bar q$, the corresponding quark and antiquark momentum densities are equal and hence {for the Mellin moments 
 one has $\langle\xi_q^{k}\rangle_{HFS}=\langle\xi_{\bar q}^{k}\rangle_{HFS}$ }(this property does not imply  the charge symmetry of the total density, given the presence of the gluon contribution, cf. Eq. \eqref{Eq:HFS}). From Eq. \eqref{Eq:gluon_term}, it follows that the gluon contribution reads
\be
  \langle \xi_g\rangle =1 -2~ \langle \xi_q\rangle_{HFS}\, .
\ee
 Then, in our model one has $\langle \xi_{g}
 \rangle=0.2$. We should note that  i) {   $\langle\xi_q\rangle>\langle\xi_q\rangle_{HFS}$}, 
 as it should be, and ii)  the massive gluons carry 20\% of the HFS momentum fraction  and contribute to the total longitudinal fraction by  6\% (recalling that $P_{HFS}=0.3$).  This result indicates that the exchanged gluons in the pion are not soft (differently from the ones considered in Ref.~\cite{Chang:2014lva} where the subtraction of the effect due to soft gluons is advocated for  getting a symmetric PDF from the LF projected BS amplitude).

 It has to be emphasized that the above analysis, made transparent by the adopted LF variables,  is valid in any gauge (both covariant gauges or the light-cone one), and the only difference is the amount of the shift one gets. The possibility to regain the full gauge-invariance by taking into account the additional gluon exchanges that could affect the interaction between the knocked-out quark and the spectator one (see, e.g., the analysis of the gauge-invariance  and the hand-bag contribution in Ref.~\cite{jaffe9602236spin,Collins:2011zzd}) will be explored elsewhere.

The real test of the  longitudinal dof  is obviously the comparison between the PDF and  the experimental data~\cite{ConwayPRD1989}. As it is shown in Ref.~\cite{dePaula:2022pcb},  after evolving $u^S(\xi)$ from an assigned initial scale of $360\,$ MeV (suggested by the inflection point of the effective running charge $\alpha_s(Q^2)$)   to  the scale of $5.2\,$GeV, as given by the reanalysis in Ref.~\cite{Wijesooriya:2005ir}, the   result  compares very satisfactorily with the   experimental data extracted by taking into account logarithmic resummation effects in the hard part of the Drell-Yan cross-section, as performed in Ref.~\cite{Aicher_PRL}. Moreover,  we have achieved  a nice agreement with other dynamical calculations, such as the Dyson-Schwinger result of Ref.~\cite{Cui:2021mom}, the basis light-front quantization  calculation of Refs.~\cite{Lan:2021wok,Lan:2022phd}, and also the recent LQCD outcomes of Ref.~\cite{Alexandrou:2021mmi}. In particular, both the overall shape  and, importantly, the  tail for $\xi\to 1$,  gives  great confidence in our formalism, and encourages the further steps we have undertaken in this work. 

\begin{figure*}[t]
\begin{center}
\includegraphics[width=8.8cm]{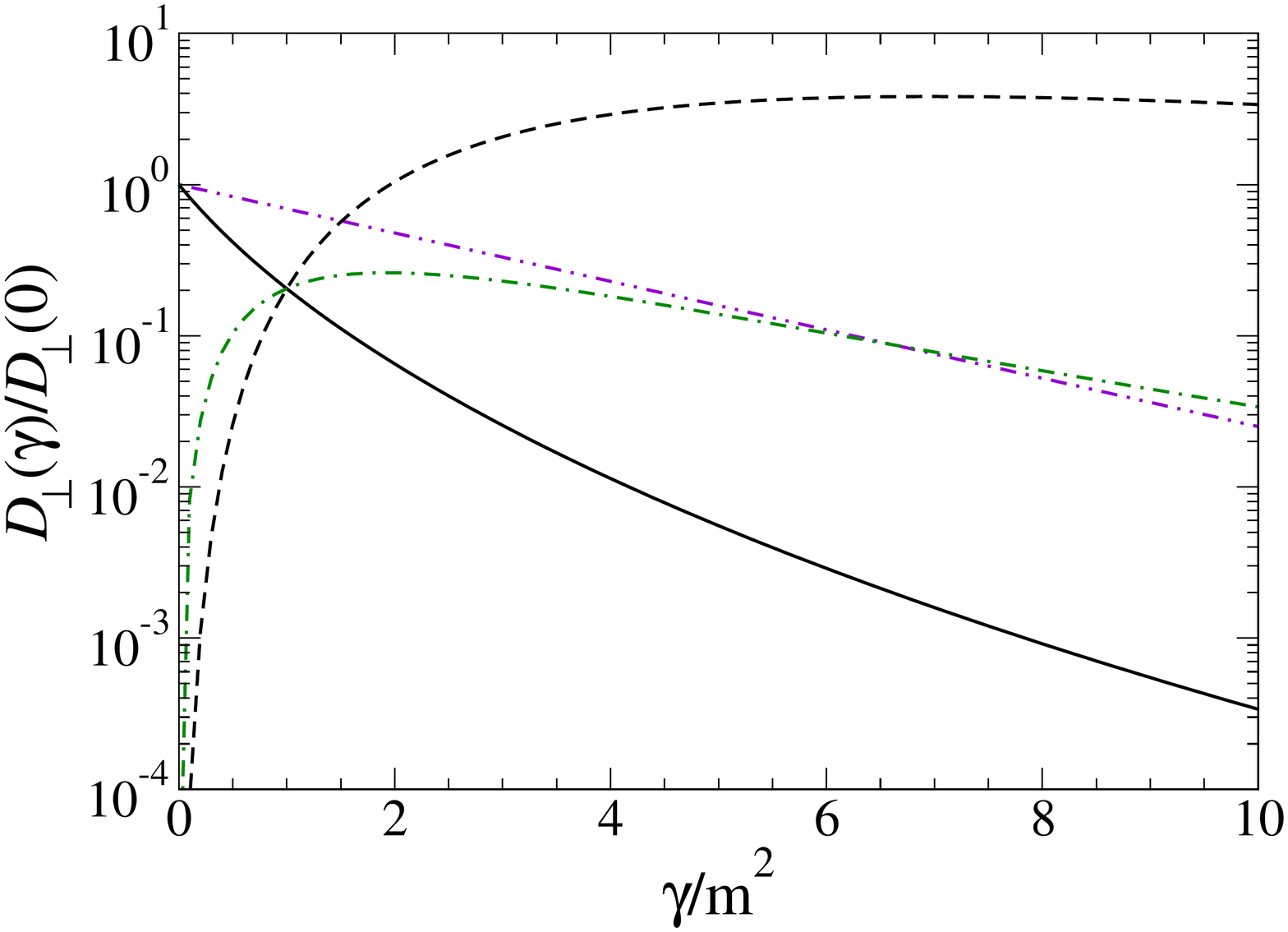}
\hspace{-0.8cm}\includegraphics[width=8.8cm]{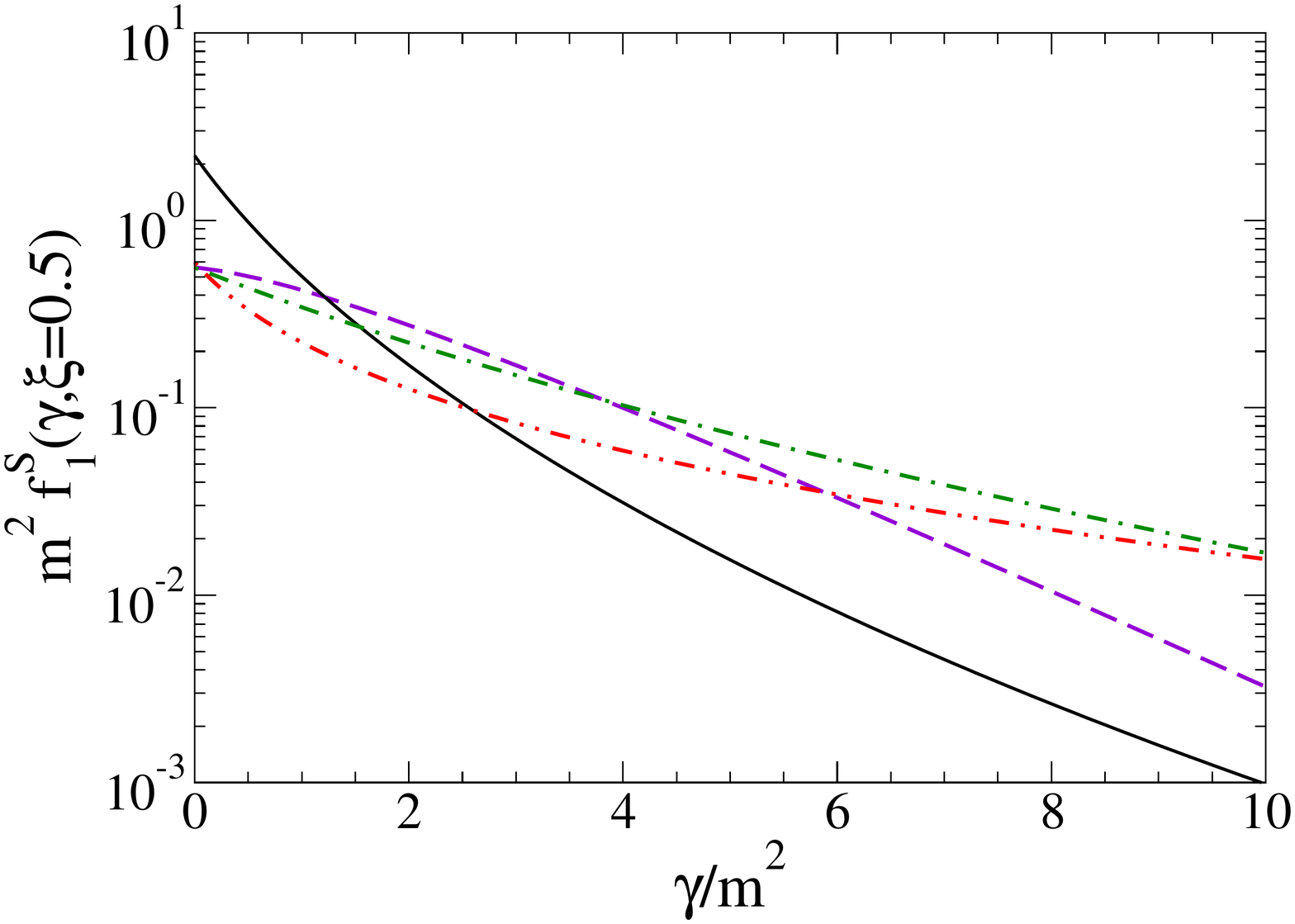}
\caption{ (Color online) {\it Left panel. } Normalized pion transverse 
distribution function, Eq.~\eqref{Eq:dperp}, vs $\gamma/m^2$. The normalization is given by $D_\perp(0)=22.945\,{\rm GeV^{-2}}$. Thick solid line: full calculation.
 Dashed line: the same as the full line, but times $(\gamma/m^2)^4$.  Dash-dotted line: the same as the full line, but times $(\gamma/m^2)^2$.
 Dash-double-dotted line: exponential form  $e^{-\gamma/(m~0.42)^2}$,
with the parameter from Table 1 of Ref.~\cite{Lorce:2016ugb}, corresponding to a Gaussian Ansatz for $f_1(\gamma,\xi)$ (see text). {\it Right Panel.} Pion unpolarized transverse-momentum  distribution $f^S_1(\gamma,\xi)$, Eq.~\eqref{Eq:f1t}, for  $\xi=0.5$.
Solid line: full calculation as in Fig.~\ref{fig:f1_pion}. Dashed line: LF constituent quark model~\cite{Pasquini:2014ppa,Lorce:2016ugb}. Dash-dotted line: LF wave function from DSE calculations~\cite{Shi:2020pqe}. Dash-Double-dotted line:  NJL model~\cite{Noguera:2015iia}. The adopted quark mass $m=255\,$MeV.}
\label{fig:dperp-f1_xi_fix}
 \end{center}
\end{figure*} 
In Fig.~\ref{fig:xpdf},  one can observe a further comparison,  involving the product $\xi\,u(\xi)$, that sheds more light on the link between the shift of the peak and the gluon dynamics taken explicitly into account in the ladder kernel of the BSE. In particular, we get a scale of $389$ MeV for  $u^q(\xi)$, the solid line  in the left panel of Fig.~\ref{fig:pdf}, by backward-evolving its first moment,  $\langle \xi_q\rangle =0.471$ (cf. Eq.~\eqref{Eq:f1_firstm}), to $0.5$,  the first moment of  $u^S(\xi)$, that has an assigned {\em hadronic} scale of $360$ MeV, as above mentioned and   thoroughly discussed in Ref.~\cite{dePaula:2022pcb}. In the right panel, the comparison at 360 MeV between $\xi\,u^S(\xi)$ and the backward-evolved $\xi\,u^q(\xi)$ shows that the effect of the interaction taken into account  in the ladder BSE  is  reproduced at large extent by applying a leading-order DGLAP evolution with an effective running charge as suggested in Ref.~\cite{Cui:2020tdf} and already applied to our PDF in Ref.~\cite{dePaula:2022pcb}.  This is not surprising once we remind that the dressing of the quark-gluon vertex, as expressed by the effective charge, is governed by the same interaction kernel  present in the BSE (i.e. the $q\bar q$ amputated  T-matrix). The left panel shows the comparison at $5.2\,$GeV between  the evolved $u^S(\xi)$, starting from the scale of $360\,$MeV, and the evolved $u^q(\xi)$, starting from the scale of $389\,$MeV. Nicely, the difference is even  smaller.

\subsection{Transverse degree of freedom}
 In the left panel of Fig.~\ref{fig:dperp-f1_xi_fix}, it is shown the transverse distribution defined by
\be 
 {\cal D}_\perp(\gamma)=\int_0^1 d\xi~f^{S}_1(\gamma,\xi)=\int_0^1 d\xi~f^{q}_1(\gamma,\xi)~.
 \label{Eq:dperp}
 \ee 
 It has to be pointed out that the integration on $\xi$ eliminates the anti-symmetric  term $f^{AS}_1(\gamma,\xi)$, and therefore one gets the same transverse distribution also by using $f^{q}_1(\gamma,\xi)$.
In order to emphasize the analysis of the general pattern, we have presented ${\cal D}_\perp(\gamma)/{\cal D}_\perp(0)$, so that the widely adopted exponential or power-like fall-off can be readily compared to our result.

In  addition, in the left panel of Fig.~\ref{fig:dperp-f1_xi_fix} one can find:  i) an exponential form ${\cal D}_{\perp}(\gamma)/{\cal D}_{\perp}(0)= e^{-\gamma/(m\,0.42)^2}$,  with the parameter given in Table 1 of Ref.~\cite{Lorce:2016ugb}, corresponding to the so-called Gaussian Ans\"{a}tz (recall $\gamma=|{\bf k}_\perp|^2$), amounting to  a  factorized form for  $f^{S}_1(\gamma,\xi)\sim u^S(\xi) e^{-\gamma/(0.42^2)}$ very often adopted   in  phenomenological studies; ii) our full results multiplied by $(\gamma/m^2)^2$ and iii) our full results multiplied by $(\gamma/m^2)^4$. Not surprisingly,  a power-like shape  provides a better approximation to the dynamically calculated $D_\perp(\gamma)$, but the proper power is different from the one expected by the action of only a one-gluon exchange, that should govern the ultraviolet behavior and lead to a $(\gamma/m^2)^2$ (as suggested by a generalized counting rule in Ref.~\cite{Ji:2003fw}).  
Indeed, the adopted    form-factor featuring the extension of the quark-gluon interaction vertex (cf. Eq.~\eqref{Eq:def1}) generates a different power-like fall-off, namely  $(\gamma/m^2)^4$, as already pointed out  in Refs.~\cite{dePaula:2016oct,dePaula:2017ikc}.
 Finally, it is worth noticing that, unlike the PDF, the two terms in  $f^S_1(\gamma,\xi)$ containing derivatives of the delta-function do not contribute, as it is discussed  at the end of  Appendix~\ref{norm_app}.

In the right panel of Fig.~\ref{fig:dperp-f1_xi_fix}, it is presented the quantitative comparison  between $f^S_1(\gamma,\xi)$ at $\xi=0.5$  and  some phenomenological outcomes from i) the  approach  based on the LF wave function obtained by using the DSE calculation in Ref.~\cite{Shi:2020pqe}; ii) the LF constituent quark-model of Ref.~\cite{Pasquini:2014ppa,Lorce:2016ugb}; iii) the  NJL model  with Pauli-Villars regulator as given in  Ref. \cite{Noguera:2015iia}. For $\gamma/m^2\to 0$, there are   remarkable differences that, indeed, are present  also on  the tails. This last feature impacts  the value of  $\langle \gamma/m^2\rangle^\frac12$, as  shown in Table~\ref{Tab:f1xi}, where, for the sake of completeness,   the value of $u^S(\xi=0.5)$ and the pion charge radius are also presented.  As can be expected, the larger the average transverse moment, the smaller the radius of charge. The current  model has the smaller $\langle \gamma/m^2\rangle^\frac12$ (of the  order of the infrared scale $\Lambda_{QCD}$,  effectively incorporated in the QCD-inspired choice of our parameters) which in turn leads to a larger charge radius, in agreement with the experimental value. 
\begin{table}[htb]
 \caption{The average value $\langle \gamma/m^2\rangle$ (with $m=0.255\,$MeV), $u^S(\xi=0.5)$ and the  pion charge radius  are presented for:  i) $f^S_1(\gamma,\xi=0.5)$ from the present approach (NIR+BSE); ii) the outcome from the LF wave function obtained by using DSE calculation~\cite{Shi:2020pqe} (LFDSE); iii) the LF constituent quark-model of Ref.~\cite{Pasquini:2014ppa,Lorce:2016ugb} (LFCQM) and iv) 
 the NJL with Pauli-Villars regulator~\cite{Noguera:2015iia}. (Recall that the most recent PDG value of the charge radius is $r^{PDG}_{ch}=0.659\pm 0.004$ {fm}~\cite{Zyla:2020zbs}).}
  \label{Tab:f1xi}
\begin{center}
 \begin{tabular}{cccc}
 \hline  \hline
   ~& $\langle \gamma/m^2\rangle^\frac12$ 
   &$u^S(\xi=0.5)$&$r_{ch}$ [fm]\\
 \hline
 NIR+BSE& 1.25 
 &1.60&0.663\\
LFDSE   &1.94 
&  1.36&0.590\\
LFCQM  & 1.65 
&  1.37&0.572 \\
NJL  & 2.02 
& 1.01&0.557\\
 \hline
 \end{tabular}
 \end{center}
 \end{table}

\begin{figure}[t]
\begin{center}
\hspace{-0.2cm}\includegraphics[width=8.3cm]{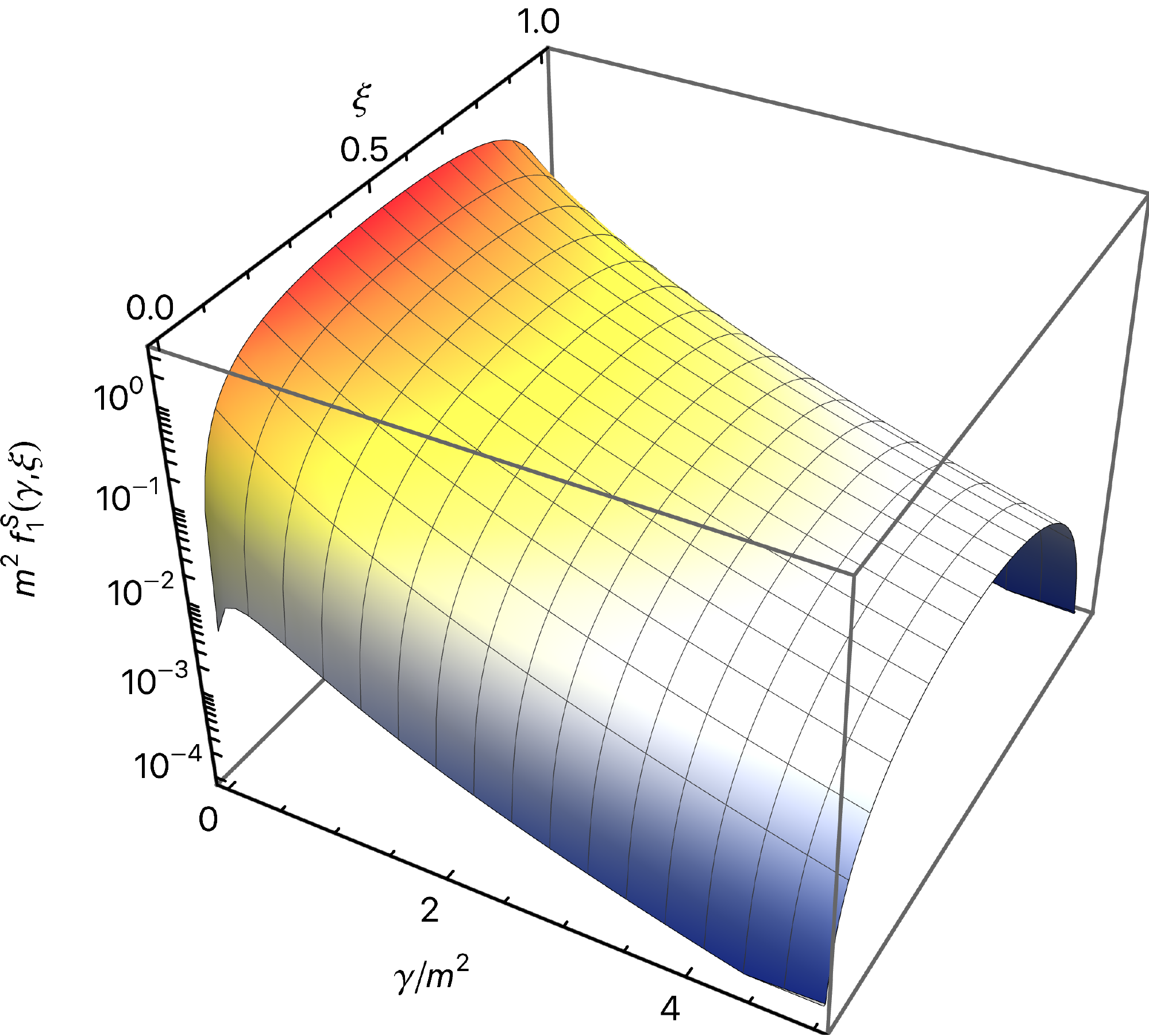}
\caption{ (Color online) Pion unpolarized transverse-momentum  distribution $f^S_1(\gamma,\xi)$, Eq.~\eqref{Eq:f1t}, at the initial scale. The normalization is $\int_0^1d\xi \int_0^\infty d\gamma ~f^S_1(\gamma,\xi)=1$.}
\label{fig:f1_pion}
 \end{center}
\end{figure}

In Fig.~\ref{fig:f1_pion}, the uTMD $f^S_1(\gamma,\xi)$ is shown in full, in order to appreciate the main features, i.e.  i) the peak at $\xi=0.5$ for running $\gamma/m^2$, ii) the vanishing values at the end-points and iii) the order of magnitude fall-off already for $\gamma/m^2> 2 $. Comparing to other approaches, one can notice the sharp  difference with the results from  the LF constituent model in Ref.~\cite{Pasquini:2014ppa} and the  LF holographic framework, like   in Refs.~\cite{Bacchetta:2017vzh,Ahmady:2019yvo} where a  double-humped structure is found  due to 
the $\xi$-dependence in the holographic wave
functions. Also the value at $\xi=0.5$ and small $\gamma/m^2$ is substantially  lower than ours (almost an order of magnitude less). Differently, the shape of our  $f^S_1(\gamma,\xi)$  is more similar, i.e. without any double-humped structure, to the one obtained in Ref.~\cite{Shi:2020pqe}, where  the pion LF-wave function   is determined from a beyond rainbow-ladder Dyson-Schwinger equations (DSE) in Euclidean space,  by exploiting the $\gamma$-dependent moments in $\xi$ and a suitable parametrization of the BS-amplitude.

\section{The subleading-twist uTMDs}
\label{Sec:subl}
\begin{figure*}[t]
\begin{center}
\includegraphics[width=8.8cm]{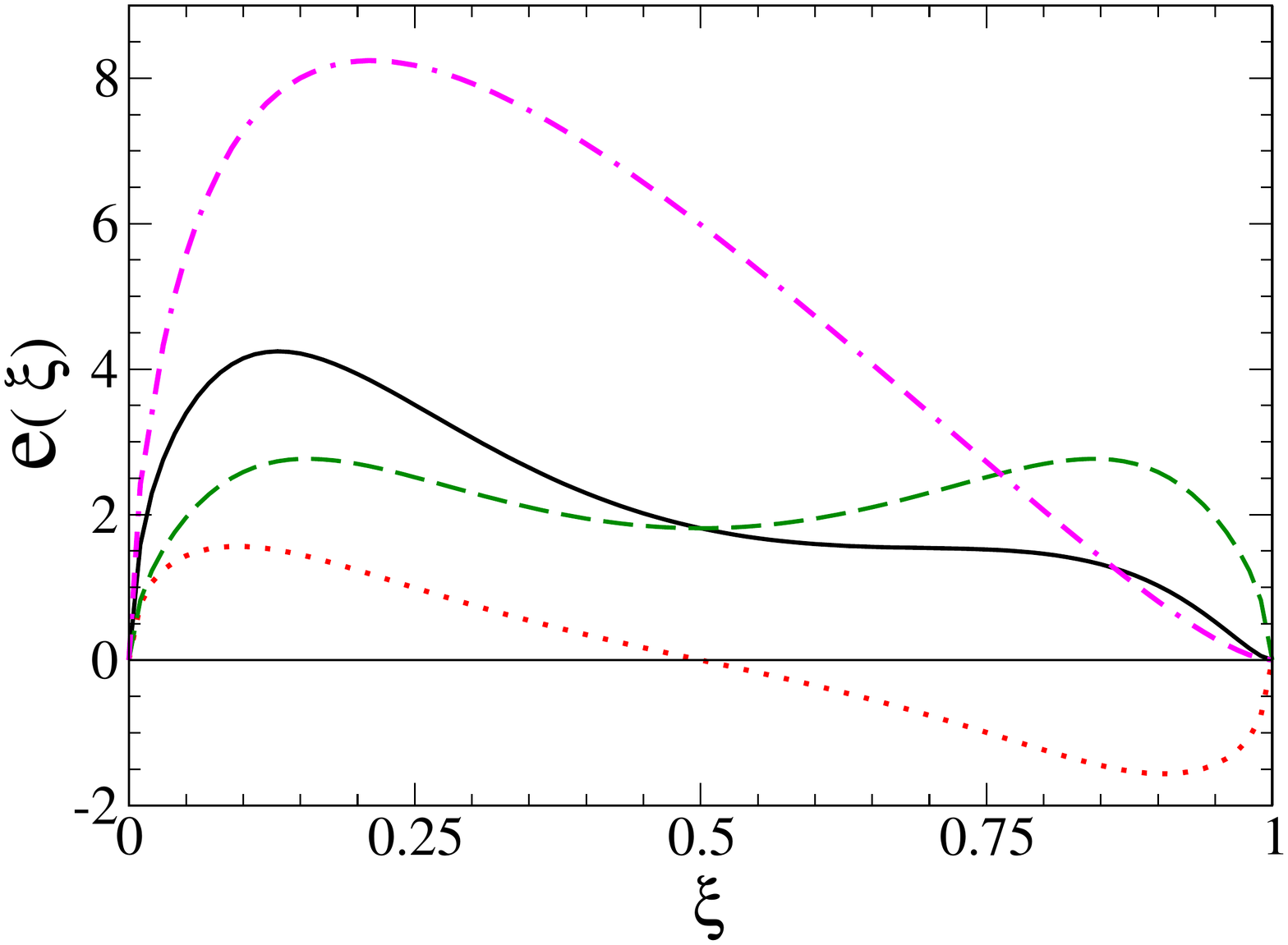}
\hspace{-0.8cm}\includegraphics[width=8.8cm]{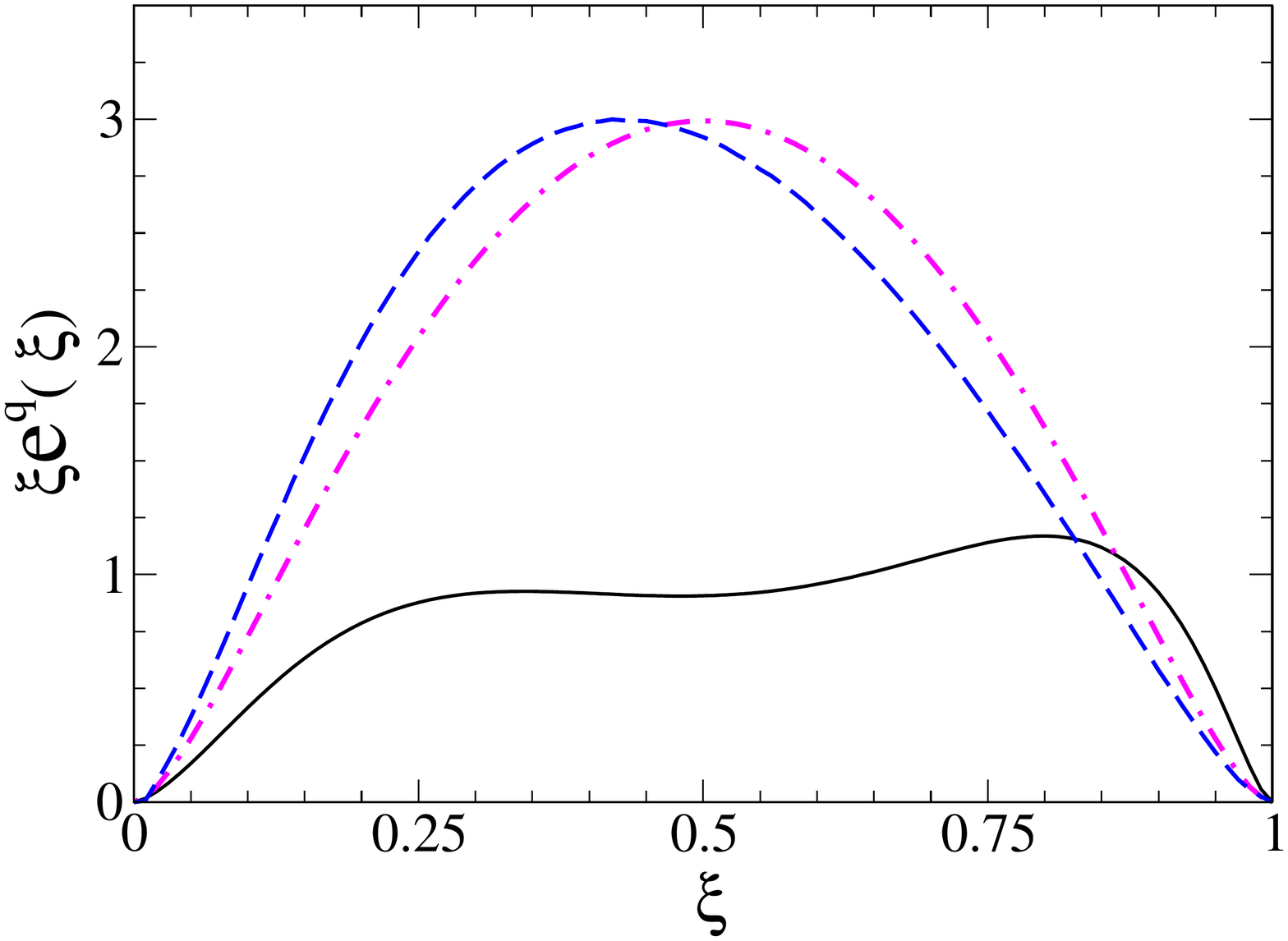}
\caption{ (Color online) {\it  Left panel.} Pion unpolarized collinear  PDFs: i)  $e^q(\xi)$ (solid line), Eq.~\eqref{Eq:eLq}, ii) $e^S(\xi)$ (dashed line)  and $e^{AS}(\xi)$ (dotted line),  Eqs.~\eqref{Eq:eLSAS}.  It is also shown $e^q_{EoM}(\xi)$ (dash-dotted line), Eq.~\eqref{Eq:eqfree}. {\it  Right panel.} 
Quark unpolarized collinear  PDFs:  $\xi~e^q(\xi)$. Solid line: full calculation as in the left panel.
Dashed line: $m/M~u^q(\xi)$, with $u^q(\xi)$ shown in the right panel of Fig. \ref{fig:pdf}. Double-dot-dashed line: $\xi~e^q_{EoM}(\xi)$, Eq. \eqref{Eq:eqfree}.
}
\label{fig:eL_pion}
 \end{center}
\end{figure*}

In this Section we present the numerical results for  (T-even)  uTMDs beyond the leading-twist. The detailed expressions can be found in the Appendix~\ref{twist34_app}, but it is useful to recall that the  decomposition in symmetric and antisymmetric combinations adopted for $f_1(\gamma,\xi)$ remains still valid, as well as the relations with  the quark and anti-quark contributions.

As introduction to the outcomes of our dynamical approach, it is worth  anticipating that the  comparison between  full calculations and    naive estimates one can infer from Eq.~\eqref{Eq:LIR} by using a valence approximation of the leading-twist $f_1(\gamma,\xi)$,  highlights the inspiring  statement one can  read in Ref.~\cite{Jaffe:1991ra}: {\em the higher-twist distributions are naturally related to
multiparton distributions}. The role of the exchanged gluons becomes definitely clear through a remarkable shift of the peak in all the sub-leading uTMD we have analyzed, as already discussed in the previous Section,  as well as through the sharp difference with the naive estimates, which exclude the effect of  the one-gluon exchange.

\subsection{Twist-3 uTMD:  $e(\gamma,\xi)$}\label{subsec:e}

In the frame where ${\bf P}_\perp=0$ and hence $P^+=M$, by using Eq.~\eqref{Eq:tmd_t2}, \eqref{calF0b_app}, \eqref{calF1b_app},  \eqref{calF2b_app} and \eqref{calF3b_app}, 
with $i=1$ and the functions $b^1_{n;\ell j}$ given in  Table~\ref{b1_tab},
one gets  the  twist-3 uTMDs $e^{S(AS)}(\gamma,\xi)$, decomposed as follows
\be 
e^{S(AS)}(\gamma,\xi)=
{\cal E}_0(\gamma,\xi;S(AS))+{\cal E}_d(\gamma,\xi;S(AS))
\nonu+{\cal E}_{2d}(\gamma,\xi;S(AS))+{\cal E}_{3d}(\gamma,\xi;S(AS))~,
\label{Eq:eS_AS}\ee 
where the functions in the rhs are given  in Appendix~\ref{twist34_app}.

\begin{figure*}[t]
    \begin{center}
    \includegraphics[width=8.8cm]{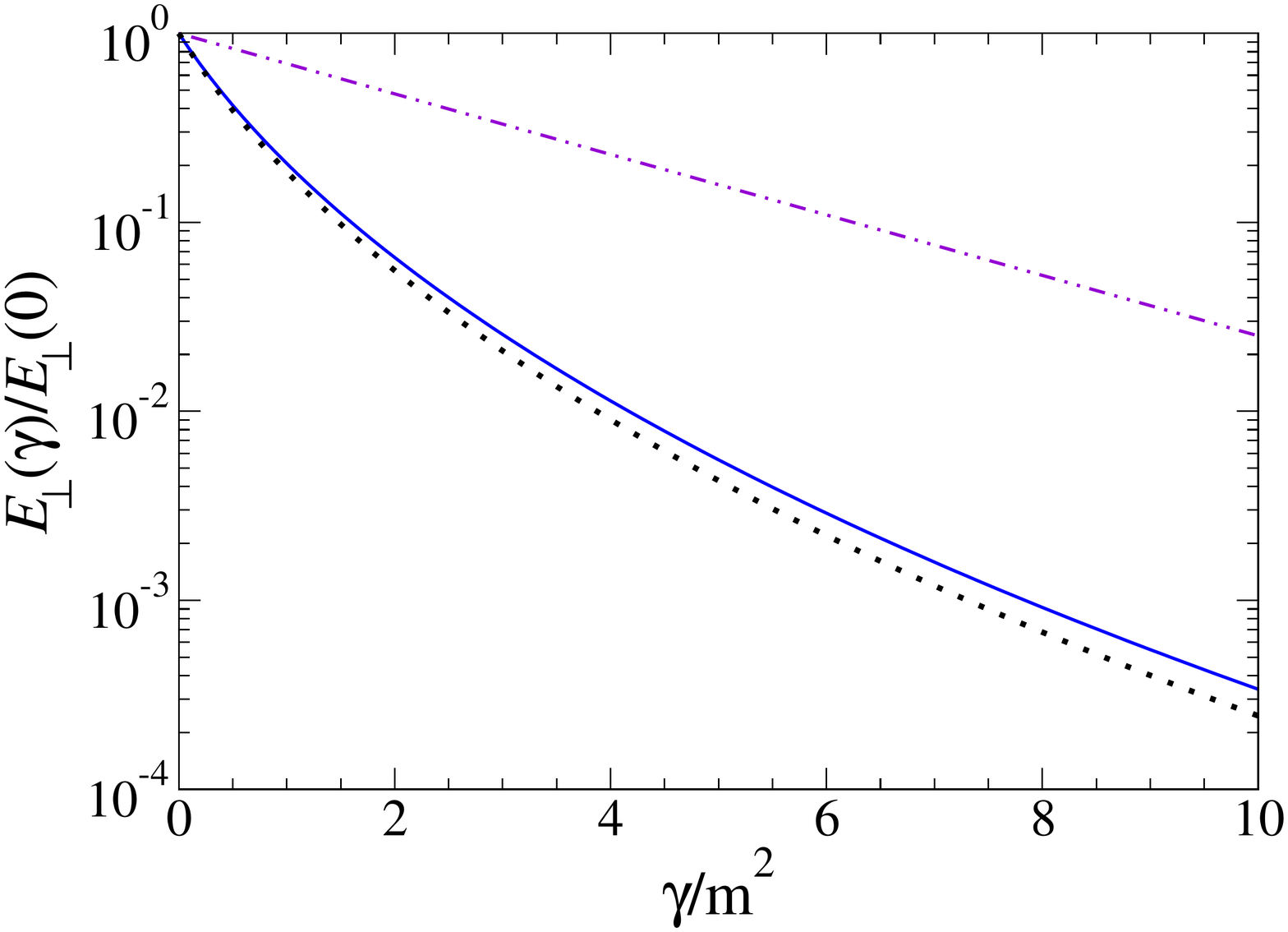}   \hspace{-0.8cm}\includegraphics[width=8.8cm]{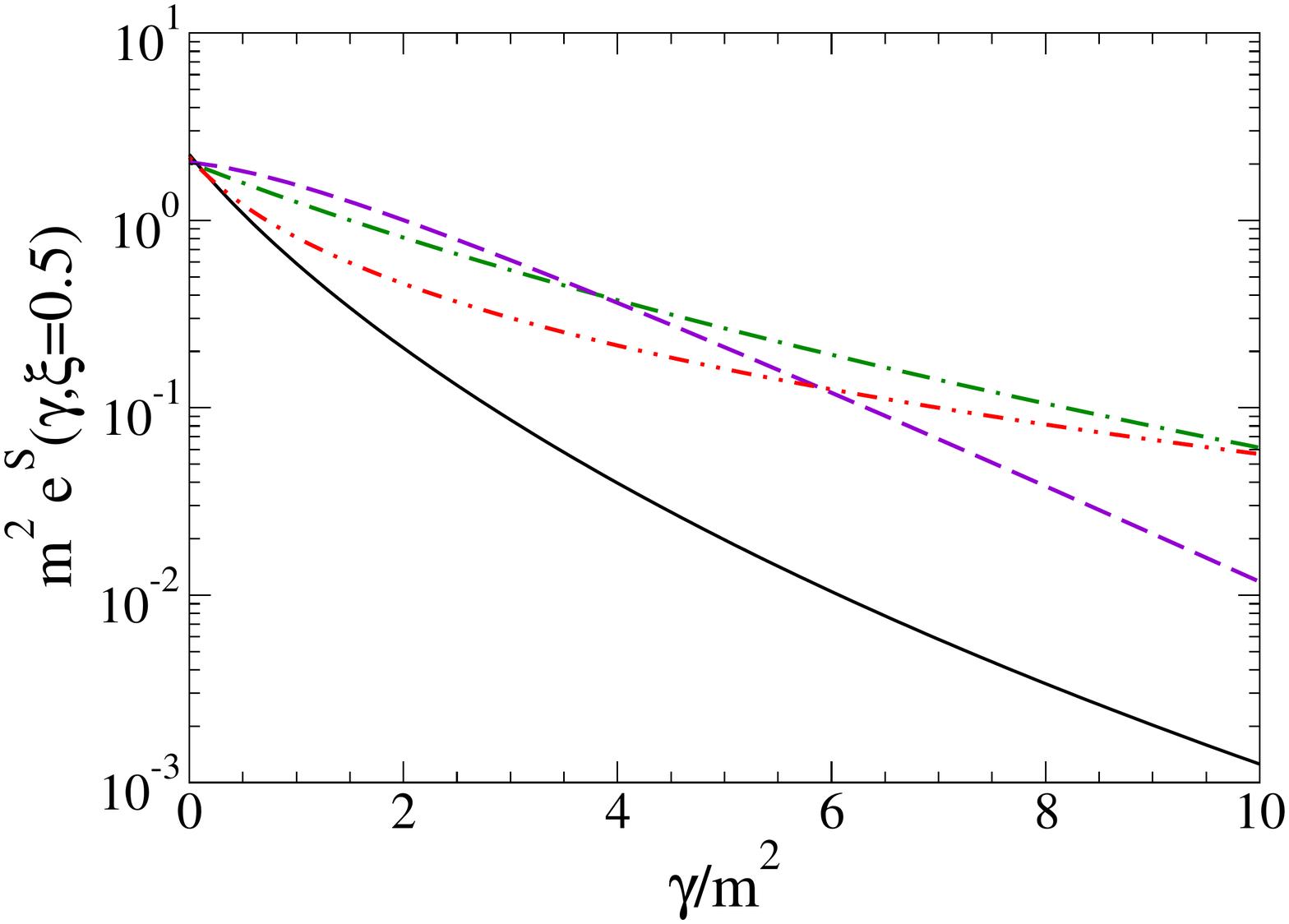}
    \caption{(Color online). {\it Left panel.} Normalized transverse distribution function $E_\perp(\gamma)/E_\perp(0)$ (cf. Eq.~\eqref{Eq:etran}). Dotted line: full calculation. Solid line: $D_\perp(\gamma)/D_\perp(0)$ for the sake of comparison. Dash-double-dotted line: the same as in the left panel of 
 Fig.~\ref{fig:dperp-f1_xi_fix}. {\it Right panel.} Pion unpolarized transverse-momentum  distribution $e^S(\gamma,\xi)$, Eq.~\eqref{Eq:eS_AS}, for  $\xi=0.5$.
Solid line: full calculation. Dashed line:  LF constituent quark model of Ref.~\cite{Pasquini:2014ppa,Lorce:2016ugb}, but multiplied by $m/(M~0.5)$ (cf. Eq.~\eqref{Eq:eqfree}). Dash-dotted line: the same as the dashed line but with the LF wave function from DSE calculations~\cite{Shi:2020pqe}. Dash-Double-dotted line: the same as the dashed line but with the  NJL model~\cite{Noguera:2015iia}. The adopted quark mass $m=255\,$MeV.}
\label{fig:etran-e_xi_fix}
    \end{center}   
\end{figure*}

\subsubsection{Longitudinal degree of freedom}

In the left panel of Fig.~\ref{fig:eL_pion}, the following  collinear PDFs
are shown
\be 
e^{(S,AS)}(\xi)=\int_0^\infty d\gamma~e^{(S,AS)}(\gamma,\xi)~.
\label{Eq:eLSAS}
\ee
and 
\be 
e^q(\xi)=e^{S}(\xi)+e^{AS}(\xi)~. 
\label{Eq:eLq}\ee
Moreover, in the spirit of Ref.~\cite{Lorce:2016ugb}, we also present the collinear PDF, $ e^q_{EoM}(\xi)$, obtained by integrating  the first line in  Eq.~\eqref{Eq:LIR}, but disregarding the gluon contribution,  
viz
\be 
e^q_{EoM}(\xi) \sim {m\over M \xi}\int_0^\infty d\gamma~ f^q_{1;EoM}(\gamma,\xi)
\nonu\sim {m\over M \xi}~{u^{LF}_{val}(\xi)\over P_{val}}~,
\label{Eq:eqfree}
\ee
where $u^{LF}_{val}(\xi)/P_{val}$,    normalized to $1$ (cf. Eq.~\eqref{Eq:norm_val}), approximates the integral of $f^q_{1;EoM}(\gamma,\xi)$.  The large difference between our $e^q(\xi)$ and   $(m/M\xi) u^{LF}_{val}(\xi)/P_{val}$ indicates the sizable role of the gluon contribution   from the HFS generated by  our dynamical model.  In addition, one should point out that the strength of $e^q(\xi)$ is spread out on the whole range of $\xi$, and not concentrated at the end-point $\xi=0$ as  QCD investigations indicate. The latter feature leads to   the singular contribution given in Eq.~\eqref{Eq:eq_sing} (see, e.g., Ref.~\cite{Efremov:2002qh},  for a detailed discussion, but notice  that  the focus is on the nucleon).

In the right panel of Fig.~\ref{fig:eL_pion}, the comparison between {$\xi\,e(\xi)$ } and  the other two approximations: i)~$(m/M) f^{q}_1(\xi)$   and ii)~$(m/M) u^{LF}_{val}(\xi)$ (cf. Eq.~\eqref{Eq:eqfree}) is carried out. The relevance of such a comparison is given by the possibility of  more directly assessing  the gluon role, since the factor $\xi$ eliminates the singular term present in the QCD analysis of $e(\xi)$,  and one  remains with the mass contribution $(m/M) f^{q}_1(\xi)$ and the term from the quark-gluon-antiquark correlator.

Still within the QCD framework (see, e.g., Ref.~\cite{Efremov:2002qh}), the    moments $\langle \xi^n\rangle_{e^q}$, for $n\le 2$,  read as follow    
\be
\int d\xi~ e^q(\xi)={\sigma_\pi\over m_{cur}}
\nonu
\int d\xi~ \xi~e^q(\xi)={m_{cur}\over M}
\nonu
\int d\xi~ \xi^2~e^q(\xi)={m_{cur}\over M} \int d\xi~ \xi~f^q_1(\xi)~,
\label{Eq:mom_eq}\ee
 and,  for $n>2$,  they  receive contributions not only from  the $(n-1)$-th moment of $f^q_1(\gamma,\xi)$ , but also from  the $n$-th moment of the twist-3 contribution pertaining to the quark-gluon-antiquark correlator. Given the highly non trivial dynamical content of the $e^q(\xi)$  moments,  it is interesting to show the results obtained  with our  dynamical model. 

\begin{table}[t]
 \caption{The moments $\langle \xi^n\rangle_{e^q}$ of the quark  twist-3 $e^q(\gamma,\xi)$ for $n<4$,  and the ratio $R(n,e^q,f^q_1)=\langle \xi^n\rangle_{e^q}/
\langle \xi^{n-1}\rangle_{f^q_1} $ (it is assumed $\langle \xi^{-1}\rangle_{f^q_1}=\langle \xi^{0}\rangle_{f^q_1}=1$, and the values of $\langle \xi^{1}\rangle_{f^q_1}=0.471$ and $\langle \xi^{2}\rangle_{f^q_1}=0.266$ have been numerically evaluated).
}
  \label{Tab:eximom}
\begin{center}
  \begin{tabular}{ccccc} \hline  \hline   
  $n$& 0& 1 & 2  & 3 \\ \hline 
  $\langle \xi^n\rangle_{e^q}$& ~2.190~~ &  ~0.814~~ & ~0.445~~ & ~0.292~~ \\ \hline 
  $R(n,e^q,f^q_1)$ & 2.190  & 0.814 & 0.943 & 1.10\\ \hline \end{tabular}
 \end{center}
 \end{table}

 \begin{figure}[t]
\begin{center}
\hspace{-0.2cm}\includegraphics[width=8.3cm]{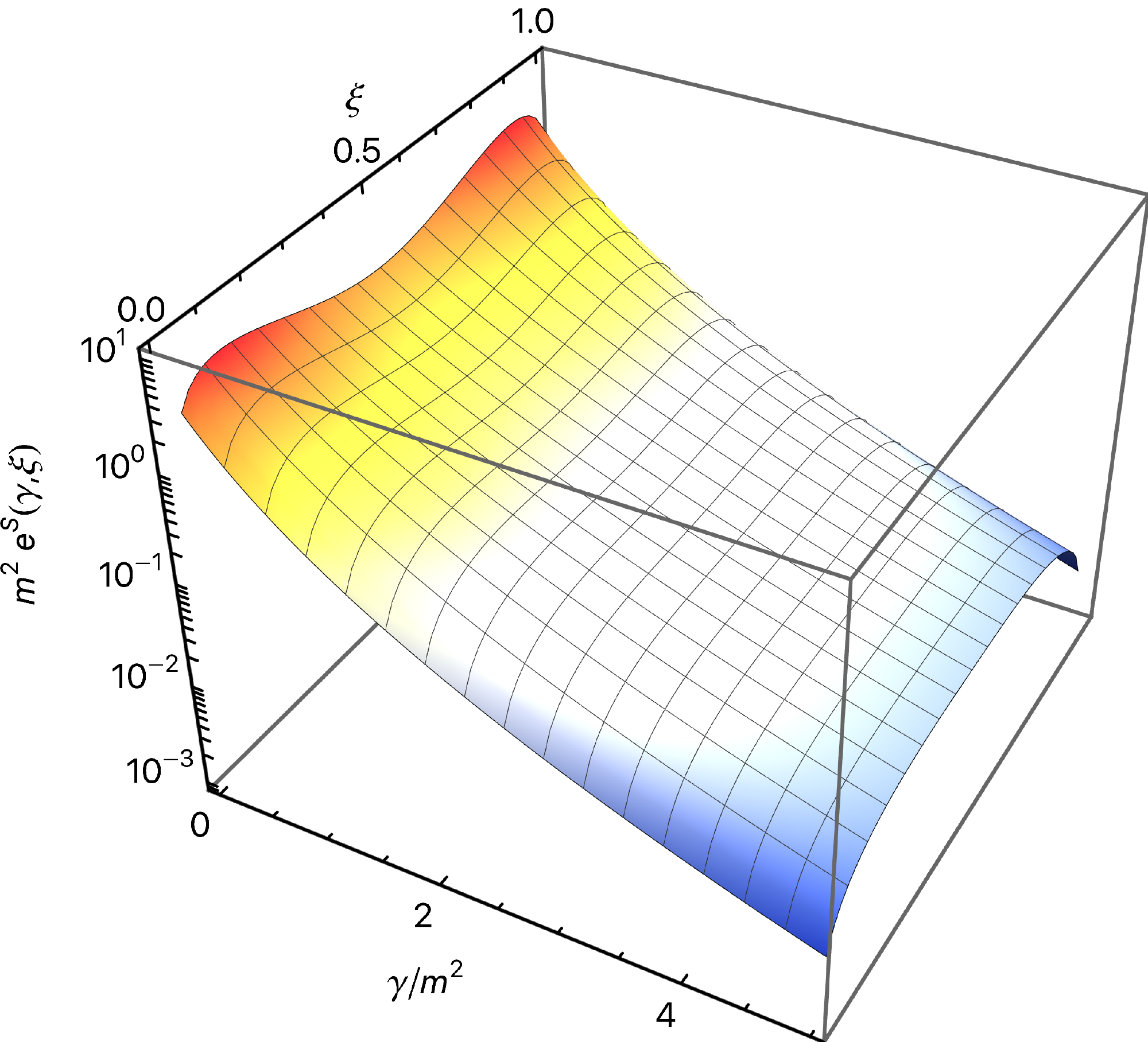}
\caption{ (Color online) Pion unpolarized transverse-momentum  distribution $e^S(\gamma,\xi)$, Eq.~\eqref{Eq:eLSAS}, at the initial scale. }
\label{fig:eS_pion}
 \end{center}
\end{figure}

 In Table~\ref{Tab:eximom}, both the moments up to $n=3$ and the ratio   $R(n,e^q,f^q_1)=\langle \xi^n\rangle_{e^q}/
\langle \xi^{n-1}\rangle_{f^q_1}$ are  presented.
In particular, as to the first two moments,
to get rid of the dependence upon $m_{cur}$ it is helpful to compare the result obtained by multiplying the zero-th and the first moment, (cf. Eq.~\eqref{Eq:mom_eq}), with final outcome $\sigma_\pi/M$. The estimate of $\sigma_\pi$ at  the
leading order of the chiral expansion leads to $\sigma_\pi/M=1/2$, as satisfactorily confirmed by the lattice calculations in  Ref.~\cite{Bali:2016lvx}, where    $\sigma^{lat}_\pi=78.2\pm 4.2\,$MeV, for $M=149.5\pm 1.3\,$MeV and $m_{cur} \sim 4.9\,$MeV.  Eliminating the current quark mass,  that is outside our approach, through the above product, we get $\sigma_\pi/M=1.78$, instead of $\sim 0.5$.  Such a conspicuous difference is surely influenced by the different distribution of the $e^q(\xi)$ strength, as already mentioned, and points to a needed enrichment of the gluon dynamics in our approach.   However, it is worth mentioning   that for a simple non relativistic constituent quark model one has    $\sigma_\pi^{NR}=2 m$, so that $\sigma_\pi^{NR}/M=3.64$ (with $m=255$ MeV the constituent mass), almost twice the result  obtained in the BS framework. 

In QCD, the ratios $R(1,e^q,f^q_1)$ and  $R(2,e^q,f^q_1)$ are equal and amount to $m_{cur}/M$ (see Eq.~\eqref{Eq:mom_eq}),   while $R(3,e^q,f^q_1)=m_{cur}/M+\Delta^3_g$, where  $\Delta^3_g$ contains  the contribution from
the twist-3 gluonic contribution.  In our calculation, the  ratios for $n=1,2$ are almost equal,  but different from the naive expectation $m/M=1.82$ with the adopted $m=255$ MeV.  The difference with the third ratio indicates the onset  of the contribution from the twist-3 gluonic term. 
 
\begin{figure*}[t]
\begin{center}
\includegraphics[width=8.8cm]{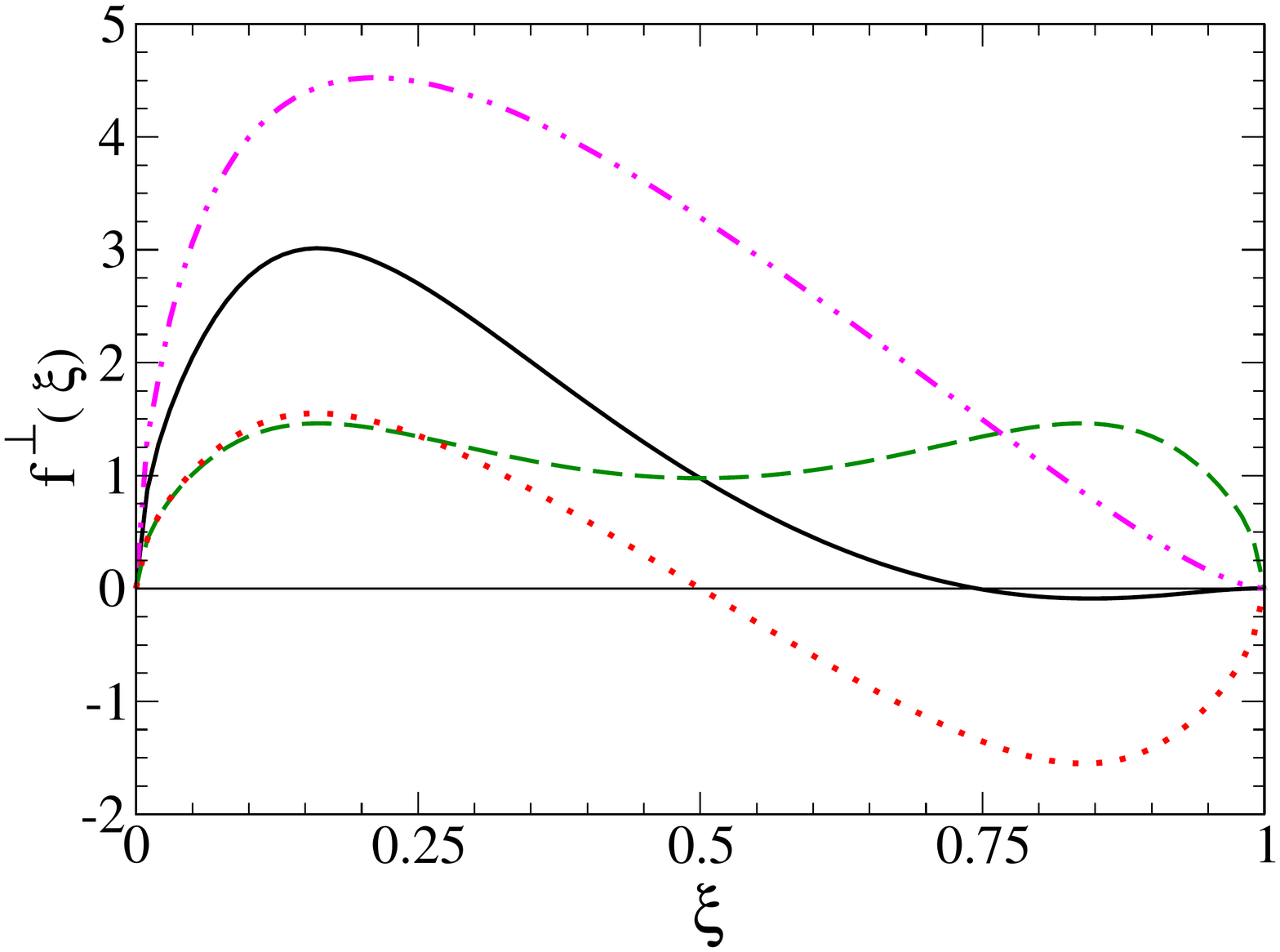}
\hspace{-0.8cm}\includegraphics[width=8.8cm]{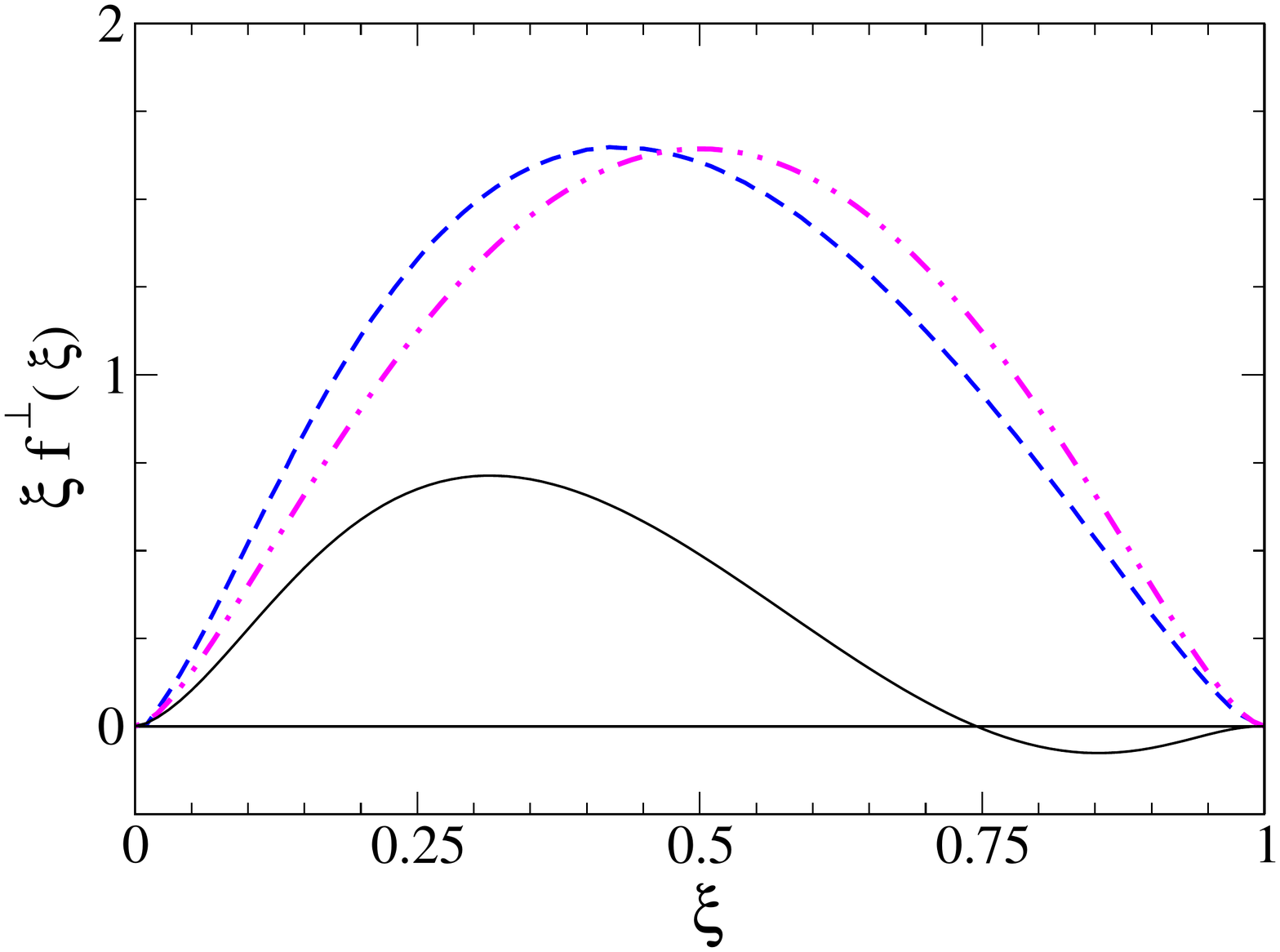}
\caption{ (Color online) {\it Left panel.}  The same as in Fig.~\ref{fig:eL_pion}, but for 
$f^{\perp q}(\xi)$, $f^{\perp S}(\xi)$ and $f^{\perp AS}(\xi)$, Eq.~\eqref{Eq:fpL}, and $f^{\perp q}_{EoM}(\xi)$ as given in Eq.~\eqref{Eq:fperpfree}.
{\it Right panel.}  Quark unpolarized collinear  PDFs $\xi\,f^{q\perp}(\xi)$. Solid line: full calculation as in left panel. 
Dashed line: $\xi\,f^{q\perp}(\xi)$ obtained by using the second line in Eq.~\eqref{Eq:LIR}  and our $f^q_1(\xi)$.
Double-dot-dashed line: the same as the dashed line but using  the valence approximation of the PDF, $u^{LF}_{val}(\xi)$, with norm equal to $1$.
}
\label{fig:fpL_pion-xifpL_pion}
 \end{center}
\end{figure*}

\subsubsection{Transverse degree of freedom}

The transverse dof can be analyzed globally by introducing the following transverse distribution function, as already accomplished with the leading-twist uTMD, viz.
\be 
E_\perp(\gamma)=\int_0^1 d\xi~ e^S(\gamma,\xi)=
\int_0^1 d\xi~ e^q(\gamma,\xi)~.
\label{Eq:etran}
\ee
In the left panel of  Fig.~\ref{fig:etran-e_xi_fix}, it is presented our calculation and the ratio $D_\perp(\gamma)/D_\perp(0)$ to show the  similar fall-off, as generated  from gluon dynamics and the form-factor featuring the quark-gluon vertex.

 A more close view of the decreasing as a function of $\gamma/m^2$ is provided by the right panel of Fig.~\ref{fig:etran-e_xi_fix}, where it is shown the comparison between our calculation of $e(\gamma,\xi=0.5)$ and the outcomes obtained by using Eq. \eqref{Eq:eqfree} with i) the LF wave function from the constituent quark model of Ref.~\cite{Pasquini:2014ppa,Lorce:2016ugb}, ii) 
the LF wave function from DSE calculations~\cite{Shi:2020pqe} and iii) the  PDF from the NJL model~\cite{Noguera:2015iia}. The differences again point to the role of the interaction in the various approaches, and highlight the relevance of an experimental investigation of the transverse dof.

In Fig.~\ref{fig:eS_pion}, the full dependence of $e^S(\gamma,\xi)$ is presented, displaying a double-hump shape that for larger $\gamma/m^2$ becomes smoother and smoother.

\begin{figure*}[t]
    \begin{center}
    \includegraphics[width=8.8cm]{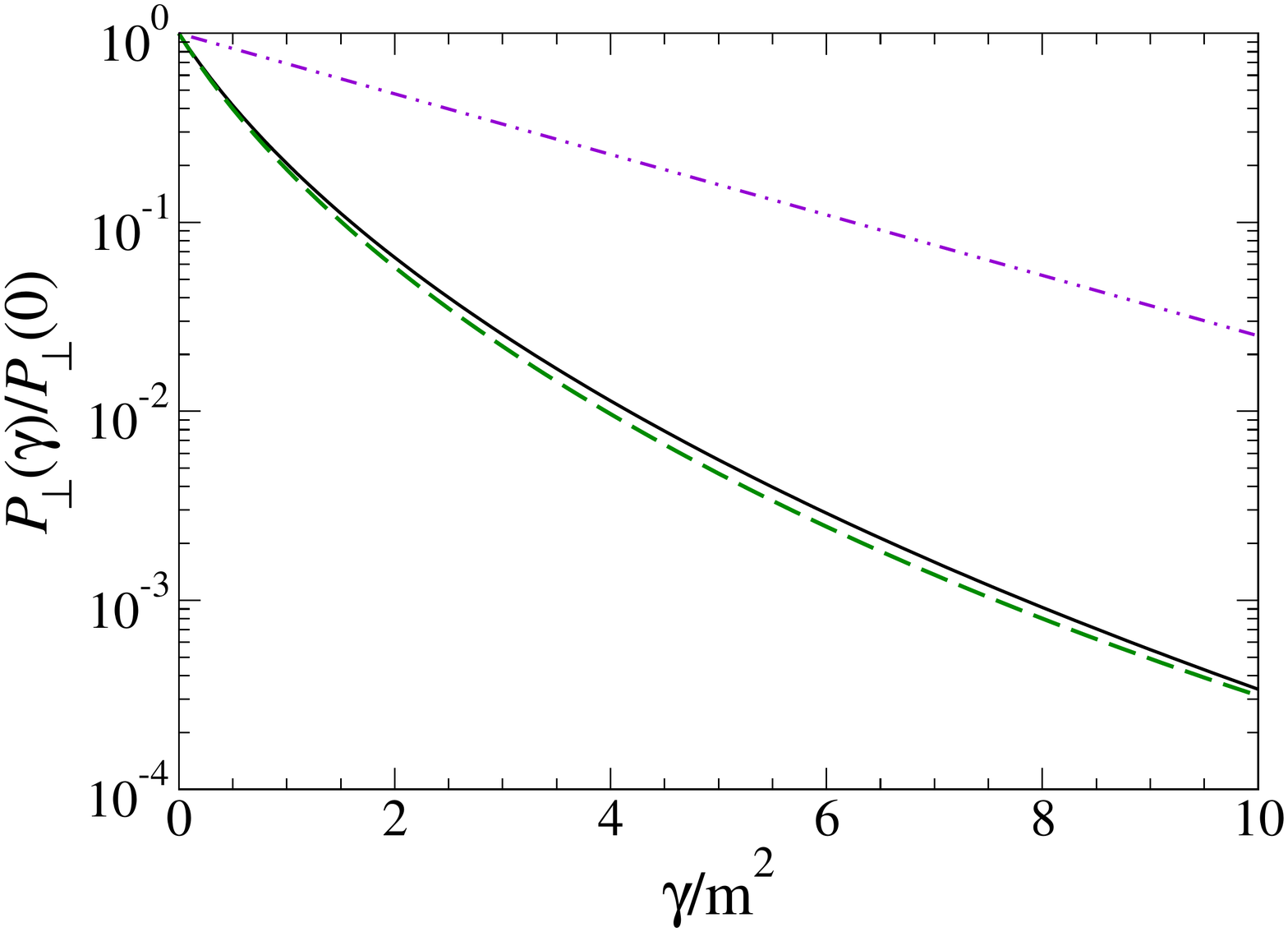}
\hspace{-0.8cm}\includegraphics[width=8.8cm]{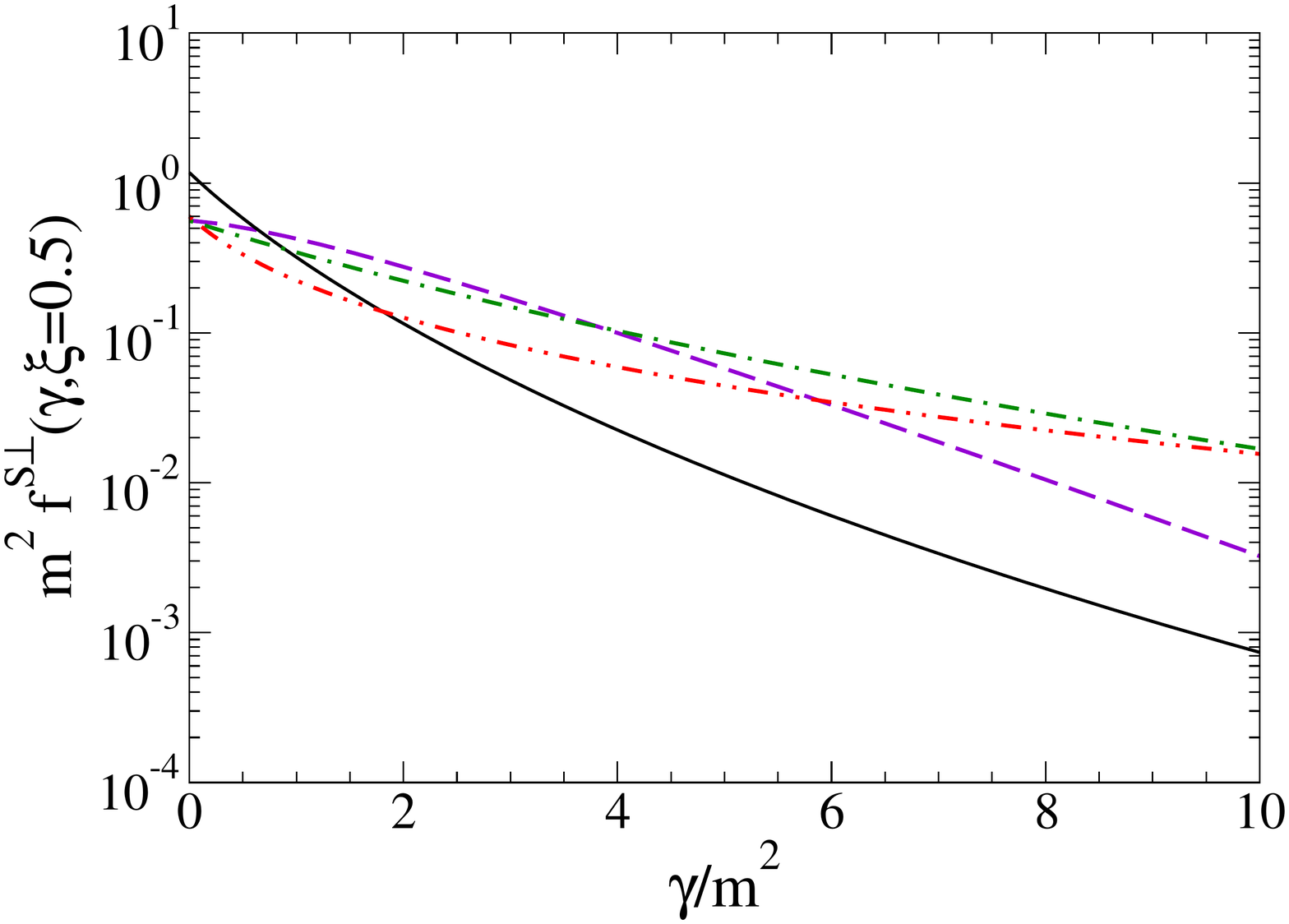}
    \caption{(Color online). {\it Left panel.} Normalized transverse distribution function $P_\perp(\gamma)/P_\perp(0)$ (cf. Eq. \eqref{Eq:fperptran}). Dotted line: full calculation. Solid line: $D_\perp(\gamma)/D_\perp(0)$ for the sake of comparison. Dash-double-dotted line: the same as in the left panel of Fig.~\ref{fig:dperp-f1_xi_fix}. 
    {\it Right panel.}  Pion unpolarized transverse-momentum  distribution $f^{S\perp}(\gamma,\xi)$, Eq.~\eqref{Eq:fperpS_AS}, for  $\xi=0.5$.
 Solid line: full calculation. Dashed line:  by using $f_1(\gamma,\xi=0.5)$ in Fig.~\ref{fig:dperp-f1_xi_fix}  from the LF constituent quark model~of Ref.~\cite{Pasquini:2014ppa,Lorce:2016ugb} (cf. the second line in  Eq.~\eqref{Eq:LIR}, without the gluonic term). Dash-dotted line:  the LF wave function from DSE calculations~\cite{Shi:2020pqe}. Dash-Double-dotted line:  the  NJL model~\cite{Noguera:2015iia}. The adopted quark mass $m=255$ MeV.}
\label{fig:fperptran}
    \end{center}   
\end{figure*}

\subsection{Twist-3 uTMD:  $f^\perp(\gamma,\xi)$}\label{subsec:fperp}

In an analogous way, for  $i=2$ and using Table~\ref{b2_tab}, one gets the   twist-3  $f^{\perp S(AS)}(\gamma,\xi)$, with the following decomposition 
\be 
f^{\perp S(AS)}(\gamma,\xi)={\cal P}_0(\gamma,\xi;S(AS))+
{\cal P}_d(\gamma,\xi;S(AS))
\nonu
+{\cal P}_{2d}(\gamma,\xi;S(AS))
\label{Eq:fperpS_AS}\ee
where the above functions are given in Appendix~\ref{twist34_app}. Notice that in this case ${\cal P}_{2d}(\gamma,\xi;S(AS))=0$.

\subsubsection{Longitudinal degree  of freedom}

In the left panel of Fig.~\ref{fig:fpL_pion-xifpL_pion}, the following collinear PDFs are shown
\be 
f^{\perp S(AS)}(\xi)=\int_0^\infty d\gamma~ f^{\perp S(AS)}(\gamma,\xi)~~,
\label{Eq:fpL}
\ee
and the corresponding quark combination.  As a reference, it is also presented $f^{\perp q}_{EoM}(\xi)$, obtained from the second line of Eq.~\eqref{Eq:LIR}, without the gluon term,  as follows
\be 
f^{\perp q}_{EoM}(\xi) \sim {1 \over \xi}
\int_0^\infty d\gamma f^{\perp q}_{EoM}(\gamma,\xi)
\sim {u^{LF}_{val}(\xi) \over \xi}~.
\label{Eq:fperpfree}
\ee
For the sake of completeness, in the right panel of Fig.~\ref{fig:fpL_pion-xifpL_pion}, the product $\xi\,f^{q\perp}(\xi)$ is compared to $f^q_1(\xi)$ and $u^{LF}_{val}(\xi)$ that represents the approximation to $f^{\perp q}_{EoM}(\xi)$ as given in Eq.~\eqref{Eq:fperpfree}. Also for $f^{q\perp}(\xi)$, the full calculation substantially differs from approximated evaluations, prompting further investigation of the gluon  contributions.

\subsubsection{Transverse degree of freedom}
Also for $f^\perp(\gamma,\xi)$, we introduce the transverse distribution function, viz.
\be 
P_\perp(\gamma)=\int_0^1 d\xi~ f^{S\perp}(\gamma,\xi)=
\int_0^1 d\xi~ f^{q\perp}(\gamma,\xi)~.
\label{Eq:fperptran}
\ee
In the left panel  of  Fig.~\ref{fig:fperptran},  a comparison, built with the same spirit as in the left panel of Fig.~\ref{fig:etran-e_xi_fix}, is shown  for the ratio $P_\perp(\gamma)/P_\perp(0)$. 

 A more detailed view of the fall-off can be gained from the right panel of Fig.~\ref{fig:fperptran}, where $f^{S\perp}(\gamma,\xi=0.5)$ is compared with the results  obtained by using i) the LF constituent quark model~of Ref. \cite{Pasquini:2014ppa,Lorce:2016ugb} (cf. the second line in  Eq. \eqref{Eq:LIR}, without the gluonic term). ii)   the LF wave function from DSE calculations~\cite{Shi:2020pqe} and iii)  the  NJL model~\cite{Noguera:2015iia}. 

\begin{figure}[h]
\begin{center}
\hspace{-0.2cm}\includegraphics[width=8.3cm]{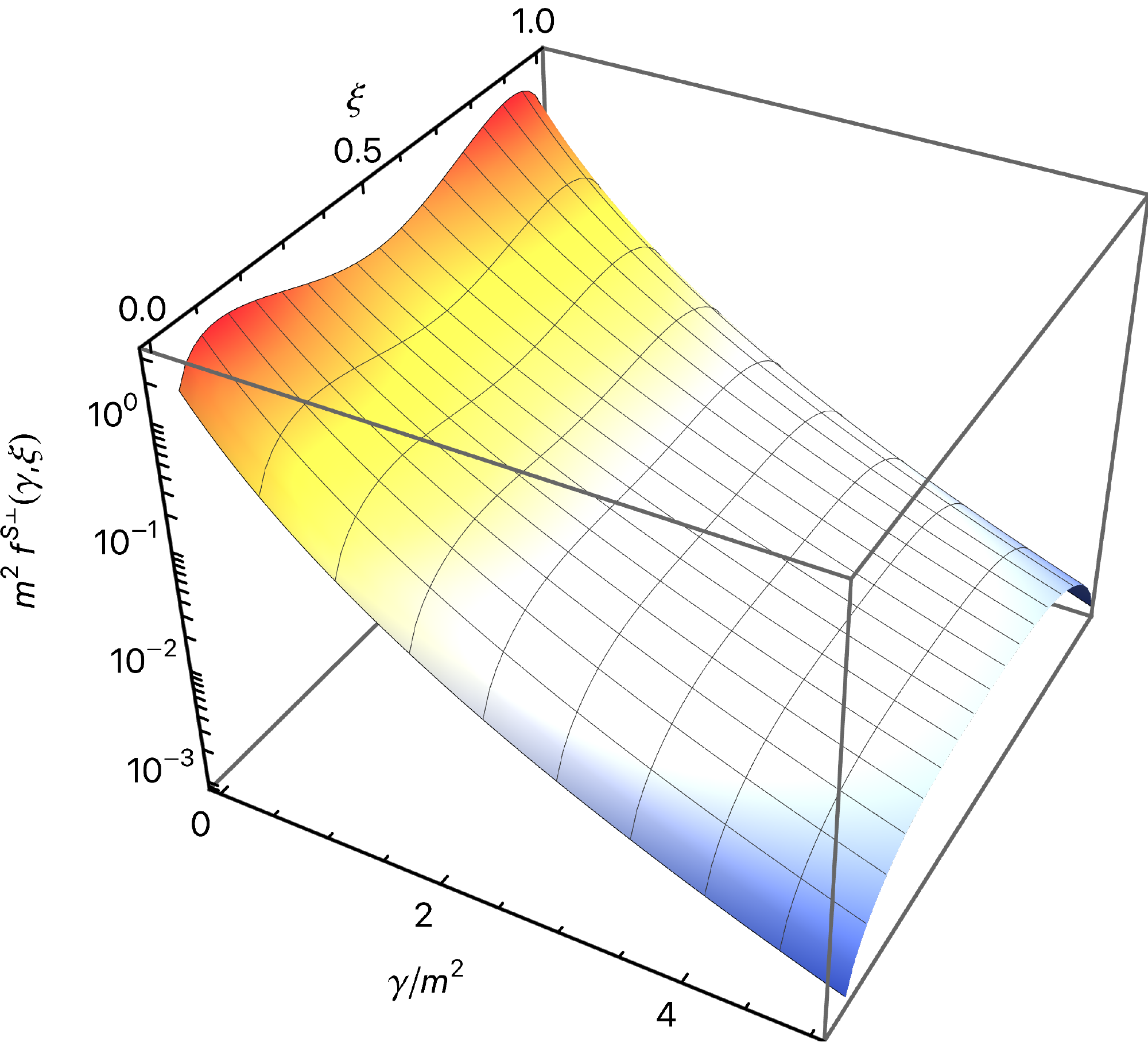}
\caption{ (Color online) Pion unpolarized transverse-momentum  distribution $f^{S\perp}(\gamma,\xi)$, Eq.~\eqref{Eq:fperpS_AS}. }
\label{fig:fperpS_pion}
 \end{center}
\end{figure}

Finally in Fig.~\ref{fig:fperpS_pion}, the full dependence of $f^{S\perp}(\gamma,\xi)$
 is shown. Also in this uTMD, the double-hump shape decreases when $\gamma/m^2$ increases.
 
 To summarize, a coherent view of the tail  in $\gamma$ is plainly provided by  Figs.  \ref{fig:dperp-f1_xi_fix}, \ref{fig:etran-e_xi_fix} and \ref{fig:fperptran}.  Namely,  the interaction taken into account in the ladder kernel  together with the extended structure of the quark-gluon vertex governs the fall-off of both the leading and subleading-twist uTMDS. { Therefore, the quantitative estimates obtained through our dynamical model, in Minkowski space, is  shown to be in a favorable position to provide insights into the interplay between transverse dof and the role of gluons.}

%

\section{Conclusions}

The twist-2 (leading)  and twist-3 (subleading) unpolarized (T-even) transverse-momentum dependent parton distribution functions have been calculated for the pion within an approach based on the solution of the Bethe-Salpeter equation in Minkowski space,  namely, within a genuinely relativistic quantum-field theory framework. We achieved this goal by  exploiting the Nakanishi integral representation of the BS-amplitude in order to get actual solution of the homogeneous BSE, in ladder approximation, through a system of integral equations  that determine the Nakanishi weight functions relevant for the problem under scrutiny \cite{dePaula:2016oct,dePaula:2017ikc,dePaula:2020qna}. After obtaining the pion electromagnetic form factor \cite{Ydrefors:2021dwa}, and the pion PDF \cite{dePaula:2022pcb}, we extended the yield of  our approach by exploring the dependence of the parton distributions upon the transverse momentum. This additional step has its-own importance in view of the planned experimental efforts to achieve a fully three-dimensional investigation of hadrons (mainly of the nucleon and, more challenging, the pion).

The relevant message one gets from our calculations is given by the essential role of the  gluon exchange, that cannot be captured by purely phenomenological model. { The joint use of the Fock expansion of the pion state, allows us to shed light on the gluonic content of the quark PDF obtained through the BS amplitude, even determining a quantitative estimate, $\sim 6\%$ of the  average longitudinal momentum fraction $\langle \xi^q\rangle$. Moreover, the latter analysis explains also the source of the small, but theoretically relevant, shift between the $u^q(\xi)$ and the PDF that fulfills the charge symmetry (an issue already investigated within the Dyson-Schwinger approach, e.g., in Ref. \cite{Chang:2014lva}, where a different interpretation was proposed).}
{As to the transverse degree of freedom,} 
 a power-like fall-off of the transverse distributions, obtained by integrating on $\xi$ the uTMDs,  is supported by the one-gluon exchange interaction that governs the ultraviolet region, according to our calculations. This outcome could suggest to reconsider  an exponential or Gaussian Ansatzes for describing the high-momentum content ($\gamma>>m^2$) of  the uTMDS. }

Summarizing, our approach can be placed among those in which the dynamics  can be studied   in Minkowski space and in some detail. Moreover,    the  additive construction of  the interaction kernel allows one to address step-by-step recognized effects, achieving an implementation of the dynamics  in a controlled way.

\label{Sec:concl}
\begin{acknowledgments}
 E. Y. gratefully thanks
INFN Sezione di Roma
 for providing the computer resources to perform all the calculations shown in this work. W. d. P. acknowledges the partial
support of CNPQ under Grants No. 438562/2018-6 and
No. 313236/2018-6, and the partial support of CAPES
under Grant No. 88881.309870/2018-01. T. F. thanks
the financial support from the Brazilian Institutions:
CNPq (Grant No. 308486/2015-3), CAPES (Finance Code 001) and FAPESP (Grants 
No. 2017/05660-0 and 2019/07767-1). E.
Y. acknowledges the support of FAPESP Grants No. 2016/25143-7 and No. 2018/21758-2. This work is a part of the
project Instituto Nacional de Ciência e Tecnologia - Física
Nuclear e Aplicações Proc. No. 464898/2014-5.
\end{acknowledgments}

\newpage

\appendix
\bwt
\section{Traces}
\label{trace_app}
In this Appendix
the traces  in Eqs. \eqref{Eq:f1t}, \eqref{Eq:et} and \eqref{Eq:fperpt}, are explicitly evaluated, presenting the expressions of the functions $a^i_{\ell,j}(k^-,\gamma,z;S(AS))$ and $b^i_{n;\ell,j}(\gamma,z;S(AS))$. For the sake of convenience, let us rewrite the generic trace entering  Eq. \eqref{Eq:tmd_t1}
\be
  Tr^{S(AS)}_i(\gamma,\xi)= -{i \over 2}
  \left\{\text{Tr}\left[ S^{-1}(k-\tfrac P2) \bar \Phi(k,P)~  {\cal O}_i ~\Phi(k,P) \right]
 +\eta^{S(AS)}_i ~\text{Tr}\left[ S^{-1}(k + \tfrac P2) \Phi(k,P)~  {\cal O}_i ~ \bar \Phi(k,P) \right]\right\}
\nonu
= \sum_{\ell j}~a^i_{\ell j}(k^-,\gamma,z;S(AS)) ~ \phi_{\ell}(k;P)~ \phi_{j}(k;P) \, ,
\label{app:TR1}\ee
with ${\cal O}^i$ and $ \eta^{S(AS)}_i$ given by
\be 
\begin{array}{ll} {\cal O}_0=\gamma^+~,&\quad \eta^{S(AS)}_0=\mp 1~,
\\ ~& ~\\
 {\cal O}_1=\mathbb{1} ~,& \quad\eta^{S(AS)}_1=\pm 1~,
\\ ~& ~\\
 {\cal O}_2={M\over |{\bf k}_\perp|^2}~{\bf k}_\perp\cdot{\bg \gamma}_\perp~,&\quad \eta^{S(AS)}_2=\pm 1~~ .
\end{array}~
\ee
To proceed one has to insert  the expression of the BS-amplitude, Eq. \eqref{Eq:BS_decomp}, and the definitions, Eqs. \eqref{Eq:def1} and \eqref{EQ:def2}, in Eq. \eqref{app:TR1}. Then one gets the results shown in Tables \ref{a0_tab}, \ref{a1_tab} and  \ref{a2_tab}, for $a^i_{\ell j}(k^-,\gamma,z; S(AS))$. It is also useful to organize the functions $a^i_{\ell j}$ in powers of $k^-$ for preparing the integration on such a  variable (cfr Appendix \ref{LFPro_app}), i.e.
\be
a^i_{\ell j}(k^-,\gamma,z;S(AS)) = 2M \Biggl[b^i_{0; \ell j}(\gamma,z;S(AS)) + b^i_{1; \ell j}(\gamma,z;S(AS)) ~{k^{-}\over 2M} 
 + b^i_{2; \ell j}(\gamma,z;S(AS)) ~\Biggl({k^{-}\over
2M}\Biggr)^2 
\nonu+ b^i_{3; \ell j}(\gamma,z;S(AS)) ~\Biggl({k^{-}\over
2M}\Biggr)^3\Biggr]~.
\ee
In Tables \ref{b0_tab}, \ref{b1_tab} and  \ref{b2_tab}, one can find  the expressions for $b^i_{n; \ell j}(\gamma,z;S(AS))$
\begin{table}
 \caption{Non vanishing coefficients  $a^0_{\ell j}(k^-,\gamma,z; S(AS))$.}
\begin{center}
 \label{a0_tab}
 \begin{tabular}{crcr}
 \hline  \hline
 $~ij~$&$a^0_{\ell j}(S)$& $\quad$ & $a^0_{\ell j}(AS)$\\
 \hline
$11$ & $2M $    & $\quad$ & $ 2Mz $ \\
$12$ & $ -8 m $ & $\quad$ & ~ \\
$13$ & $  - 2 \, m \, z-4{m\over M} \, k^{-} $  & $\quad$ & ~ \\
$14$ & $- {8\over M} \gamma - M \, z^2-2 \, z \, k^{-} $ & $\quad$ & $ -Mz-2k^-$ \\
$22$ & $ 2M$ & $\quad$ & $ -4k^- $ \\
$23$ &  $ M \, z+2 \, k^{-}  $ & $\quad $  &$ -8{\gamma\over M} -2 zk^- -{4\over M} (k^-)^2 $  \\
$24$ & ~& $\quad$ & $2mz +4 {m\over M} k^-$\\
$33$ & $ {M\over 2}\Bigl(4{\gamma\over M^2}  + {z^2\over 4} \Bigr)+ {z\over 2} k^{-}+{1\over 2M} \, (k^{-})^2  $
& $\quad$ & $ -\Bigl( 4 {\gamma\over M^2}+{z^2\over 4}\Bigr)k^- -{z\over M} (k^-)^2 -{1\over M^2} (k^-)^3$ \\
$34$ &~ & $\quad$ &$2m \Bigl(4{\gamma\over M^2}+{z^2\over 4} \Bigr)  +2 {m\over M} zk^- +2 {m\over M^2} (k^-)^2$\\
$44$ & ${M\over 2}\Bigl(4{\gamma\over M^2}  + {z^2\over 4} \Bigr)+ {z\over 2} k^{-}+{1\over 2M} \, (k^{-})^2  $ 
& $\quad$  & $ {M\over 2} ~z \Bigl(4 {\gamma\over M^2} +{z^2\over 4}\Bigr) +{z^2\over 2} k^- +{z\over 2M} (k^-)^2$ \\ \\
 \hline
 \end{tabular}
 \end{center}
 \end{table} 
 \begin{table}
 \caption{{Non vanishing coefficients  $a^1_{\ell j}(k^-,\gamma,z; S(AS))$.}}
\begin{center}
 \label{a1_tab}
 \begin{tabular}{crcr}
 \hline  \hline
 $~ij~$&$a^1_{\ell j}(S)$& $\quad$ & $a^1_{\ell j}(AS)$\\
\hline
$11$ & $-4m$ & $\quad$ & ~ \\
$12$ & $4M$ & $\quad$ & $ 2Mz-4k^- $\\
$13$ &  ~  & $\quad$ & $-{2M}\Bigl(4{\gamma\over M^2}  + {z^2\over 4} \Bigr)- 2z k^{-}-{2\over M} \, (k^{-})^2 $ \\
$22$ & $-4m$ & $\quad$ & ~ \\
$24$ & $-{2M}\Bigl(4{\gamma\over M^2}  + {z^2\over 4} \Bigr)- 2z k^{-}-{2\over M} \, (k^{-})^2 $  & $\quad$ & ~   \\
$33$ &  $m\Bigl(4{\gamma\over M^2}  + {z^2\over 4} \Bigr)+z{m\over M} k^{-}+{m\over M^{{2}}} \, (k^{-})^2 $ & $\quad$ & ~    \\
$34$ & $z {M\over 2}\Bigl(4{\gamma\over M^2}+{z^2\over 4} \Bigr)  - \Bigl(4{\gamma\over M^2}-{z^2\over 4} \Bigr)k^- - {z\over2 M} (k^-)^2-{1\over M^2} (k^-)^3$ & $\quad$ & $M\Bigl(4{\gamma\over M^2}  + {z^2\over 4} \Bigr)+zk^- +{1\over M} (k^-)^2$\\
$44$ & $m\Bigl(4{\gamma\over M^2}  + {z^2\over 4} \Bigr)+z{m\over M} k^{-}+{m\over M^{{2}}} \, (k^{-})^2 $ & $\quad$ & ~    \\
 \\
 \hline
 \end{tabular}
 \end{center}
 \end{table} 
 \begin{table}
 \caption{{Non vanishing coefficients  $a^2_{\ell j}(k^-,\gamma,z; S(AS))$.}}
\begin{center}
 \label{a2_tab}
 \begin{tabular}{crcr}
 \hline  \hline
 $~ij~$&$a^2_{\ell j}(S)$& $\quad$ & $a^2_{\ell j}(AS)$\\
 \hline
$11$ & $ -4M $  & $\quad$ & ~ \\
$13$ &  ~   & $\quad$ & $8 m $  \\
$14$ & $4 M$ & $\quad$ & $ 2 zM $ ${-4 k^{-}}$ \\
$22$ & $ 4 M $    & $\quad$ & ~ \\
$23$ & $-2 Mz + 4 k^-$  & $\quad$ &  $  - 4 \,M $   \\
$24$ & $ -8 m $  & $\quad$ & ~ \\
$33$ & $ M \Bigl( 4 {\gamma\over M^2}+{z^2\over 4}\Bigr) + z k^- +{1\over M} (k^-)^2$    & $\quad$ & ~ \\
${44}$ & $ - M \Bigl( 4 {\gamma\over M^2}+{z^2\over 4}\Bigr) - z k^- -{1\over M} (k^-)^2$  & $\quad$ & ~ \\
\\
 \hline
 \end{tabular}
 \end{center}
 \end{table} 
  
\begin{table}
 \caption{Non vanishing coefficients  $b^0_{n;\ell j}(\gamma,z; S(AS))$.}
\begin{center}
 \label{b0_tab}
 \begin{tabular}{cccccccc}
 \hline  \hline
 $~ij~$&$b^0_{0;\ell j}(S)$&$b^0_{1;\ell j}(S)$&$b^0_{2;\ell j}(S)$
 &$b^0_{0;\ell j}(AS)$&$b^0_{1;\ell j}(AS)$&$b^0_{2;\ell j}(AS)$&$b^0_{3;\ell j}(AS)$\\
 \hline
$11$ & $1 $ & $ 0 $ & $ 0 $& $z$&$ 0 $&$ 0 $&$ 0 $\\
$12$ & $ -4 m/M $ & $ 0 $ & $ 0 $& $0$&$ 0 $&$ 0 $&$ 0 $\\
$13$ & $ - zm /M$  & $ - 4 m/M $ & $ 0 $& $0$&$ 0 $&$ 0 $&$ 0 $\\
$14$ & $-4{ \gamma/M^2} -  {z^2/ 2} $ & $- 2 \, z$ & $ 0 $& $-z/2$&$ -2 $&$ 0 $&$ 0 $\\
$22$ & $ 1$ & $ 0 $ & $ 0 $& $0$&$ -4 $&$ 0 $&$ 0 $\\
$23$ & $ {z/2} $ & $ 2 $  & $ 0 $& $-4\gamma/M^2$&$ -2z $&$ -8 $&$ 0 $\\
$24$  & $0$ & $ 0$ & $ 0  $& $z m/M$&$ 4 m/M $&$ 0 $&$ 0 $\\
$33$ & ${ \gamma/M^2}+{ z^2/16}$ & $ {z/2}$ & $ 1  $& $0$&$ -\Bigl(4 \gamma/M^2+ z^2/4\Bigr) $&$ -2z $&$ -4 $\\
$34$  & $0$ & $ 0$ & $ 0  $& $\left(m/{{M}}\right) \Bigl(4 \gamma/M^2+ z^2/4\Bigr) $&$ 2z m/M $&$ 4 m/M $&$ 0 $\\
$44$ & ${ \gamma/M^2}+{ z^2/16}$ & $ {z/2}$ & $ 1  $& $(z/4)\Bigl(4 \gamma/M^2+ z^2/4\Bigr)  $&$ z^2/2 $&$ z $&$ 0 $\\
 \hline
 \end{tabular}
 \end{center}
 \end{table}

\begin{table}
 \caption{{Non vanishing coefficients  $b^1_{n;\ell j}(\gamma,z; S(AS))$}.}
\begin{center}
 \label{b1_tab}
 \begin{tabular}{cccccccc}
 \hline  \hline
 $~ij~$&$b^1_{0;\ell j}(S)$&$b^1_{1;\ell j}$(S)&$b^1_{2;\ell j}(S)$&$b^1_{3;\ell j}(S)$
 &$b^1_{0;\ell j}(AS)$&$b^1_{1;\ell j}(AS)$&$b^1_{2;\ell j}(AS)$\\
 \hline
$11$ & $-2m/M$&$ 0 $&$ 0 $&$ 0 $ & $0 $ & $ 0 $ & $ 0 $ \\
$12$ & $2$&$ 0 $&$ 0 $&$ 0 $ & $z $ & $ -4 $ & $ 0 $ \\
$13$ & $0$&$ 0 $&$ 0 $&$ 0 $ & $ -4 {\gamma / M^2} - {z^2/ 4} $ & $ -2 z $ & $ -4 $ \\
$22$ & $-2m/M$&$ 0 $&$ 0 $&$ 0 $ & $0 $ & $ 0 $ & $ 0 $ \\
$24$ & $-\Bigl(4\gamma/M^2 +z^2/4\Bigr)$&$ -2z $&$ -4 $&$ 0 $ & $0 $ & $ 0 $ & $ 0 $\\
$33$ & $(m/2M)~\Bigl(4\gamma/M^2 +z^2/4\Bigr)$&$z m/M $&$ 2 m/M $&$ 0 $ & $0 $ & $ 0 $ & $ 0 $\\
$34$ & $(z/4) \-\Bigl(4\gamma/M^2 +z^2/4\Bigr)$&$ -\Bigl(4\gamma/M^2 -z^2/4\Bigr) $&$ -z $&$ -4 $ & $ 2 {\gamma / M^2} + {z^2 / 8} $  & $ z $ & $ 2 $\\
$44$ & $(m/2M)~\Bigl(4\gamma/M^2 +z^2/4\Bigr)$&$z m/M $&$ 2 m/M $&$ 0 $ & $0 $ & $ 0 $ & $ 0 $\\
 \hline
 \end{tabular}
 \end{center}
 \end{table}
 \begin{table}
 \caption{{Non vanishing coefficients  $b^2_{n;\ell j}(\gamma,z; S(AS))$.}}
\begin{center}
 \label{b2_tab}
 \begin{tabular}{cccccc}
 \hline  \hline
 $~ij~$&$b^2_{0;\ell j}(S)$& $b^2_{1;\ell j}(S)$ &$b^2_{2;\ell j}(S)$
 &$b^2_{0;\ell j}(AS)$&$b^{{2}}_{1;\ell j}(AS)$ \\
 \hline
 $11$ & $-2$&$ 0 $&$ 0 $ & $0 $ & $0$ \\
 $13$ & $0$&$ 0 $&$ 0 $ & $4 (m/ M) $ & $0$\\
 $14$ & $2$&$ 0 $&$ 0 $ & $ z $ & ${-4}$ \\
 ${22}$ & ${2}$&$ 0$&$ 0 $ & $0 $ & $0$ \\
 $23$ & $-z$&$ 4$&$ 0 $ & $-2 $ & $0$ \\
 $24$ & $-4m/M$&$ 0 $&$ 0 $ & $0 $ & $0$\\
 $33$ & $(1/2)\Bigl(4\gamma/M^2 +z^2/4\Bigr)$&$ z $&$ 2 $ & $0 $ & $0$\\
 $44$ & $-(1/2)\Bigl(4\gamma/M^2 +z^2/4\Bigr)$&$ -z $&$ -2 $ & $0 $ & $0$\\
 \hline
 \end{tabular}
 \end{center}
 \end{table}

\section{The light-front projection}
\label{LFPro_app}
The Appendix is devoted to the integration over the variable $k^-$ in Eq. \eqref{Eq:Filj}, that for the sake of clarity we rewrite
\be 
F^i_{\ell j}(\gamma, z;S(AS))= \int_{-\infty}^\infty {d k^{-}\over 2 \pi} ~a^i_{\ell j}(k^{-}, \gamma,z;S(AS)) ~\phi_{\ell}(k,P) \, \phi_j(k,P)
\nonu 
=2M\int_{-\infty}^\infty {d k^{-}\over 2 \pi} ~\phi_\ell(k,P) \, \phi_j(k,P)
 \times\Biggl[b^i_{0; \ell j}(\gamma,z;S(AS)) + b^i_{1; \ell j}(\gamma,z;S(AS)) {k^{-}\over 2M} 
 + b^i_{2; \ell j}(\gamma,z;S(AS)) \Biggl({k^{-}\over
2M}\Biggr)^2
\nonu +b^i_{3; \ell j}(\gamma,z;S(AS)) \Biggl({k^{-}\over
2M}\Biggr)^3\Biggr] ~,
\label{app:Filj}\ee
where the quantities $a^i_{\ell j}(k^{-}, \gamma,z;S(AS))$ and $b^i_{n; \ell j}(\gamma,z;S(AS))$ are given in Appendix \ref{trace_app}.

The first step (see also Refs. \cite{dePaula:2020qna,Ydrefors:2021dwa,dePaula:2022pcb}) is to introduce the NIR of $\phi_\ell (k,P)$, Eq. \eqref{Eq:NIR}, and then apply the Feynman parametrization as follows
\be
\phi_\ell (k,P) \phi_{j} (k,P)=30 \int_0^1dv\int_{-1}^{+1} dz' \int_0^{\infty} d\gamma'
\int_{-1}^{+1} dz''
\int_0^{\infty} d\gamma''~
{v^2(1-v)^2~g_\ell(\gamma', z')~g_{j}(\gamma'',
z'')\over \Bigl[k^- \alpha -\beta
+i \epsilon\Bigr]^6}~,
\label{app:phi}\ee
where
\be
\alpha={M\over 2} ~\Bigl[ \lambda(v)-z\Bigr]~,
\nonu
\beta(z\lambda(v))
=\gamma+\kappa^2+{M^2\over 4}z \lambda(v) +v\gamma'
+(1-v) \gamma''~,
\nonu
\lambda(v)=v z' +(1-v)z''~.
\ee
Hence, the following general expression, one can straightforwardly deduce from well-known result in Ref. \cite{YanIV} (corresponding to the case $m=0$) is useful for performing the relevant integrals
\be
\int_{-\infty}^\infty {d k^{-}\over 2\pi} ~
{ (k^{-})^{m} \over \Bigl[\alpha~k^{-} - \beta + i \epsilon\Bigr]^n}
=i~{(n-m-2)!\over (n-1)!}{(-1)^{m+1}\over\Bigl[-\beta + i \epsilon\Bigr]^{n-m-1}} ~  \delta^{(m)}(\alpha)
\label{app:Yan}\ee
where $\delta^{(m)}(\alpha)=\partial^m \delta(\alpha)/\partial \alpha^m$.
Finally, combining the results in Eqs. \eqref{app:phi} and \eqref{app:Yan},
one gets
\be 
F^i_{\ell j} (\gamma, z;S(AS)) = -{24 i}\int_{0}^{1} dv ~ v^{2}(1-v)^2  \int_{-1}^{+1} dz' \int_0^{\infty} d\gamma' 
\int_{-1}^{+1} dz'' \int_0^{\infty} d\gamma'' ~{\cal G}_{\ell j}(\gamma', z'; \gamma'', z'';\kappa^2)  \nonu
\times 
\Biggl\{ {b^i_{0; \ell j}(\gamma,z;S(AS))~\delta(\tilde\alpha) \over [-\beta(z\lambda(v))+i\epsilon]^5} - 
{1\over 4M^2}{b^i_{1; \ell j}(\gamma,z;S(AS))~ \delta' (\tilde\alpha) \over [-\beta(z\lambda(v))+i\epsilon]^4}
 + {1\over 12M^4}  {b^i_{2; \ell j}(\gamma,z;S(AS)) ~\delta^{''} (\tilde\alpha)\over  [-\beta(z\lambda(v))+i\epsilon]^3} 
\nonu- {1\over 24M^6}  {b^i_{3; \ell j}(\gamma,z;S(AS)) ~\delta^{'''} (\tilde\alpha)\over  [-\beta(z\lambda(v))+i\epsilon]^2}\Biggr\}~,
\ee
where 
\be
{\cal G}_{\ell j}(\gamma', z'; \gamma'', z'';\kappa^2) =g_\ell(\gamma', z';\kappa^2)~g_j(\gamma'', z'';\kappa^2)
\nonu 
\tilde\alpha = {2\over M} \alpha=\lambda(v)-z~,
\ee
and the derivatives of the delta function is with respect to $\tilde \alpha$. Recalling that $\partial \tilde \alpha/\partial z =-1$, one can also write
\be 
F^i_{\ell j} (\gamma, z;S(AS)) = -{24 i}\int_{0}^{1} dv ~ v^{2}(1-v)^2
\int_{-1}^{+1} dz' \int_0^{\infty} d\gamma' 
\int_{-1}^{+1} dz'' \int_0^{\infty} d\gamma'' ~{\cal G}_{\ell j}(\gamma', z'; \gamma'', z'';\kappa^2) 
\nonu \times~
\Biggl\{ {b^i_{0; \ell j}(\gamma,z;S(AS))~\delta(\lambda(v)-z) \over [-\beta(z\lambda(v))+i\epsilon]^5} 
 + 
{1\over 4M^2}{b^i_{1; \ell j}(\gamma,z;S(AS)) \over [-\beta(z\lambda(v))+i\epsilon]^4}~{\partial\over\partial z} \delta (\lambda(v)-z) 
 \nonu+ {1\over 12M^4}  {b^i_{2; \ell j}(\gamma,z;S(AS))  \over  [-\beta(z\lambda(v))+i\epsilon]^3} ~{\partial^2\over \partial z^2}\delta (\lambda(v)-z)
 +{1\over 24M^4}  {b^i_{3; \ell j}(\gamma,z;S(AS))  \over  [-\beta(z\lambda(v))+i\epsilon]^2} ~{\partial^3\over \partial z^3}\delta (\lambda(v)-z)
\Biggr\}~,  
\label{app:Fijk2}\ee
Finally, by using
\be
f(z) {\partial^m\over \partial z^m}\delta(\lambda(v)-z)=
\sum_{k=0}^m ~ c_{m k}~{\partial^k\over \partial z^k}\Bigl[f^{(m-k)}(z)~
\delta(\lambda(v)-z)\Bigr]~,\ee  
where the coefficient $c_{m k}$ can be obtained by repeatedly  applying  the Leibniz rule for the product of functions and $f^{(m-k)}$ indicates the $(m-k)$-th derivative (with $ f^{(0)}(z)\equiv f(z)$), one recasts Eq. \eqref{app:Fijk2}  in a form more suitable for the further elaboration. In practice,  one  trades  derivatives on the delta functions with  derivatives on the functions $b^i_{n; ij}(\gamma,z;S(AS)) $, getting
\be
F^i_{\ell j} (\gamma, z;S(AS)) =-24  i ~\int_{0}^{1} dv ~ v^{2}(1-v)^2  \int_{-1}^{+1} dz' \int_0^{\infty} d\gamma' \int_{-1}^{+1} dz'' \int_0^{\infty} d\gamma'' ~{\cal G}_{\ell j}(\gamma', z'; \gamma'', z'';\kappa^2)  \nonu
\times 
\Biggl\{ \delta (\lambda(v)-z)
\Biggl[{b^i_{0; \ell j}(\gamma,z;S(AS)) ~\over [-\beta(z\lambda(v))+i\epsilon]^5}  
-{1\over 4M^2}
{\partial\over \partial z}\Bigl( {b^i_{1; \ell j}(\gamma,z;S(AS)) \over
[-\beta(z\lambda(v))+i\epsilon]^4}\Bigr)
+ {1 \over 12M^4}  {\partial^2\over \partial z^2}
\Bigl({ b^i_{2; \ell j}(\gamma,z;S(AS)) \over  [-\beta(z\lambda(v))+i\epsilon]^3}\Bigr)
\nonu
- {1 \over 24M^6} {\partial^3\over \partial z^3}
\Bigl({ b^i_{3; \ell j}(\gamma,z;S(AS))\over  [-\beta(z\lambda(v))+i\epsilon]^2}\Bigr)
\Biggr]\nonu+ 
{1\over 4M^2}  {\partial\over \partial z}
\Biggl[{b^i_{1; \ell j}(\gamma,z;S(AS)) ~\delta (\lambda(v)-z)\over
[-\beta(z\lambda(v))+i\epsilon]^4}-{2\over 3 M^2}  
  {\partial \over \partial z } \Bigl({b^i_{2; \ell j}(\gamma,z;S(AS)) \over  [-\beta(z\lambda(v))+i\epsilon]^3} \Bigr) 
 \delta (\lambda(v)-z)     
 \nonu
 +{1\over 2 M^4}  
  {\partial^2 \over \partial z^2 } \Bigl({b^i_{3; \ell j}(\gamma,z;S(AS) \over  [-\beta(z\lambda(v))+i\epsilon]^2}\Bigr)~\delta (\lambda(v)-z)\Biggr]  
   \nonu
 + {1 \over 12M^4} {\partial^2\over \partial z^2} 
 \Biggl[{ b^i_{2; \ell j}(\gamma,z;S(AS)) \, \delta (\lambda(v)-z)\over  [-\beta(z\lambda(v))+i\epsilon]^3} -{3\over 2 M^2}  
  {\partial \over \partial z } \Bigl({b^i_{3; \ell j}(\gamma,z;S(AS)) \over  [-\beta(z\lambda(v))+i\epsilon]^2} \Bigr) \delta (\lambda(v)-z) \Biggr]
  \nonu
 + {1 \over 24M^6} {\partial^3\over \partial z^3} 
 \Biggl[{ b^i_{3; \ell j}(\gamma,z;S(AS)) \, \delta (\lambda(v)-z)\over  [-\beta(z\lambda(v))+i\epsilon]^2}\Biggr] 
  \Biggr\} ~.
\ee
Hence, by taking into account the  expressions of $ b^i_{2; \ell j}(\gamma,z;S(AS))$ and $ b^i_{3; \ell j}(\gamma,z;S(AS))$, given in Tables \ref{b0_tab}, \ref{b1_tab} and  \ref{b2_tab},  one can   drop some  derivatives. In particular the second derivative of $b^i_{2; \ell j}(\gamma,z;S(AS))$ and all the derivatives of $b^i_{3; \ell j}(\gamma,z;S(AS))$, obtaining 
\be 
F^i_{\ell j} (\gamma, z;S(AS)) = -24i\Bigl[{\cal F}^{i}_{0\ell j} (\gamma, z;S(AS))+
{\cal F}^{i}_{1;\ell j} (\gamma, z;S(AS))+{\cal F}^{i}_{2;\ell j} (\gamma, z;S(AS))+{\cal F}^{i}_{3;\ell j} (\gamma, z;S(AS))\Bigr] \nonu
\ee 
where
\be 
{\cal F}^{i}_{0,\ell j} (\gamma, z;S(AS))=
\int_{0}^{1} dv ~ v^{2}(1-v)^2
\int_{-1}^{+1} dz' \int_0^{\infty} d\gamma' 
\int_{-1}^{+1} dz'' \int_0^{\infty} d\gamma'' ~{\cal G}_{\ell j}(\gamma', z'; \gamma'', z'';\kappa^2) 
\nonu
\times ~{\delta (\lambda(v)-z)~\over [-\beta(z^2)+i\epsilon]^5}~\Biggl\{b^i_{0; \ell j}(\gamma,z;S(AS))
 +{1\over 4}~\Biggl[{\beta(z^2)\over M^2}~{ \partial \over\partial z}b^i_{1; \ell j}(\gamma,z;S(AS))-z~b^i_{1; \ell j}(\gamma,z;S(AS))\Biggr] 
 \nonu 
 + {1\over 16 } \Biggl[z^2~b^i_{2; \ell j}(\gamma,z;S(AS))-2 z{\beta(z^2)\over M^2}{\partial \over \partial z} b^i_{2; \ell j}(\gamma,z;S(AS)\Biggr]
 \nonu-{z^3\over 64} ~b^i_{3; \ell j}(S(AS))  \Biggr\}~,
\label{calF0_app}\ee 
\be
{\cal F}^{i1}_{1;\ell j} (\gamma, z;S(AS))= 
{1\over 8M^2}  {\partial\over \partial z}\Biggl\{\int_{0}^{1} dv ~ v^{2}(1-v)^2
\int_{-1}^{+1} dz' \int_0^{\infty} d\gamma' 
\int_{-1}^{+1} dz'' \int_0^{\infty} d\gamma'' ~{\cal G}_{\ell j}(\gamma', z'; \gamma'', z'';\kappa^2)
\nonu \times~{ \delta (\lambda(v)-z)\over  [-\beta(z^2)+i\epsilon]^4}~\Biggl[2b^i_{1; \ell j}(\gamma,z;S(AS)) 
-z~b^i_{2; \ell j}(\gamma,z;S(AS)) +{4\over 3} ~{\beta(z^2)\over M^2} {\partial\over \partial z} b^i_{2; \ell j}(\gamma,z;S(AS))
\nonu +{3 \over 8}z^2~b^i_{3; \ell j}(S(AS))
 \Biggr]\Biggr\}~,
 \label{calF1_app}
 \ee
 \be 
 {\cal F}^{i}_{2;\ell j} (\gamma, z;S(AS))={1\over 12M^4} {\partial^2\over \partial z^2} \Biggl\{\int_{0}^{1} dv ~ v^{2}(1-v)^2
\int_{-1}^{+1} dz' \int_0^{\infty} d\gamma' 
\int_{-1}^{+1} dz'' \int_0^{\infty} d\gamma'' ~{\cal G}_{\ell j}(\gamma', z'; \gamma'', z'';\kappa^2)
 \nonu \times ~{ \delta (\lambda(v)-z)\over  [-\beta(z^2)+i\epsilon]^3} ~\Bigl[b^i_{2; \ell j}(\gamma,z;S(AS)) -{3\over 4} z~b^i_{3; \ell j}(S(AS)) \Bigr]
\Biggr\}~,
\label{calF2_app}\ee
and 
\be
{\cal F}^{i}_{3;\ell j} (\gamma, z;S(AS))={b^i_{3; \ell j}(S(AS))\over 24M^6} {\partial^3\over \partial z^3} \Biggl\{\int_{0}^{1} dv ~ v^{2}(1-v)^2
\int_{-1}^{+1} dz' \int_0^{\infty} d\gamma' 
\int_{-1}^{+1} dz'' \int_0^{\infty} d\gamma'' ~{\cal G}_{\ell j}(\gamma', z'; \gamma'', z'';\kappa^2)
 \nonu \times ~{ \delta (\lambda(v)-z)\over  [-\beta(z^2)+i\epsilon]^2} ~ 
\Biggr\}
~.
\label{calF3_app}\ee

Collecting the above results, Eq. \eqref{Eq:tmd_t1} becomes
\be 
{\cal T}^{S(AS)}_i(\gamma,\xi)
=i{N_c \over 8}~{1 \over (2\pi)^2}
\times~ \sum_{\ell j} \int_{-1}^1 dz \, \delta( z - \left(1- 2\xi\right)) \, F^i_{\ell j}(\gamma, z;S(AS))=
\nonu
={3N_c \over  (2\pi)^2}~\sum_{\ell j}~\Bigl[{\cal F}^{i}_{0;\ell j}(\gamma,z;S(AS))+{\cal F}^{i}_{1;\ell j}(\gamma,z;S(AS))+{\cal F}^{i}_{2;\ell j}(\gamma,z;S(AS))
+{\cal F}^{i}_{3;\ell j}(\gamma,z;S(AS))\Bigr]~.
\label{calT_app}\ee
It is also useful for getting more explicit expressions to perform the integral on $v$ in the Eqs. \eqref{calF0_app}, \eqref{calF1_app} and  \eqref{calF2_app}.
This can be accomplished by using  the following result
\be
\int_{0}^{1} dv ~ v^{2}(1-v)^2~\delta[vz'+(1-v)z''-z]
=v_0^2(1-v_0)^2~{\Theta(v_0)~\Theta(1-v_0)\over |z'-z''|}
=v_0^2(1-v_0)^2~\Delta(z,z',z'')
\label{vint_app}\ee 
with
\be 
\Delta(z,z',z'')=
{\Theta(z'-z)\Theta(z-z'')-\Theta(z''-z)\Theta(z-z')\over z'-z''}~,
\quad 
 v_0={z-z''\over z'-z''}~.
 \ee
 The combination of the theta-functions implements the constraint $0\le v_0\le 1$. Moreover, notice that i) simultaneously changing the signs of $z$, $z'$ and $z''$ the function $\Delta(z,z',z'')$ does not change sign, this reflects the symmetry  with respect $\xi=0.5$, as implemented through the charge symmetry in Eq. \eqref{Eq:tmd_t1};
 ii) $\Delta(z,z',z'')$ is even under the exchange $z''\to z'$ and in the limit $z''-z'=\epsilon \to 0$ one has
 \be 
 \lim_{\epsilon \to 0}\Delta(z,z',z'+\epsilon)=
 \delta(z-z')~,
 \ee 
 so that the singularity can be addressed without particular problems.
 
By taking into account Eq. \eqref{vint_app}, and the symmetries with respect to the transformation $z'\to z''$  and $\gamma'\to \gamma''$, one gets
\be 
{\cal F}^{i}_{0;\ell j} (\gamma, z;S(AS))
 =2\int_0^{\infty} d\gamma' \int_0^{\infty} d\gamma''
\int_{-1}^{+1} dz'\int_{-1}^{+1} dz''  ~v_0^{2}(1-v_0)^2~
{\Theta(z'-z)\Theta(z-z'')\over z'-z''}
\nonu
\times ~{\bar {\cal G}_{\ell j}(\gamma', z'; \gamma'', z'';\kappa^2) 
~\over [-\beta_0(z^2)+i\epsilon]^5}~
\Biggl\{b^i_{0; \ell j}(\gamma,z;S(AS))
 +{1\over 4}~\Biggl[{\beta(z^2)\over M^2}~{ \partial \over\partial z}b^i_{1; \ell j}(\gamma,z;S(AS))-z~b^i_{1; \ell j}(\gamma,z;S(AS))\Biggr] 
 \nonu 
 + {1\over 16 } \Biggl[z^2~b^i_{2; \ell j}(\gamma,z;S(AS))-2 z{\beta(z^2)\over M^2}{\partial \over \partial z} b^i_{2; \ell j}(\gamma,z;S(AS)\Biggr]
 \nonu-{z^3\over 64} ~b^i_{3; \ell j}(S(AS)) \Biggr] \Biggr\}
~,
\label{calF0b_app}\ee
\be
{\cal F}^{i}_{1;\ell j} (\gamma, z;S(AS))
 ={1\over 4M^2}  {\partial\over \partial z}\Biggl\{\int_0^{\infty} d\gamma' \int_0^{\infty} d\gamma'' 
\int_{-1}^{+1} dz'  
\int_{-1}^{+1} dz''  v_0^{2}(1-v_0)^2~{\Theta(z'-z)\Theta(z-z'')\over z'-z''}
\nonu\times~{ \bar{\cal G}_{\ell j}(\gamma', z'; \gamma'', z'';\kappa^2)\over  [-\beta_0(z^2)+i\epsilon]^4}
\Biggl[2b^i_{1; \ell j}(\gamma,z;S(AS)) 
-z~b^i_{2; \ell j}(\gamma,z;S(AS)) +{4\over 3} ~{\beta(z^2)\over M^2} {\partial\over \partial z} b^i_{2; \ell j}(\gamma,z;S(AS))
\nonu +{3 \over 8}z^2~b^i_{3; \ell j}(S(AS))
 \Biggr]
\Biggr\}~,
 \label{calF1b_app}
 \ee
 \be 
 {\cal F}^{i}_{2;\ell j} (\gamma, z;S(AS))
={1\over 6M^4} {\partial^2\over \partial z^2} \Biggl\{ 
\int_0^{\infty} d\gamma' \int_0^{\infty} d\gamma''\int_{-1}^{+1} dz' 
\int_{-1}^{+1} dz''  ~v_0^{2}(1-v_0)^2~{\Theta(z'-z)\Theta(z-z'')\over z'-z''}
 \nonu \times~
 {\bar  {\cal G}_{\ell j}(\gamma', z'; \gamma'', z'';\kappa^2)\over  [-\beta_0(z^2)+i\epsilon]^3} 
\Bigl[b^i_{2; \ell j}(\gamma,z;S(AS)) -{3\over 4} z~b^i_{3; \ell j}(S(AS)) \Bigr]
\Biggr\}~, \nonu
\label{calF2b_app}\ee
and
\be 
 {\cal F}^{i}_{3;\ell j} (\gamma, z;S(AS))= {b^i_{3; \ell j}(S(AS))\over 12M^6}{\partial^3\over \partial z^3} \Biggl\{ 
\int_0^{\infty} d\gamma' \int_0^{\infty} d\gamma''\int_{-1}^{+1} dz' 
\int_{-1}^{+1} dz''  ~v_0^{2}(1-v_0)^2~{\Theta(z'-z)\Theta(z-z'')\over z'-z''}
 \nonu \times~
 {\bar  {\cal G}_{\ell j}(\gamma', z'; \gamma'', z'';\kappa^2)\over  [-\beta_0(z^2)+i\epsilon]^2}\Biggr\}
 \label{calF3b_app}\ee
 where the functions $b^i_{n;\ell j}(\gamma,z;S(AS))$  are given in the Tables of Appendix~\ref{trace_app} and 
\be 
\beta_0(z^2)=\gamma+\kappa^2+z^2 {M^2\over 4} +v_0\gamma'+(1-v_0)\gamma''~,
\nonu
\bar{\cal G}_{\ell j}(\gamma', z'; \gamma'', z'';\kappa^2)={g_\ell(\gamma',z';\kappa^2)g_j(\gamma'',z'';\kappa^2) 
+g_\ell(\gamma'',z'';\kappa^2)g_j(\gamma',z';\kappa^2)\over 2}~ \, , \nonu
v_0={z-z''\over z'-z''}~.
\label{calgm_app}\ee   
Also notice  that for a bound state one has 
 \be
\beta_0(z^2)=\gamma+\kappa^2 +{M^2\over 4}z^2+v_0\gamma'+(1-{v_0})\gamma''\ge m^2- {M^2\over 4} (1-z^2)\ge \kappa^2>0~,
 \ee 
 and therefore no poles are associated to such a quantity. It should be pointed out that the presence of the theta-functions, that ensure $0\le v_0\le 1$, prevents singular behaviors, shrinking the area of integration in the  space $z'\otimes z''$, when $z'\to z''$.
Interestingly, in the Appendix \ref{norm_app}, it is shown that only  ${\cal F}^{i0}_{\ell j} (\gamma, z)$, i.e.  without derivative of  the delta-function, contributes to the norm of the twist-2 uTMD $f_1(\gamma,\xi)$.

\section{The leading-twist uTMD $f^{S(AS)}_1(\gamma,\xi)$}
\label{LTMD_app}
In this Appendix, the symmetric and anti-symmetric combinations of the quark and antiquark  leading-twist uTMDs are explicitly given and their relevant features discussed.

By specializing  the expressions in Eq.~\eqref{Eq:tmd_t2}, 
one can write 
 \be 
f^{S(AS)}_1(\gamma,\xi)={\cal I}_N(\gamma,\xi;S(AS))+{\cal I}_d(\gamma,\xi;S(AS))+{\cal I}_{2d}(\gamma,\xi;S(AS))+{\cal I}_{3d}(\gamma,\xi;S(AS))
\ee 
where the  four contributions are obtained from Eqs. \eqref{calF0b_app}, \eqref{calF1b_app},  \eqref{calF2b_app} and \eqref{calF3b_app}, respectively.  Inserting  the functions $b^0_{n;\ell j}(\gamma,z;S)$  listed in  in the first three columns of Table \ref{b0_tab} in Appendix \ref{trace_app}, one gets the following non vanishing symmetric contributions, viz.
\be 
{\cal I}_N(\gamma,\xi;S)
 =~{3N_c\over 2\pi^2}
\int_{-1}^{+1} dz' \int_0^{\infty} d\gamma' \int_{-1}^{+1} dz'' 
\int_0^{\infty} d\gamma''  
 v_0^2(1-v_0)^2~{\Theta(z'-z)~\Theta(z-z'')
\over (z'-z'')~ [-\beta_0(z^2)+i\epsilon]^5}
\nonu
\times~
\Biggl\{\Bigl[\bar{\cal G}_{11}(\gamma', z';\gamma'', z'')+
 \bar{\cal G}_{22}(\gamma', z';\gamma'', z'')
-4{m\over M}\bar{\cal G}_{12}(\gamma', z';\gamma'', z'')\Bigr]
\nonu
+
{\beta_0(z^2)+8\gamma\over 8M^2}
 ~
\Bigl[\bar{\cal G}_{33}(\gamma', z';\gamma'', z'')
+\bar{\cal G}_{44}(\gamma', z';\gamma'', z'')
-4\bar{\cal G}_{14}(\gamma', z';\gamma'', z'')
\Bigr]\Biggr\}~,
\label{calINcS_app}\ee
\be
{\cal I}_d(\gamma,\xi;S)
 =-~{3N_c\over 4\pi^2 M^2 }~{\partial\over \partial z}\Biggl\{ \int_{-1}^{+1} dz' \int_0^{\infty} d\gamma'
  \int_{-1}^{+1} dz'' \int_0^{\infty} d\gamma''  ~
 v_0^2(1-v_0)^2~{\Theta(z'-z)~\Theta(z-z'')
\over (z'-z'')~
[-\beta_0(z^2)+i\epsilon]^4}
\nonu
\times ~\Bigl[2{m\over M}~\bar{\cal G}_{13}(\gamma', z';\gamma'', z'')+z ~\bar{\cal G}_{14}(\gamma', z';\gamma'', z'')
-\bar{\cal G}_{23}(\gamma', z';\gamma'', z'')
 \Bigr]\Biggr\}~,
 \label{calIdcS_app}
 \ee
 and
 \be 
 {\cal I}_{2d}(\gamma,\xi;S) 
={N_c\over 8\pi^2 M^4}~{\partial^2\over \partial z^2}\Biggl\{  \int_{-1}^{+1} dz' \int_0^{\infty} d\gamma' 
 \int_{-1}^{+1} dz'' \int_0^{\infty} d\gamma''   ~   
 v_0^2(1-v_0)^2~{\Theta(z'-z)~\Theta(z-z'')
\over (z'-z'')~  [-\beta_0(z^2)+i\epsilon]^3}
 \nonu
\times~\Bigl[\bar{\cal G}_{33}(\gamma', z';\gamma'', z'')+\bar{\cal G}_{44}(\gamma', z';\gamma'', z'')
 \Bigr]\Biggr\}~,
\label{calI2dcS_app}
\ee
 with $\beta_0(z^2)$,  $\bar{\cal G}_{\ell j}$ and $v_0$ given in Eq.~\eqref{calgm_app}. 
The symmetry property under the transformation $z\to -z$ can be easily demonstrated, recalling also that under the exchange $z'\to -z''$ and $\gamma'\to \gamma''$ the functions  $\bar{\cal G}_{\ell j}(\gamma', z';\gamma'', z'')$ do not change, since  the 
 NWFs $g_i(\gamma,z;\kappa^2)$ are even for $i=1,2,4$ and odd for $i=3$. Moreover,  under $z\to -z$ and $z'\to -z''$ one also has $v_0\to (1-v_0)$, so that $\beta_0(z^2)$ remains unchanged, as well as $\Theta(z'-z)~\Theta(z-z'')
/ (z'-z'')$.

The anti-symmetric combinations are
\be 
{\cal I}_N(\gamma,\xi;AS)
 ={3 \, N_c \over 2\pi^2 }~  
\int_{-1}^{+1} dz' \int_0^{\infty} d\gamma' \int_{-1}^{+1} dz'' 
\int_0^{\infty} d\gamma'' ~{\Theta(z'-z)\Theta(z-z'')\over z'-z''}  \nonu
\times 
~ {v_0^{2}(1-v_0)^2\over [- {{\beta_0}}(z^2)+i\epsilon]^5}~
\Biggl\{ z \, \bar{\cal G}_{11}(\gamma', z';\gamma'', z'') + z \,   \bar{\cal G}_{22}(\gamma', z';\gamma'', z'') 
\nonu
+ {\beta_0(z^2)+ 8 \gamma\over 2 M^2} \Bigl[- \bar{\cal G}_{23}(\gamma', z';\gamma'', z'') 
+ {z \over 4} \bar {\cal G}_{33}(\gamma', z';\gamma'', z'') \nonu
+ {m \over M } \bar{\cal G}_{34}(\gamma', z;\gamma'', z'') + {z \over 4} \bar{\cal G}_{44}(\gamma', z';\gamma'', z'')
\Bigr]\Biggr\}~,
\label{calI0dcAS_app}
\ee
\be 
{\cal I}_{d}(\gamma,\xi;AS)
  =  {N_c\over 2 M^2 \pi^2} {\partial\over \partial z} ~\Biggl\{
   \int_{-1}^{+1} dz' \int_0^{\infty} d\gamma'
  \int_{-1}^{+1} dz'' \int_0^{\infty} d\gamma'' ~{\Theta(z'-z)\Theta(z-z'')\over z'-z''} ~  {v_0^{2}(1-v_0)^2~\over
[-{{\beta_0}}(z^2) + i\epsilon]^4} \nonu
\times \Biggl[ - {3\over 2} \,   \bar {\cal G}_{14}(\gamma', z';\gamma'', z'') -  3 \, \bar {\cal G}_{22}(\gamma', z';\gamma'', z'') + 3  {z \over 2}   \bar {\cal G}_{23}(\gamma', z';\gamma'', z'') + 3  {m \over M}  \bar  {\cal G}_{24}(\gamma', z';\gamma'', z'') \nonu
- {\beta_0(z^2)+  3 \, \gamma\over M^2 }  \bar {\cal G}_{33}(\gamma', z';\gamma'', z'') 
+ {\beta_0(z^2)\over 2 \, M^2 }  \bar {\cal G}_{44}(\gamma', z';\gamma'', z'')\Biggr]
\Biggr\}
 \label{calI1dcAS_app}
\ee
\be 
{\cal I}_{2d}(\gamma,\xi;AS)
 ={N_c \over 8 \pi^2 M^4}~{\partial^2\over \partial z^2} ~ \Biggl\{
  \int_{-1}^{+1} dz' \int_0^{\infty} d\gamma' 
 \int_{-1}^{+1} dz'' \int_0^{\infty} d\gamma'' ~{\Theta(z'-z)\Theta(z-z'')\over z'-z''} ~{ v_0^{2}(1-v_0)^2 \over  [-{{\beta_0}}(z^2)+i\epsilon]^3}
 \nonu
\times ~   
 \Biggl[ -8  \, \bar {\cal G}_{23}(\gamma', z';\gamma'', z'') + z \,  \bar {\cal G}_{33}(\gamma', z';\gamma'', z'') + 4 {m\over M} \bar {\cal G}_{34}(\gamma', z';\gamma'', z'')  
 + z  \, \bar {\cal G}_{44}(\gamma', z';\gamma'', z'')  \Biggr]\Biggr\}
 \label{calI2dcAS_app}
\ee
\be 
{\cal I}_{3d}(\gamma,\xi;AS)
 =- {N_c \over 4\pi^2 M^6}~{\partial^3\over \partial z^3} ~\Biggl\{ 
   \int_{-1}^{+1} dz' \int_0^{\infty} d\gamma' 
 \int_{-1}^{+1} dz'' \int_0^{\infty} d\gamma''  ~{\Theta(z'-z)\Theta(z-z'')\over z'-z''}  \nonu
\times ~   
 { v_0^{2}(1-v_0)^2\over  [-{{\beta_0}}(z^2)+i\epsilon]^2}
 ~ \bar{\cal G}_{33}(\gamma', z';\gamma'', z'')\Biggr\}
 \label{calI3dcAS_app}
\ee
The anti-symmetry with respect to the transformation $z\to -z$ can be easily shown by using the properties listed below Eq. \eqref{calI2dcS_app}.
\ewt
\subsection{The normalization of $f^S_1(\gamma,\xi)$}
\label{norm_app}
While the integration on $\xi$ and $\gamma$ of $f^{AS}_1(\gamma,\xi)$ trivially yields zero, since the  anti-symmetry in $z$  translates in an anti-symmetry in  $\xi$ with respect to $\xi=1/2$, it is interesting to analyze   how to recover the  normalization of $f^S_1(\gamma,\xi)$, once the BS-amplitude is properly normalized as  in Eq. \eqref{Eq:norm}. To proceed in the most easy way, let us perform a step backward, and reinsert the dependence upon $\delta(z-\lambda(v))$ in  Eqs.  \eqref{calINcS_app},  \eqref{calIdcS_app} and   \eqref{calI2dcS_app} by using Eq. \eqref{vint_app}.
Then one has
\bwt
\be 
{\cal I}_N(\gamma,\xi;S)=
~{3N_c\over 4\pi^2}
\int_{0}^{1} dv ~ v^{2}(1-v)^2  
\int_{-1}^{+1} dz' \int_0^{\infty} d\gamma' \int_{-1}^{+1} dz'' 
\int_0^{\infty} d\gamma''  \nonu
\times 
~ {\delta (\lambda(v)-z)~\over [-\beta_0(z^2)+i\epsilon]^5}~
\Biggl\{\Bigl[{\cal G}_{11}(\gamma', z';\gamma'', z'')+
 {\cal G}_{22}(\gamma', z';\gamma'', z'')
-4{m\over M}{\cal G}_{12}(\gamma', z';\gamma'', z'')\Bigr]
\nonu
+
{\beta_0(z^2)+8\gamma\over 8M^2}
 ~
\Bigl[{\cal G}_{33}(\gamma', z';\gamma'', z'')
+{\cal G}_{44}(\gamma', z';\gamma'', z'')
-4{\cal G}_{14}(\gamma', z';\gamma'', z'')
\Bigr]\Biggr\}~,
\label{calINa_app}
\ee
 \be
 {\cal I}_d(\gamma,\xi;S)
=-~{3N_c\over 8\pi^2 M^2 }~{\partial\over \partial z}\Biggl\{\int_{0}^{1} dv ~
 v^{2}(1-v)^2  \int_{-1}^{+1} dz' \int_0^{\infty} d\gamma'
  \int_{-1}^{+1} dz'' \int_0^{\infty} d\gamma''  \nonu
\times ~
 {\delta (\lambda(v)-z)~\over
[-\beta_0(z^2))+i\epsilon]^4}~\Bigl[2{m\over M}~{\cal G}_{13}(\gamma', z';\gamma'', z'')+z ~{\cal G}_{14}(\gamma', z';\gamma'', z'')
-{\cal G}_{23}(\gamma', z';\gamma'', z'')
 \Bigr]\Biggr\}
 \label{calIda_app}\ee
 and
\be
 {\cal I}_{2d}(\gamma,\xi;S)
 ={N_c\over 16\pi^2 M^4}~{\partial^2\over \partial z^2}\Biggl\{\int_{0}^{1} dv ~ 
 v^{2}(1-v)^2  \int_{-1}^{+1} dz' \int_0^{\infty} d\gamma' 
 \int_{-1}^{+1} dz'' \int_0^{\infty} d\gamma''  \nonu
\times ~   
 { \delta (\lambda(v)-z)\over  [-\beta_0(z^2)+i\epsilon]^3}
 ~\Bigl[{\cal G}_{33}(\gamma', z';\gamma'', z'')+{\cal G}_{44}(\gamma', z';\gamma'', z'')
 \Bigr]\Biggr\}~.
\label{calI2da_app}\ee

Performing the integration on $\gamma$ and $\xi=(1-z)/2$, one gets the following results.
 From  Eq. \eqref{calINa_app}, one recovers the  standard normalization of the BS-amplitude in ladder approximation (cf. Eq. (12) in Ref. \cite{dePaula:2020qna}), viz.
  \be
\int_{-\infty}^\infty d\xi \int_0^\infty d\gamma~{\cal I}_N(\gamma,\xi;S)
=-{3N_c\over 32\pi^2 }
~\int_{0}^{1} dv ~ v^{2}(1-v)^2  
\int_{-1}^{+1} dz' \int_0^{\infty} d\gamma' \int_{-1}^{+1} dz'' 
\int_0^{\infty} d\gamma''  \nonu
\times 
~ {1~\over [\kappa^2+{M^2\over 4}  z^2+v \gamma'+(1-v)\gamma'']^4}~
\Biggl\{\Bigl[{\cal G}_{11}(\gamma', z';\gamma'', z'')+
 {\cal G}_{22}(\gamma', z';\gamma'', z'')
-4{m\over M}{\cal G}_{12}(\gamma', z';\gamma'', z'')
\Bigr]
\nonu
+{\kappa^2+{M^2\over 4}z^2+v\gamma' +(1-v)\gamma''\over 2M^2}~
\Bigl[{\cal G}_{33}(\gamma', z';\gamma'', z'')
+{\cal G}_{44}(\gamma', z';\gamma'', z'')
-4{\cal G}_{14}(\gamma', z';\gamma'', z'')
\Bigr]\Biggr\}~,
\label{calINb_app}\ee
while the other two terms do not contribute.  In fact, let us first  integrate on $z$ and take into account that in $ \delta (\lambda(v)-z)$ one has $1\ge
\lambda(v)\ge -1$. One gets for Eq. \eqref{calIda_app}
\be
\int_{-\infty}^\infty d\xi \int_0^\infty d\gamma~ {\cal I}_d(\gamma,\xi;S)
 =-{3\over 16\pi^2 M^2 }~\int_0^\infty d\gamma\Biggl[\int_{0}^{1} dv ~
 v^{2}(1-v)^2  \int_{-1}^{+1} dz' \int_0^{\infty} d\gamma'
  \int_{-1}^{+1} dz'' \int_0^{\infty} d\gamma''  \nonu
\times ~{\delta (\lambda(v)-z)~\over
[-\beta(z^2)+i\epsilon]^4}~
 \Bigl[2{m\over M}~{\cal G}_{13}(\gamma', z';\gamma'', z'')+z ~{\cal G}_{14}(\gamma', z';\gamma'', z'')
-{\cal G}_{23}(\gamma', z';\gamma'', z'')
 \Bigr]\Biggr]_{z=-\infty}^{z=+\infty}=0~.
 \label{calIdb_app}\ee  
For Eq \eqref{calI2da_app} one has
\be
\int_{-\infty}^\infty d\xi \int_0^\infty d\gamma~ {\cal I}_{2d}(\gamma,\xi;S)
 ={1\over 32\pi^2 M^4}~\int_0^\infty d\gamma \Biggl[{\partial\over \partial z}\int_{0}^{1} dv ~ 
 v^{2}(1-v)^2  \int_{-1}^{+1} dz' \int_0^{\infty} d\gamma' 
 \int_{-1}^{+1} dz'' \int_0^{\infty} d\gamma''  \nonu
\times ~   
 { \delta (\lambda(v)-z)\over  [-\beta(z\lambda(v))+i\epsilon]^3}
 ~\Bigl[{\cal G}_{33}(\gamma', z';\gamma'', z'')+{\cal G}_{44}(\gamma', z';\gamma'', z'')
 \Bigr]\Biggr]_{z=-\infty}^{z=+\infty}=
 \nonu
 ={1\over 32\pi^2 M^4}~\int_0^\infty d\gamma \Biggl[{\partial\over \partial z}
  \int_{-1}^{+1} dz' \int_0^{\infty} d\gamma' 
 \int_{-1}^{+1} dz'' \int_0^{\infty} d\gamma'' ~ {v^{2}_0(1-v_0)^2 \over z'-z''} \nonu
\times ~  
 { \Theta(z'-z)\Theta(z-z'')-\Theta(z''-z)\Theta(z-z')\over 
  [-\beta_0(z^2)+i\epsilon]^3}
 ~\Bigl[{\cal G}_{33}(\gamma', z';\gamma'', z'')+{\cal G}_{44}(\gamma', z';\gamma'', z'')
 \Bigr]\Biggr]_{z=-\infty}^{z=+\infty} ~,
\label{calI2db_app}
\ee
where in the last step Eq. \eqref{vint_app} has been used. Moreover, by explicitly performing the derivative on $z$, given by (recall  $\beta_0(z^2)
 =\gamma+\kappa^2+z^2{M^2/ 4}+v_0\gamma' +(1-v_0)\gamma''$)
\be
{\partial\over \partial z} 
\Biggl[ {v^{2}_0(1-v_0)^2\over [-\beta_0(z^2)+i\epsilon]^3}
\Bigl(\Theta(z'-z)\Theta(z-z'')-\Theta(z''-z)\Theta(z-z')\Bigr) \Biggr]=
\nonu
={\partial\over \partial z} 
\Biggl[ {v^{2}_0(1-v_0)^2\over [-\beta_0(z^2)+i\epsilon]^3}\Biggr]~
\Bigl(\Theta(z'-z)\Theta(z-z'')-\Theta(z''-z)\Theta(z-z')\Bigr) 
 + 
 {v^{2}_0(1-v_0)^2\over [-\beta_0(z^2)+i\epsilon]^3}
\nonu \times~\Bigl(-\delta(z'-z)\Theta(z-z'')+ \Theta(z'-z)\delta(z-z'') 
+\delta(z''-z)\Theta(z-z')-\Theta(z''-z)\delta(z-z')\Bigr) \Biggr]=
\nonu
={\partial\over \partial z} 
\Biggl[ {v^{2}_0(1-v_0)^2\over [-\beta_0(z^2)+i\epsilon]^3}\Biggr]~
\Bigl(\Theta(z'-z)\Theta(z-z'')-\Theta(z''-z)\Theta(z-z')\Bigr) 
\nonu + 
 {v^{2}_0(1-v_0)^2\over [-\beta_0(z^2)+i\epsilon]^3}
\Bigl(-\delta(z'-z)+ \delta(z-z'') \Bigr) \Biggr]~~,
\label{deriv_app}\ee 
 one can straightforwardly  see that the derivative  vanishes for $z=\pm \infty$, being $z'$ and $z''$  $\in[-1,1]$ 
\ewt
Hence
\be
\int_{-\infty}^\infty d\xi \int_0^\infty d\gamma~ {\cal I}_{2d} (\gamma,\xi;S)=0 \, . 
\label{calI2dc_app}\ee
Two comments are in order. First, the leading-twist uTMD is vanishing outside the range $\xi\in[0,1]$, and hence one can restrict the integration on $z$ between $[-1,1]$. It is easy to prove that the same results can be obtained also in this case, recalling that $z'$ and $z''$ are in the same range, and in the last line of Eq. \eqref{deriv_app} one has $v_0=(z-z')/(z'-z'')$ and $1-v_0=(z''-z)/(z'-z'')$.
Second, the integrand in Eq. \eqref{calIdb_app} and \eqref{calI2db_app} should lead to  contributions to the transverse distribution 
\be 
D_\perp(\gamma)=\int_0^\infty d\xi~f^S_1(\gamma,\xi)~~,
\ee
but  from the above results one can see that they are vanishing.
\section{The parton distribution function and  the leading-twist uTMD}
\label{pdf_app}
By integrating $f^{S}_1(\gamma,\xi)$ on $\gamma$ one gets the symmetric parton distribution function $u^S(\xi)$. In particular, one has
\be
u^S(\xi)=\int_0^\infty d\gamma~f^S_1(\gamma,\xi)
\nonu 
=u^S_N(\xi)+u^S_d(\xi)+u^S_{2d}(\xi)
\ee
where the three contributions are obtained  by integrating  on $\gamma$ of the three quantities ${\cal I}_N(\gamma,\xi;S)$, ${\cal I}_d(\gamma,\xi;S)$ and ${\cal I}_{2d}(\gamma,\xi;S)$ given in Eqs. \eqref{calINa_app}, \eqref{calIda_app} and \eqref{calI2da_app}, respectively. By using the result in Eq. \eqref{vint_app} and the integrals
\bwt
\be
\int_0^\infty d\gamma~{1\over [-\beta_0(z^2)+i\epsilon]^n}
= {(-1)^n\over n-1} ~
{1\over [D(z,v_0,\gamma',\gamma'')]^{n-1}}~,
\nonu 
\int_0^\infty d\gamma~{\gamma\over [-\beta_0(z^2)+i\epsilon]^4}
={1\over 6} ~{1\over [D(z,v_0,\gamma',\gamma'')]^2}~,
\nonu
\int_0^\infty d\gamma~{\gamma\over [-\beta_0(z^2)+i\epsilon]^5}
=
-{1\over 12}~{1\over [D(z,v_0,\gamma',\gamma'')]^3}~,
\ee
where 
\be
D(z,v_0,\gamma',\gamma'')=\kappa^2+{M^2\over 4}  z^2+v_0 \gamma'+(1-v_0)\gamma''~, \nonu
\ee
one writes
\be
u^S_N(\xi)=\int_0^\infty d\gamma~{\cal I}_N(\gamma,\xi;S)
=-{3\over 8\pi^2 }  
\int_{-1}^{+1} dz'  
\int_0^{\infty} d\gamma'\int_{-1}^{+1} dz'' 
\int_0^{\infty} d\gamma''  ~
{\Theta(z'-z)\Theta(z-z'')\over z'-z''}
~ {v^2_0(1-v_0)^2\over [D(z,v_0,\gamma',\gamma'')]^4 }
 \nonu \times~
 \Biggl\{\Bigl[\bar{\cal G}_{11}(\gamma', z';\gamma'', z'')
 +
 \bar{\cal G}_{22}(\gamma', z';\gamma'', z'')
-4{m\over M}\bar{\cal G}_{12}(\gamma', z';\gamma'', z'')\Bigr]
\nonu +{D(z,v_0,\gamma',\gamma'')\over 2M^2}
 ~
\Bigl[\bar{\cal G}_{33}(\gamma', z';\gamma'', z'')
+\bar{\cal G}_{44}(\gamma', z';\gamma'', z'')
-4\bar{\cal G}_{14}(\gamma', z';\gamma'', z'')\Bigr]
\Biggr\}~,
\label{uN_app}\ee
\be
 u^S_d(\xi)=\int_0^\infty d\gamma~{\cal I}_d(\gamma,\xi;S)
=-{1\over 4\pi^2 M^2 }
~{\partial\over \partial z}
   \int_{-1}^{+1} dz' 
  \int_0^{\infty} d\gamma'
  \int_{-1}^{+1} dz'' \int_0^{\infty} d\gamma''
  ~{\Theta(z'-z)\Theta(z-z'')\over z'-z''}
 \nonu \times~
 { v^2_0(1-v_0)^2\over [D(z,v_0,\gamma',\gamma'')]^3}~
 \Bigl[2{m\over M}~\bar{\cal G}_{13}(\gamma', z';\gamma'', z'')
+z ~\bar{\cal G}_{14}(\gamma', z';\gamma'', z'')
-\bar{\cal G}_{23}(\gamma', z';\gamma'', z'')\Bigr]~,
 \label{ud_app}\ee
and  
\be
 u^S_{2d}(\xi)=\int_0^\infty d\gamma~{\cal I}_{2d}(\gamma,\xi;S)=
 -{1\over 16\pi^2 M^4}
  ~
 {\partial^2\over \partial z^2}
 \int_{-1}^{+1} dz' 
  \int_0^{\infty} d\gamma' 
 \int_{-1}^{+1} dz'' \int_0^{\infty} d\gamma''~{\Theta(z'-z)\Theta(z-z'')\over z'-z''}   
\nonu ~\times { v^2_0(1-v_0)^2\over  
 [D(z,v_0,\gamma',\gamma'')]^2 }\Bigl[ \bar{\cal G}_{33}(\gamma', z';\gamma'', z'')
 +\bar{\cal G}_{44}(\gamma', z';\gamma'', z'') 
 \Bigr]~.
\label{u2d_app}\ee
\ewt

If the BS-amplitude has the standard  normalization \cite{Lurie}, after integrating $u^S_N(\xi)$ one gets
\be
\int_0^{+1}d\xi~ u^S(\xi)=\int_0^{+1}d\xi~ u^S_N(\xi)=1
\ee
from i) Eq. \eqref{calINb_app}, \eqref{calIdb_app} and \eqref{calI2dc_app}   
and ii) Eq. (12) in Ref. \cite{dePaula:2020qna}.

 The anti-symmetric PDF $u^{AS}(\xi)$ is given by
\bwt 
\be 
u^{AS}(\xi)=\int_0^\infty d\gamma~f^{AS}_1(\gamma,\xi)
=u^{AS}_N(\xi)+u^{AS}_d(\xi)+u^{AS}_{2d}(\xi)+u^{AS}_{3d}(\xi)
\ee
where
{
\be 
u^{AS}_N(\xi)=\int_0^\infty d\gamma~{\cal I}_N(\gamma,\xi;AS)=
-{3 \, N_c \over 8\pi^2 }~  
\int_{-1}^{+1} dz' \int_0^{\infty} d\gamma' \int_{-1}^{+1} dz'' 
\int_0^{\infty} d\gamma'' ~{\Theta(z'-z)\Theta(z-z'')\over z'-z''}  \nonu
\times 
~ {v_0^{2}(1-v_0)^2\over [D(z,v_0,\gamma',\gamma'')]^4}~
\Biggl\{ z \, \bar{\cal G}_{11}(\gamma', z';\gamma'', z'') + z \,   \bar{\cal G}_{22}(\gamma', z';\gamma'', z'') 
+ 2{D(z,v_0,\gamma',\gamma'')\over  M^2} 
\nonu \times ~\Bigl[{z \over 4} \bar {\cal G}_{33}(\gamma', z';\gamma'', z'') + {z \over 4} \bar{\cal G}_{44}(\gamma', z';\gamma'', z'')- \bar{\cal G}_{23}(\gamma', z';\gamma'', z'') 
+ {m \over M } \bar{\cal G}_{34}(\gamma', z;\gamma'', z'') 
\Bigr]\Biggr\}~,\ee
\be 
u^{AS}_d(\xi)=\int_0^\infty d\gamma~{\cal I}_d(\gamma,\xi;AS)=
{N_c\over 6 M^2 \pi^2} {\partial\over \partial z} ~\Biggl\{
   \int_{-1}^{+1} dz' \int_0^{\infty} d\gamma'
  \int_{-1}^{+1} dz'' \int_0^{\infty} d\gamma'' ~{\Theta(z'-z)\Theta(z-z'')\over z'-z''}
  \nonu \times ~  {v_0^{2}(1-v_0)^2~\over
[D(z,v_0,\gamma',\gamma'')]^3} ~
 \Biggl[ - {3\over 2} \,   \bar {\cal G}_{14}(\gamma', z';\gamma'', z'') \nonu
 -  3 \, \bar {\cal G}_{22}(\gamma', z';\gamma'', z'') + 3  {z \over 2}   \bar {\cal G}_{23}(\gamma', z';\gamma'', z'') + 3  {m \over M}  \bar  {\cal G}_{24}(\gamma', z';\gamma'', z'') 
- 3{D(z,v_0,\gamma',\gamma'') \over M^2 }  \bar {\cal G}_{33}(\gamma', z';\gamma'', z'') \nonu
+ 3{D(z,v_0,\gamma',\gamma'') \over 4 \, M^2 }  \bar {\cal G}_{44}(\gamma', z';\gamma'', z'')\Biggr]
\Biggr\}~,\ee
\be 
u^{AS}_{2d}(\xi)=\int_0^\infty d\gamma~{\cal I}_{2d}(\gamma,\xi;AS)=
-{N_c \over 16 \pi^2 M^4}~{\partial^2\over \partial z^2} ~ \Biggl\{
  \int_{-1}^{+1} dz' \int_0^{\infty} d\gamma' 
 \int_{-1}^{+1} dz'' \int_0^{\infty} d\gamma'' ~{\Theta(z'-z)\Theta(z-z'')\over z'-z''} \nonu
\times ~   
 { v_0^{2}(1-v_0)^2 \over  [D(z,v_0,\gamma',\gamma'')]^2}
 ~\Biggl[ -8  \, \bar {\cal G}_{23}(\gamma', z';\gamma'', z'') + z \,  \bar {\cal G}_{33}(\gamma', z';\gamma'', z'') + 4 {m\over M} \bar {\cal G}_{34}(\gamma', z';\gamma'', z'')  
  \nonu 
 + z  \, \bar {\cal G}_{44}(\gamma', z';\gamma'', z'')  \Biggr]\Biggr\}
~,\ee
and
\be 
u^{AS}_{3d}(\xi)=\int_0^\infty d\gamma~{\cal I}_{3d}(\gamma,\xi;AS)=
- {N_c \over 4\pi^2 M^6}~{\partial^3\over \partial z^3} ~\Biggl\{ 
   \int_{-1}^{+1} dz' \int_0^{\infty} d\gamma' 
 \int_{-1}^{+1} dz'' \int_0^{\infty} d\gamma''  ~{\Theta(z'-z)\Theta(z-z'')\over z'-z''}  \nonu
\times ~   
 { v_0^{2}(1-v_0)^2\over  D(z,v_0,\gamma',\gamma'')}
 ~ \bar{\cal G}_{33}(\gamma', z';\gamma'', z'')\Biggr\} 
 ~.\ee}
\section{Twist-3 unpolarized TMDs}
\label{twist34_app}
The Appendix presents the explicit expressions of the twist-3 and twist-4 uTMDs, obtained from Eq. \eqref{Eq:tmd_t2} and Eqs. \eqref{calF0b_app}, \eqref{calF1b_app}, \eqref{calF2b_app} and \eqref{calF3b_app}, by using the Tables \ref{b1_tab} and \ref{b2_tab}.
In particular for the twist-3 $e^{S(AS)}(\gamma,\xi)$, i.e. for $i=1$ in eq. \eqref{Eq:tmd_t2}, one has
\be
e^{S(AS)}(\gamma,\xi)=
{\cal E}_0(\gamma,\xi;S(AS)) + {\cal E}_d(\gamma,\xi;S(AS)) + {\cal E}_{2d}(\gamma,\xi;S(AS))  +  {\cal E}_{3d}(\gamma,\xi;S(AS)) 
\ee
where the symmetric combinations are
\be 
{\cal E}_0(\gamma,\xi;S)=  {3N_c\over 2\pi^2}
\int_0^{\infty} d\gamma' \int_0^{\infty} d\gamma''
\int_{-1}^{+1} dz'\int_{-1}^{+1} dz''  ~v_0^{2}(1-v_0)^2~
{\Theta(z'-z)\Theta(z-z'')\over (z'-z'')~[-\beta_0(z^2)+i\epsilon]^5}
\nonu
\times 
~ \{ -2 {m\over M} \bar{\cal G}_{11}(\gamma', z';\gamma'', z'') + 2  \bar{\cal G}_{12}(\gamma', z';\gamma'', z'') -2 {m\over M}  \bar{\cal G}_{22}(\gamma', z';\gamma'', z'') 
\nonu +
{8\gamma +\beta_0(z^2)\over 2 M^2} \Bigl[- \bar{\cal G}_{24}(\gamma', z';\gamma'', z'') 
+ {m \over 2 M}\bar {\cal G}_{33}(\gamma', z';\gamma'', z'')
+ {z \over 2 } \bar{\cal G}_{34}(\gamma', z;\gamma'', z'') + {m \over 2 \, M} \bar{\cal G}_{44}(\gamma', z';\gamma'', z'')
\Bigr]\Biggr\} ~,
\label{calE0S_app} \nonu
\ee
\be
{\cal E}_d(\gamma,\xi;S)
  = - {N_c\over 4 M^4 \pi^2} {\partial\over \partial z} ~
   \int_{-1}^{+1} dz' \int_0^{\infty} d\gamma'
  \int_{-1}^{+1} dz'' \int_0^{\infty} d\gamma''~v_0^{2}(1-v_0)^2 ~{\Theta(z'-z)\Theta(z-z'')\over (z'-z'') [-\beta_0(z^2) + i\epsilon]^4} \nonu
\times ~ \Bigl[6\gamma +\beta_0(z^2) \Bigr]  ~
  \bar {\cal G}_{34}(\gamma', z';\gamma'', z'')~,
 \label{calE1S_app}
 \ee
 \be 
 {\cal E}_{2d}(\gamma,\xi;S) 
 ={N_c \over 4 \pi^2 M^4}~{\partial^2\over \partial z^2} ~ 
  \int_{-1}^{+1} dz' \int_0^{\infty} d\gamma' 
 \int_{-1}^{+1} dz'' \int_0^{\infty} d\gamma'' ~v_0^{2}(1-v_0)^2 ~{\Theta(z'-z)\Theta(z-z'')\over (z'-z'') [-\beta_0(z^2)+i\epsilon]^3} \nonu
\times ~   
\Bigl[ -2  \, \bar{\cal G}_{24}(\gamma', z';\gamma'', z'') +  {m\over M} ~ \bar{\cal G}_{33}(\gamma', z';\gamma'', z'')
+ z  \, \bar{\cal G}_{34}(\gamma', z';\gamma'', z'') +  {m\over M}   \, \bar{\cal G}_{44}(\gamma', z';\gamma'', z'') \Bigr]~,
\label{calE2S_app}
\ee
and 
\be
 {\cal E}_{3d}(\gamma,\xi;S)
 =- {N_c \over 4\pi^2 M^6}~{\partial^3\over \partial z^3} ~ 
   \int_{-1}^{+1} dz' \int_0^{\infty} d\gamma' 
 \int_{-1}^{+1} dz'' \int_0^{\infty} d\gamma'' ~ v_0^{2}(1-v_0)^2 \nonu\times ~{\Theta(z'-z)\Theta(z-z'')\over (z'-z'') [-{\beta_0}(z^2)+i\epsilon]^2}  \bar{\cal G}_{34}(\gamma', z';\gamma'', z'') 
\label{calE3S_app}\ee

The anti-symmetric combinations are
\be 
{\cal E}_0(\gamma,\xi;AS)
 ={3N_c\over 2\pi^2}
\int_0^{\infty} d\gamma' \int_0^{\infty} d\gamma''
\int_{-1}^{+1} dz'\int_{-1}^{+1} dz''  ~v_0^{2}(1-v_0)^2~
{\Theta(z'-z)\Theta(z-z'')\over (z'-z'')~[-\beta_0(z^2)+i\epsilon]^5}
\nonu
\times ~\Biggl\{2z \bar {\cal G}_{12}(\gamma', z'; \gamma'', z'') 
 -{8\gamma +\beta_0(z^2)\over 4M^2}~\Bigl[2\bar {\cal G}_{13 }(\gamma', z'; \gamma'', z'')
 -\bar {\cal G}_{34 }(\gamma', z'; \gamma'', z'')\Bigr]
 \Biggr\}~, 
\label{calE0AS_app}\ee
\be
{\cal E}_d(\gamma,\xi;AS)
 =-{3N_c\over 2\pi^2M^2}  {\partial\over \partial z}\Biggl\{\int_0^{\infty} d\gamma' \int_0^{\infty} d\gamma'' 
\int_{-1}^{+1} dz'  
\int_{-1}^{+1} dz''  v_0^{2}(1-v_0)^2~{\Theta(z'-z)\Theta(z-z'')\over (z'-z'')~[-\beta_0(z^2)+i\epsilon]^4}
\nonu\times~ \bar{\cal G}_{12}(\gamma', z'; \gamma'', z'')\Biggr\}~,
 \label{calE1AS_app}
\ee
\be 
 {\cal E}_{2d}(\gamma,\xi;AS) 
-{N_c\over 4\pi^2M^4}~ {\partial^2\over \partial z^2} \Biggl\{
\int_0^{\infty} d\gamma' \int_0^{\infty} d\gamma''\int_{-1}^{+1} dz' 
\int_{-1}^{+1} dz''  ~v_0^{2}(1-v_0)^2~{\Theta(z'-z)\Theta(z-z'')\over (z'-z'')~[-\beta_0(z^2)+i\epsilon]^3}
\nonu \times ~\Bigl[2 \bar{\cal G}_{13}(\gamma', z'; \gamma'', z'')-\bar{\cal G}_{34}(\gamma', z'; \gamma'', z'')\Bigr] 
\Biggr\}~,
\label{calE2AS_app}
\ee
and 
\be
 {\cal E}_{3d}(\gamma,\xi;AS) = 0~~.
\ee
\subsection{The twist-3 uTMD $f^\perp(\gamma,\xi) $}
For $i=2$, one has  the twist-3 uTMD $f^\perp(\gamma,\xi) $, with the following decomposition
\be 
f^{\perp {S(AS)}}(\gamma,\xi)={\cal P}_0(\gamma,\xi;S(AS))+{\cal P}_d(\gamma,\xi;S(AS))+{\cal P}_{2d}(\gamma,\xi;S(AS)) + {\cal P}_{3d}(\gamma,\xi;S(AS))~.
\ee
The symmetric contributions are given by
\be
{\cal P}_0(\gamma,\xi;S)
= -{3 \, N_c \over \pi^2 }~ 
\int_{-1}^{+1} dz' \int_0^{\infty} d\gamma' \int_{-1}^{+1} dz'' 
\int_0^{\infty} d\gamma''~v_0^{2}(1-v_0)^2  ~{\Theta(z'-z)\Theta(z-z'')\over (z'-z'')[- {{\beta_0}}(z^2)+i\epsilon]^5}
\nonu \times~  \Biggl\{  \bar{\cal G}_{11}(\gamma', z';\gamma'', z'') -  \bar {\cal G}_{14}(\gamma', z';\gamma'', z'') 
-  \bar{\cal G}_{22}(\gamma', z';\gamma'', z'') + z\,  \bar {\cal G}_{23}(\gamma', z';\gamma'', z'') 
\nonu
+  {2 m\over M}  \bar {\cal G}_{24}(\gamma', z';\gamma'', z'') 
- {8\gamma +\beta_0(z^2)\over 8 M^2} \Bigl[ 
\bar{\cal G}_{33}(\gamma', z';\gamma'', z'') -  \bar{\cal G}_{44}(\gamma', z';\gamma'', z'')
\Bigr]\Biggr\}~,
\nonu
\ee
\be
 {\cal P}_d(\gamma,\xi;S) ={ 3 \, N_c  \over 2 \pi^2 \, M^2} {\partial\over \partial z} ~
 \int_{-1}^{+1} dz' \int_0^{\infty} d\gamma'
  \int_{-1}^{+1} dz'' \int_0^{\infty} d\gamma'' ~v_0^{2}(1-v_0)^2 ~{\Theta(z'-z)\Theta(z-z'')\over (z'-z'')[-\beta_0(z^2) + i\epsilon]^4}  \nonu \times ~\bar{\cal G}_{23}(\gamma', z';\gamma'', z'') ~,\nonu
\ee
\be
 {\cal P}_{2d}(\gamma,\xi;S)
 = {N_c \over 4 \pi^2 M^4}~{\partial^2\over \partial z^2} ~ 
   \int_{-1}^{+1} dz' \int_0^{\infty} d\gamma' 
 \int_{-1}^{+1} dz'' \int_0^{\infty} d\gamma'' ~v_0^{2}(1-v_0)^2 ~{\Theta(z'-z)\Theta(z-z'')\over (z'-z'') [-{{\beta_0}}(z^2)+i\epsilon]^3} \nonu
\times ~  
 ~\Bigl[ \bar{\cal G}_{33}(\gamma', z';\gamma'', z'') - \bar {\cal G}_{44}(\gamma', z';\gamma'', z'') \Bigr]~,
\ee
and
\be
{\cal P}_{3d}(\gamma,\xi;S)
= 0
\ee
The anti-symmetric contributions read
\be 
{\cal P}_0(\gamma,\xi;AS)
 ={3N_c\over \pi^2}\int_0^{\infty} d\gamma' \int_0^{\infty} d\gamma''
\int_{-1}^{+1} dz'\int_{-1}^{+1} dz''  ~v_0^{2}(1-v_0)^2~
{\Theta(z'-z)\Theta(z-z'')\over (z'-z'')~ [-\beta_0(z^2)+i\epsilon]^5}
\nonu
\times ~
~\Biggl[2{m\over M}
 \bar {\cal G}_{13}(\gamma', z'; \gamma'', z'')
 +  \, z
 \bar {\cal G}_{14}(\gamma', z'; \gamma'', z'')
 -
 \bar {\cal G}_{23}(\gamma', z'; \gamma'', z'')\Biggr] ~,
\label{calP0_app}\ee
and 
\be
{\cal P}_d(\gamma,\xi;AS)
 =~-{3N_c\over 2\pi^2M^2}  {\partial\over \partial z}\Biggl\{\int_0^{\infty} d\gamma' \int_0^{\infty} d\gamma'' 
\int_{-1}^{+1} dz'  
\int_{-1}^{+1} dz''  v_0^{2}(1-v_0)^2~{\Theta(z'-z)\Theta(z-z'')\over (z'-z'')~[-\beta_0(z^2)+i\epsilon]^4}
\nonu\times~ \bar{\cal G}_{14}(\gamma', z'; \gamma'', z'')\Biggr\}~,
\label{calP1_app}\ee
and 
\be
{\cal P}_{2d}(\gamma,\xi;AS) = {\cal P}_{3d}(\gamma,\xi;AS) = 0 ~.
\ee
\ewt



\end{document}